\documentclass[aps,prf,preprint,groupedaddress]{revtex4-2}
\usepackage{amsmath,amssymb,graphicx,graphics,hyperref,xcolor,float,enumitem}

\begin{document}

\title{Green function and singularities in Stokes flow confined by  cylindrical walls}

\author{Giuseppe Procopio}
\email[]{giuseppe.procopio@uniroma1.it}

\affiliation{Dipartimento di Ingegneria Chimica Materiali Ambiente, Sapienza Università di Roma, via Eudossiana 18, Rome 00184, Italy}

\date{\today}

\begin{abstract}
In this article, the Green function for the Stokes flow in the interior, exterior, and annular regions bounded by cylindrical walls is derived as a function of the pole position and expressed in an invariant form at both the field and pole points. Specifically, the Green function, assuming no-slip boundary conditions, is obtained using a cylindrical harmonic expansion of the Stokes flow within the bitensorial formulation introduced in \cite{procopio_bitensorial_2022}. This formulation allows us to obtain higher-order singularities within the same domains and under the same boundary conditions, such as the confined {\em Couplet} and {\em Stresslet}, by simply differentiating the Green function at its pole. Moreover, the confined {\em Sourcelet} {\color{black}(or point source)} and its associated multipoles are derived from the Green function through a new method that enforces the reciprocal properties of the Stokes flow. The resulting singularities are then employed to address hydrodynamic problems involving active and passive colloids interacting {\color{black} with cylindrical and planar walls, such as sedimenting particles in an annular cylindrical region and between two parallel plane walls, and the attractive or repulsive hydrodynamic forces exerted by cylindrical boundaries on microswimmers.}
\end{abstract}

\maketitle

\section{Introduction}
Stokes flows inside or outside cylindrical boundaries represent hydrodynamic systems of great interest in physical, chemical, and biological sciences and their technological applications. 
The Poiseuille flow in a cylindrical capillary and the Taylor-Couette flow between two concentric rotating cylinders are among the most paradigmatic flows in hydrodynamics.
However, many hydrodynamic systems delimited by cylindrical boundaries involve more complex flows than simple Poiseuille and Taylor-Couette flows, especially whenever colloids are suspended in the fluid. This is the case of several classical equipment for treating colloidal suspensions \cite{happel_1983,guazzelli_2012}, as well as of more recent microdevices, such as those based on Deterministic Lateral Displacement (DLD) \cite{hochstetter_2020} (in which particles are suspended in the fluid around cylindrical pillars) or Hydrodynamic Chromatography (HDC) \cite{striegel_2012} (in which particles move across a microcapillary). Cylindrical geometries are also a valuable model for the hydrodynamics of particles transported in porous media \cite{edam_2011} or in fibrous media such as hydrogels \cite{hansing_2018,mangal_2021}.
Moreover, many experimental setups designed to study the effects of confinement, such as on Brownian motion \cite{mo_2015,mo_2015diss}, microswimming \cite{sipos_2015,secchi_2020,thery_2025}, and colloidal transport \cite{leighton_1987,wang_2025}, employ cylindrical boundaries as one of the simplest confinement configurations.

In order to fully grasp the numerous hydrodynamic phenomena arising in physical systems involving cylindrical walls, it is fundamental to understand the Stokes hydrodynamic interactions with suspended colloids in their proximity. It is well-known that an effective strategy for mathematically modeling Stokes flows caused by suspended colloids is to employ singular solutions of the Stokes problem \cite{kim_1991,pozrikidis_1992}.
For instance, the far field of a sedimenting particle in the fluid can be represented by the unbounded Green function of the Stokes problem, also known as the {\em Stokeslet}.
 The antisymmetric part of the {\em  \color{black} Stokeslet dipole} (rather the so called {\em Couplet} or {\em Rotlet} \cite{blake_fundamental_1974}) provides the far-field generated by a rotating particle, whereas the symmetric part ({\em Stresslet} or {\em Strainlet} \footnote{The dichotomy Couplet and Stresslet have been introduced by Batchelor in \cite{batchelor_1970} with reference to the disturbance flow due to an infinitesimal immersed particle in a linear flow. Specifically, the Couplet refers to the disturbance flow due to a non rotating particle in the ambient flow (hence, introducing an additional torque to the fluid), and the Stresslet refers to the disturbance flow due to a rigid particle in the fluid (hence, introducing and additional stress). Alternatively to the term Couplet, the term Rotlet have been introduced, in the field of active particles, by Chwan and Wu \cite{chwang} for indicating the flow induced by a rotating infinitesimal particle in the fluid. Following this logic, it would be more appropriate to refer to a singularity as a Strainlet when it describes the infinitesimal deformation introduced by an active particle in the fluid, in dichotomy with its antisymmetric counterpart, the Rotlet.}) provides the disturbance flow due to the presence of a particle in a shear ambient flow \cite{lorentz_1907,einstein_1906,einstein_1911,burgers_1938,batchelor_1970} (see the monographs by Kim and Karrila \cite{kim_1991} and by Pozrikidis \cite{pozrikidis_1992} for detailed discussions).
Furthermore, the first derivative of the Green function at its pole effectively models the far field generated by active particles such as microswimmers \cite{pedley_1990,pedley_1992,drescher_fluid_2011,lauga_2020}. 

There is a vast class of geometries of confinement for which the associated Green function and some of the higher order singularities are available in the literature for different boundary conditions. For example, Stokes singularities 
are reported for the fluid bounded by an infinite plane wall \cite{blake_note_1971,blake_fundamental_1974,bickel_hindered_2007,
lopez_dynamics_2014,procopio_bitensorial_2022}, by a semi-infinite plane wall \cite{hasimoto_effect_1983}, between two parallel planes \cite{liron_stokes_1976,hackborn_asymmetric_1990,bhattacharya_image_2002,
mathijssen_hydrodynamics_2016,daddi-moussa-ider_state_2018,fortune_biophysical_2024}, between two planes forming a corner \cite{sono_slow_1976,sano_effect_1978,kim_effect_1983,dauparas_leading-order_2018,sprenger_microswimming_2023}, internal and external to a sphere \cite{oseen_neuere_1927,fuentes_mobility_1988,shail_stokes_1988,
fuentes_mobility_1989,usha_flow_1993,maul_image_1994,shaik_point_2017,
chamolly_stokes_2020,sprenger_towards_2020}, in a cone \cite{kim_slow_1979,kim_slow_1980,blinova_stokes_2014}, between two parallel disks \cite{daddi-moussa-ider_axisymmetric_2020} and near a slender body \cite{tanasijevic_hydrodynamic_2021}. 
Stokes Green functions for the internal, external and annular region of cylindrical walls were also addressed in some previous works \cite{hasimoto_slow_1976,blake_generation_1979,
md_shamsul_alam_slow_1980,fukumoto_slow_1985,liron_stokes_1978}, with some studies accounting for wall elasticity \cite{daddi-moussa-ider_hydrodynamic_2017} and the presence of additional boundaries \cite{blake_generation_1979}.
{\color{black}
However, the existing solutions are still unsatisfactory for many practical and theoretical applications. Specifically, the solutions reported in \cite{hasimoto_slow_1976} for the domain internal to a cylinder and in \cite{md_shamsul_alam_slow_1980} for the domain external to a cylinder are limited to providing only the value of the regular part of the Green function at its own pole (given its utility in studying hydrodynamic interactions between walls and colloids). The solution for the domain internal to a cylinder provided in \cite{liron_stokes_1978} is not expressed in terms of the pole coordinates, and its full expression cannot be straightforwardly derived. Moreover, it contains an inaccuracy that is difficult to detect without redoing the calculations (this is briefly discussed at the end of Section \ref{sec_green_func_1_cil}). In \cite{fukumoto_slow_1985}, the solution internal to an annular cylindrical region is provided, but the limiting cases of a single internal or external cylinder are neither reported nor discussed. In addition, as discussed in Section \ref{sec_green_func_1_cil}, the solution in the domain internal to a cylinder cannot be obtained by taking the limit of vanishing inner cylinder radius in the annular configuration, as this represents a singular limit.
Finally, the solution reported in \cite{blake_generation_1979,daddi-moussa-ider_hydrodynamic_2017} are restricted to axisymmetric configurations. 
Although singularities other than the Green function for flows near cylindrical walls have not been previously reported in the literature (apart from specific axisymmetric cases \cite{blake_generation_1979}), many physical problems require such a broader set of singularities.}
This need is also reflected in the growing recent interest in extending the list of known singular solutions for various confinement geometries, particularly in the study of biological locomotion at the microscale \cite{sprenger_microswimming_2023, mathijssen_hydrodynamics_2016, fortune_biophysical_2024, chamolly_stokes_2020}.
Furthermore, as recently shown in \cite{procopio_hinchkim_2024,procopio_theory_2024}, the knowledge of all the $n$th order singularities associated with a confined fluid can provide the full hydromechanics of a body immersed in it. 

Although a large class of higher order singularities could, in principle, be obtained by differentiating the Green function at its pole, in most of the aforementioned cases concerning different domains, they were obtained independently, reiterating the same strategy used to obtain the Green function with the appropriate boundary conditions. This is mainly due to the difficulty of differentiating the available expressions 
for the Green functions in the literature, where the pole is often considered as a fixed parameter and each orientation in the space is treated separately by solving three different problems.
As shown in \cite{procopio_bitensorial_2022} (by general argumentations) and in \cite{mathijssen_hydrodynamics_2016} (in considering the specific case of a fluid bounded by planar walls), in order to have a differentiable Green function at its pole, it is necessary to consider its bi-invariant nature both at the field and at the pole points even in the case Cartesian coordinate systems were employed. 
Employing a bi-invariant form of the Blake's Green function in the semi-space \cite{blake_note_1971} allowed also
to mathematically handle singularities lying along a generically oriented segment for calculating hydrodynamic resistance onto a spheroid near a plane wall \cite{procopio_theory_2024}
and  to overcome the limit prospected by Liron and Mochon in \cite{liron_stokes_1976} of obtaining the Green function between two parallel planes by multiple reflections \cite{mathijssen_hydrodynamics_2016}.
In addition to the need for representing singular solution in a differentiable form at the pole, a formalism able to describe at the same time both trajectories of active microparticles and the flows due to their motion in the fluid could represent a valuable mathematical tool for better understanding recent experiments on the dynamics of microswimmers near curved surfaces \cite{takagi_hydrodynamic_2014,sipos_hydrodynamic_2015,
secchi_effect_2020,jalaal_interfacial_2022} (for instance, to describe phenomena such as surface trapping \cite{spagnolie_geometric_2015} and optimal microswimmer navigation \cite{daddi-moussa-ider_hydrodynamics_2021}).

One of the objectives of this work is to obtain explicit general expressions for the Green function of the Stokes flow in the domain bounded by cylindrical walls, valid for any value of the cylinder radii, mathematically tractable and functional to the applications, overcoming the main limitations of the existing solutions which are not explicitly reported (as in the cases of \cite{hasimoto_slow_1976,md_shamsul_alam_slow_1980, blake_generation_1979}) or contain some inconsistencies \cite{liron_stokes_1978}.
Furthermore, the present article aims to provide expressions for the Green function and for the higher-order singularities (not available in the literature, such as Couplet, Stresslet, point source and point source dipole) bounded by cylindrical surfaces which are bi-invariant at both the field and pole points, hence allowing to derive any other singularity by simple differentiation and/or combination of the reported solutions. 

With the aim of considering the bi-invariant nature of any bounded singularity in Stokes flow, the bitensorial calculus has been introduced in \cite{procopio_bitensorial_2022} for representing the Stokes flow singularities by distinguishing the entries of bitensors at the field and the pole points and providing a simple and clearer notation for the application of integro-differential operators and for representing singularities lying or moving along curved manifolds. 
In this article, the Green function solution for a fluid bounded externally and/or internally by infinitely long cylindrical boundaries is provided in a bitensorial form, and higher-order singular solutions are derived by its differentiation at the pole or by using the bitensorial properties provided in \cite{procopio_bitensorial_2022}. 
More specifically, the method developed in \cite{brenner_slow_1958} (see also \cite[pp. 71-78]{happel_1983}), for representing Stokes solutions in cylindrical coordinate systems, and used in \cite{liron_stokes_1978} for obtaining the Green function for the fluid internal to cylindrical walls, is recast in a bitensorial form and employed to impose no-slip boundary conditions at the cylindrical surfaces.

The article is organized as follows. In Section \ref{setting}, we set the Green problem for Stokes flow in the bitensorial form, considering a generic coordinate system for the field point and the pole point. Furthermore, higher-order singularities obtained by differentiating the Green function are expressed in bitensorial form. In Section \ref{source_bounded}, it is shown that, by using the bitensorial formalism, it is possible to obtain the singular solution of the Stokes flow due to a concentrated fluid source or sink  (referred to as the {\em Sourcelet}), from the pressure field associated to the Green function for any given bounded domain{\color{black}, provided no-slip boundary conditions are assumed}.
In Section \ref{sec_stokeslet}, the expression for the unbounded Stokeslet in the cylindrical coordinate system at both the field and pole point is provided.
In Section \ref{sec_green_function}, the Green function in the annular, internal and external region of a cylinder is provided and its hydrodynamic aspects (such as the singular limit as the radius of the internal cylinder tends to zero, the back flow due to a sedimenting particle and the difference in the flow between a particle moving near to a cylinder or near to a sphere) are discussed.
In Section \ref{sec_dipole}, {\color{black} the Stokeslet dipole (also referred to as a {\em Stokes doublet} \cite{blake_note_1971} or {\em Stokeslet Doublet} \cite{pozrikidis_1992})} is obtained by differentiating the Stokeslet at its pole. By computing the antisymmetric and symmetric parts, the Couplet and Stresslet near cylindrical boundaries are obtained. In Section \ref{sec_sourcelet_sourcedipole}, the velocity fields due to a Sourcelet and its dipole (referred to as the {\em Sourcelet Dipole}) are provided.
 {\color{black} 
In Section \ref{sec_applications}, the derived singularities are applied to study hydrodynamic interactions between particles and cylindrical walls. Specifically, the drag on sedimenting particles in annular, internal, and external cylindrical domains is analyzed through the regular part of the Green function at its pole. Additionally, the hydrodynamic forces experienced by microswimmers near cylindrical boundaries are evaluated for different orientations, revealing attractive and repulsive effects depending on the microswimmer orientation and position relative to the walls. Equivalent results are obtained for the domain between two parallel planes by taking the limit of the annular region as the cylinder curvature approaches zero.}

\section{Setting of the confined Stokes problem}
\label{setting}
Before focusing on the specific case of cylindrical boundaries,
let us start by considering a generic bounded Stokes problem and formulate the associated singular problems by using invariant expressions . Therefore,
consider the Stokes problem in the domain of the fluid $V_f$ with viscosity $\mu$ and no-slip boundary conditions at the boundaries $\partial V_f$
\begin{equation}
\left\{
\begin{array}{l}
\mu  \, \Delta_x\,  v^b ({\pmb x}) - \nabla^b {p} ({\pmb x}) = -f^b ({\pmb x}) 
\\
[5pt]

\nabla_b\, v^b ({\pmb x})= 0   \quad \text{for}\quad {\pmb x} \in V_f
\\
[5pt]

v^b ({\pmb x})= 0  \quad \text{for}\quad {\pmb x} \in \partial V_f
\end{array} \right.
\label{eq1}
\end{equation}
where $v^b({\pmb x})$, with $b... = 1,2,3$, represent the contravariant entries of the velocity field of the fluid at the point ${\pmb x}$ in a generic coordinate system $Ox^b$ with origin at a point $O$, $f^b$ the contravariant entries for an external force field ${\pmb f}({\pmb x})$ in the same coordinate system, $p({\pmb x})$ the pressure field and the operators $\Delta_x$, $\nabla_b$ and $\nabla^b$ the Laplacian operator, the covariant and contravariant derivative with respect to the coordinates of the point ${\pmb x}$ respectively. In eqs. (\ref{eq1}) and throughout this work, we employ the Einstein summation convention for repeated indices of tensorial quantities and adopt the standard tensor notation, with distinction between covariant and contravariant indices \cite{synge_tensor_1978}.
 
Next, consider a second point, referred to as the pole point in the remainder, belonging to the domain of the fluid  ${\pmb \xi} \in V_f$, with entries expressed in a coordinate system $\Omega\, \xi^\beta$ having origin at the point $\Omega$, with $\beta=1,2,3$. 
By the distributional properties of the tridimensional Dirac delta function $\delta({\pmb x},{\pmb \xi})$  \cite{poisson_motion_2011,kanwal_generalized_2004},
it is possible to express the external force field $f^b({\pmb x})$
in terms of its value at the point ${\pmb \xi}$ as
\begin{equation}
f^b({\pmb x})=
\int_{V_f} g^b_{\, \,  \beta}({\pmb x},{\pmb \xi})\, f^\beta ({\pmb \xi})\,  \delta({\pmb x},{\pmb \xi})
\, \sqrt{g({\pmb \xi})} \,
d^3\xi
\label{eq2}
\end{equation}
where $d^3\xi= d\xi^1\, d\xi^2\, d\xi^3$, $\delta({\pmb x},{\pmb \xi})=\delta(x^1-\xi^1)\, \delta(x^2-\xi^2)\, \delta(x^2-\xi^2)/\sqrt{g({\pmb \xi})}$, with
$g({\pmb \xi})=\rm{det}\left[g_{\alpha\, \beta}(\xi) \right] $ being the determinant
of the metric tensor $g_{\alpha \beta}({\pmb \xi})$ of the coordinate system $\Omega \xi^\alpha$,
and $g^b_{\,\, \beta}({\pmb x},{\pmb \xi}) $
is the parallel propagator transporting a vector from the point ${\pmb \xi}$ to the point ${\pmb x}$. 
In the present case, where an Euclidean space is considered, it is possible to
define the parallel propagator introducing a generic Cartesian coordinate system
 $\{Y_i\} \equiv (Y_1,Y_2,Y_3)$, then
\begin{equation}
g^b_{\,\, \beta}({\pmb x},{\pmb \xi})= \dfrac{\partial x^b}{\partial  Y_i}\,  \dfrac{\partial Y_i}{\partial \xi^\beta}
\label{eq3}
\end{equation}
In the more general case in which a curved space is considered (for instance in the case the flow was defined along a membrane),  $\{Y_i\} \equiv (Y_1,Y_2,Y_3)$ is a local Cartesian coordinate system parallel transported from ${\pmb \xi}$ to ${\pmb x}$ . 
In eqs. (\ref{eq2}), (\ref{eq3}) and in what follows, we adopted a notation with the following index rules: (i) indexes associated with entries in a generic coordinate system of the field point ${\pmb x}$ are represented by the Latin letters $a,b,c=1,2,3$; (ii) indexes associated with entries in a generic coordinate system in at the pole point ${\pmb \xi}$ are represented by Greeks letters $\alpha,\beta, \gamma, ...=1,2,3$; (iii) Cartesian coordinate systems are represented by the letters $h,i,j,k,l=1,2,3$.

The solution of eqs. (\ref{eq1}) is provided by the bitensorial kernels $G^b_{\, \, \beta}({\pmb x},{\pmb \xi})$
 and
$P_{\beta}({\pmb x},{\pmb \xi})$
according to the Ladyzhenskaya volume potential expressions for the solution of non-homogeneous Stokes problems \cite{ladyzhenskaia_mathematical_2014,kim_1991}
\begin{equation}
v^b({\pmb x})=\dfrac{1}{8\pi \mu} \int_{V_f} G^b_{\,\, \beta}({\pmb x},{\pmb \xi}) \, f^\beta({\pmb \xi})\, \sqrt{g({\pmb \xi})}\, d^3\xi ;
\quad
p({\pmb x})=\dfrac{1}{8 \pi} \int_{V_f} P_{\,\, \beta}({\pmb x},{\pmb \xi}) \, f^\beta({\pmb \xi})\, \sqrt{g({\pmb \xi})} 	\, d^3\xi 
\label{eq4}
\end{equation}
By inserting eqs. (\ref{eq2}) and (\ref{eq4}) into eq. (\ref{eq1}), the bitensorial formulation for the Green problem of the Stokes flow reads
\begin{equation}
\left\{
\begin{array}{l}
 \, \Delta_x\, G^b_{\,\, \beta} ({\pmb x},{\pmb \xi}) - \nabla^b P_{\beta}({\pmb x},{\pmb \xi}) = -8 \pi\,  g^b_{\,\, \beta}({\pmb x},{\pmb \xi}) \,\delta({\pmb x},{\pmb \xi})
\\
[5pt]

\nabla_b\, G^b_{\,\, \beta} ({\pmb x},{\pmb \xi}) = 0   \quad \text{for}\quad {\pmb x} \in V_f
\\
[5pt]

G^b_{\,\, \beta} ({\pmb x},{\pmb \xi}) = 0  \quad \text{for}\quad {\pmb x} \in \partial V_f
\end{array} \right.
\label{eq5}
\end{equation}
The Green function $G^b_{\,\, \beta}({\pmb x},{\pmb \xi})$ can be expressed as a superposition of two terms
\begin{equation}
G^b_{\,\, \beta} ({\pmb x},{\pmb \xi}) =
S^b_{\,\, \beta} ({\pmb x},{\pmb \xi}) 
+
W^b_{\,\, \beta} ({\pmb x},{\pmb \xi}) 
\label{eq6}
\end{equation}
where $S^b_{\,\, \beta}({\pmb x},{\pmb \xi})$ is 
the Green function of the Stokes flow in the entire domain $\mathbb{R}^3$, denoted as unbounded {Stokeslet} \cite{pozrikidis_1992,kim_1991}, and the remaining regular part $W^b_{\,\, \beta}({\pmb x},{\pmb \xi})$ is a solution of the problem
\begin{equation}
\left\{
\begin{array}{l}
 \, \Delta_x\, W^b_{\,\, \beta} ({\pmb x},{\pmb \xi}) - \nabla^b Q_{\beta}({\pmb x},{\pmb \xi}) = 0
\\
[5pt]

\nabla_b\, W^b_{\,\, \beta} ({\pmb x},{\pmb \xi}) = 0   \quad \text{for}\quad {\pmb x} \in V_f
\\
[5pt]
W^b_{\,\, \beta} ({\pmb x},{\pmb \xi}) = -S^b_{\,\, \beta} ({\pmb x},{\pmb \xi})   \quad \text{for}\quad {\pmb x} \in \partial V_f
\end{array} \right.
\label{eq7}
\end{equation}
Similarly, $n$th order singularities can be expressed as the superposition of the associated free-space singularity and of a regular part
\begin{equation}
\nabla_{\beta_n} ... \nabla_{\beta_1} G^b_{\,\, \beta} ({\pmb x},{\pmb \xi}) =
\nabla_{\beta_n} ... \nabla_{\beta_1} S^b_{\,\, \beta} ({\pmb x},{\pmb \xi}) 
+
\nabla_{\beta_n} ... \nabla_{\beta_1} W^b_{\,\, \beta} ({\pmb x},{\pmb \xi}) 
\label{eq8}
\end{equation}
the $0$th order (i.e. for $n=0$ in eq. (\ref{eq8})) being equal to the not-differentiated Green function.

Since the expression of the Stokeslet and of the higher order 
free-space singularities is well known \cite{lorentz_1907,oseen_neuere_1927,kim_1991,pozrikidis_1992} (see the Appendix \ref{App_Stokeslet} for the bitensorial expressions), the problem of obtaining a confined $n$th order singularity $ \nabla_{\beta_n} ... \nabla_{\beta_1} G^a_{\,\, \beta} ({\pmb x},{\pmb \xi})  $ reduces to obtaining its associated regular part $\nabla_{\beta_n} ... \nabla_{\beta_1} W^a_{\,\, \beta} ({\pmb x},{\pmb \xi}) $. 
\section{Relation between Source singularities and the Green function in a bounded fluid domain}
\label{source_bounded}
Beside the class of singular Stokes flows originated by a concentrated momentum generation term and represented by eqs. (\ref{eq8}), a different class of singular Stokes flows can be obtained by relaxing the divergence constraint with a concentrated mass generation term at the pole point ${\pmb \xi}$. This second class cannot be obtained by simply differentiating the Green function at its pole and cannot even be defined in domains bounded by closed rigid surfaces, since it would violate the impermeability constraint.
Since we are not considering a closed surface of the fluid domain, but infinitely long cylinders, 
it is possible to obtain a singular solution $M^b({\pmb x},{\pmb \xi})$, referred to as the {Sourcelet}, such that $\nabla_b \, M^b({\pmb x},{\pmb \xi})= -4\pi \, \delta({\pmb x},{\pmb \xi})$
without violating the impermeability condition at the boundaries of the cylinders' surfaces. More specifically, $M^b({\pmb x},{\pmb \xi})$ correspond to the Stokes flow due to a source (or sink) of fluid concentrated in a point ${\pmb \xi}$, solution of the system
\begin{equation}
\left\{
\begin{array}{l}
\Delta_x \, M^b({\pmb x},{\pmb \xi})- \nabla^b \, \phi({\pmb x},{\pmb \xi})= -4\pi \, \nabla^a\, \delta({\pmb x},{\pmb \xi})
\\
[5pt]
\nabla_b \, M^b({\pmb x},{\pmb \xi})= -4\pi \, \delta({\pmb x},{\pmb \xi})
\\
[5pt]
M^b({\pmb x},{\pmb \xi})= 0 \quad {\pmb x} \in \partial V_f
\end{array} \right.
\label{eq9}
\end{equation}
Higher order singular Source
Stokes singularities are obtained by differentiating $M^b({\pmb x},{\pmb \xi})$ at its pole. 
In this case, it is possible to show that the source solution is strictly related to the pressure of the Green function. In fact, considering the system eqs. (\ref{eq5}) exchanging
the notation of the points so that ${\pmb x} \leftrightarrow {\pmb \xi}$
and considering the reciprocity relations $G^{\, \, b}_{\alpha}({\pmb \xi},{\pmb x})\,=\, G^b_{\,\, \alpha}({\pmb x},{\pmb \xi})$, and $g^{\, \, b}_{\beta}({\pmb \xi},{\pmb x}) = g^b_{\,\, \beta}({\pmb x},{\pmb \xi})$ \cite{procopio_bitensorial_2022}, we obtain the dual system
\begin{equation}
\left\{
\begin{array}{l}
 \, \Delta_{\xi}\, G^b_{\,\, \beta} ({\pmb x},{\pmb \xi}) - \nabla_\beta P^{b}({\pmb \xi},{\pmb x}) = -8 \pi\,  g^b_{\,\, \beta}({\pmb x},{\pmb \xi}) \,\delta({\pmb x},{\pmb \xi})
\\
[5pt]

\nabla^\beta\, G^b_{\,\, \beta} ({\pmb x},{\pmb \xi}) = 0   \quad \text{for}\quad {\pmb x} \in V_f
\\
[5pt]

G^b_{\,\, \beta} ({\pmb x},{\pmb \xi}) = 0  \quad \text{for}\quad {\pmb \xi} \in \partial V_f
\end{array} \right.
\label{eq10}
\end{equation}
Considering the identity $\Delta_x \Delta_{\xi}\, G^b_{\,\, \beta} ({\pmb x},{\pmb \xi}) = \Delta_\xi \Delta_{x}\, G^b_{\,\, \beta} ({\pmb x},{\pmb \xi})$, and comparing eq. (\ref{eq5}) with eq. (\ref{eq10}), the relation
\begin{equation}
\nabla_\beta \Delta_x P^{b}({\pmb \xi},{\pmb x}) - 
 \nabla^b  \Delta_\xi P_{\beta}({\pmb x},{\pmb \xi})=0
\label{eqeq11}
\end{equation}
is obtained.
It is possible to express the pressure $ P_{\beta}({\pmb x},{\pmb \xi})$ as the pressure due to the unbounded Green function (see the Appendix \ref{App_Stokeslet}, eq. (\ref{eqB7})) and the pressure $Q_\beta({\pmb x},{\pmb \xi})$ entering eq. (\ref{eq7}), associated to the regular part of the Green function. This leads to
\begin{equation}
\nabla_\beta \Delta_x P^{b}({\pmb \xi},{\pmb x}) - 
 \nabla^b  \Delta_\xi Q_{\beta}({\pmb x},{\pmb \xi})= 8 \pi \,\nabla^b\
 \, \nabla_\beta \delta({\pmb x},{\pmb \xi})
\label{eqeq12}
\end{equation}
Since $G^b_{\, \, \beta}({\pmb x},{\pmb \xi})=0$ for ${\pmb x} \in \partial V_f$, from the momentum balance in eq. (\ref{eq10}) $P^{b}({\pmb \xi},{\pmb x})$ is a constant at the boundaries, which can be always set equal to zero. Furthermore, by applying the divergence operator to the point ${\pmb x}$ of the momentum balance in eq. (\ref{eq10}), the relation $\nabla_b \, P^{b}({\pmb \xi},{\pmb x})\, =\, 8\pi\, \delta({\pmb x},{\pmb \xi}) $ follows. Finally, defining $\phi({\pmb x},{\pmb \xi})$ such that $\nabla_\beta \phi({\pmb x},{\pmb \xi})\,=\, - \Delta_\xi Q_\beta({\pmb x},{\pmb \xi})$, we obtain that the "dual" pressure $P^b({\pmb \xi},{\pmb x})$ solve the Stokes system of equations
\begin{equation}
\left\{
\begin{array}{l}
 \, \Delta_{x}\, P^b ({\pmb \xi},{\pmb x}) + \nabla^b \phi({\pmb x},{\pmb \xi}) = 8 \pi\,  \nabla^b \,\delta({\pmb x},{\pmb \xi})
\\
[5pt]

\nabla_b\, P^b ({\pmb \xi},{\pmb x}) = 8\pi  \,\delta({\pmb x},{\pmb \xi})   \quad \text{for}\quad {\pmb x} \in V_f
\\
[5pt]
P^b ({\pmb \xi},{\pmb x}) = 0  \quad \text{for}\quad {\pmb x} \in \partial V_f
\end{array} \right.
\label{eq13}
\end{equation}
Comparing eqs. (\ref{eq13}) with eqs. (\ref{eq9}), the relation between the bounded Source singularity and the pressure associated with the Green function of the bounded Stokes flow is
\begin{equation}
M^b({\pmb x},{\pmb \xi})\, =\, - \dfrac{P^b ({\pmb \xi}, {\pmb x})}{2}
\label{eq14}
\end{equation}
By differentiating $M^b({\pmb x},{\pmb \xi})$ at its pole, the higher order Source singularities are obtained. As in the case of singularities coming from the differentiation of the Green function,
it is possible to express $M^b({\pmb x},{\pmb \xi})$ and the 
higher order source singularities as the corresponding unbounded singularity and the associated regular part. Considering the pressure associated to the Stokeslet reported in the Appendix \ref{App_Stokeslet} eq. (\ref{eqB7}), the bounded $n$th order Source singularities read
\begin{equation}
\nabla_{\beta_n} ... \nabla_{\beta_1} M^b ({\pmb x},{\pmb \xi}) =
\nabla_{\beta_n} ... \nabla_{\beta_1} \left( \dfrac{r^b}{r^3} \right)
-\dfrac{1}{2}
\nabla_{\beta_n} ... \nabla_{\beta_1} Q^b ({\pmb \xi},{\pmb x}) 
\label{eq15}
\end{equation}
From eq. (\ref{eq15}), all the $n$th order Source singularities can expressed in terms of the pressure associated to the Green function. Considering also eq. (\ref{eq8}), it is possible to state that, once the solution for Green function in a given domain $V_f$ is known, it is possible to derive any Stokes singularity in $V_f$ by differentiation at the pole.
{\color{black} This result holds strictly for Green functions with no-slip boundary conditions. For boundary conditions other than no-slip, represented for instance by a generic linear operator $\mathcal{L}_x[\,\,]$ acting on the field point ${\pmb x}$, the boundary conditions in the problem eqs. (\ref{eq13}) cannot be expressed as
\begin{equation}
\mathcal{L}_x\left[
P^b ({\pmb \xi},{\pmb x}) \right]= 0  \quad \text{for}\quad {\pmb x} \in \partial V_f
\end{equation}
and, hence, the identity eq. (\ref{eq14}) no longer holds. Therefore, whenever boundary conditions other than no-slip are assumed, it will be generally necessary to compute $M^b({\pmb x},{\pmb \xi})$ separately from the Green function in order to obtain all the Source singularities by its differentiation.}

\section{Stokeslet representation in the cylindrical-cylindrical coordinate system}
\label{sec_stokeslet}

\begin{figure}
\includegraphics[scale=0.25]{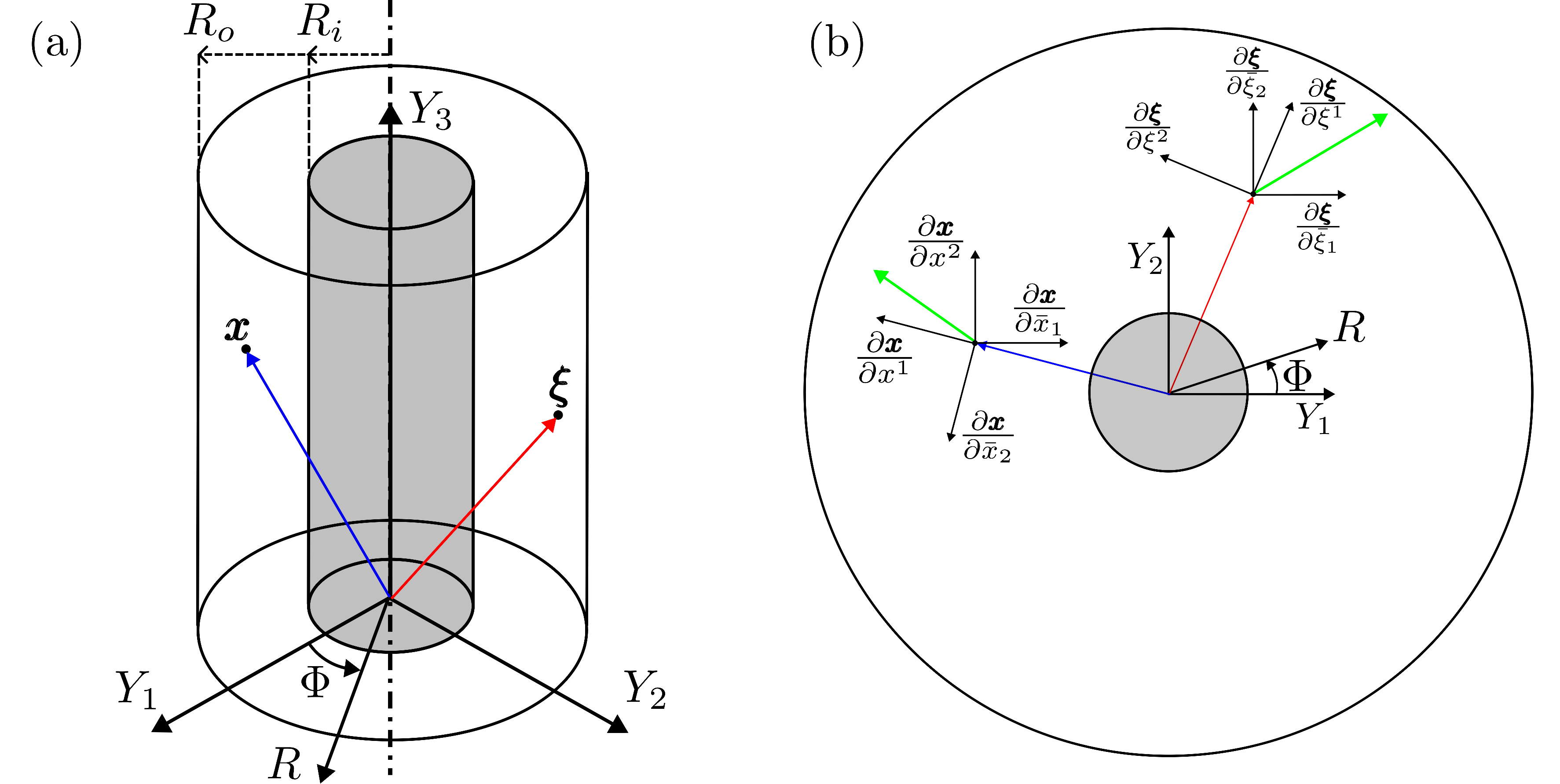}
\caption{Schematic representation of the geometry of the system consisting in the annular region between an internal cylinder with radius $R_i$ and an external cylinder with radius $R_o$. In panel (a), the absolute Cartesian coordinate system $(Y_1,Y_2,Y_3)$ and the absolute cylindrical coordinate system $(R,\Phi,Y_3)$ are represented; the position of the field point ${\pmb x}$ and of the pole point ${\pmb \xi}$ are indicated by blue and red arrows respectively.  In panel (b) {\color{black}(top view)}, the absolute and local coordinate systems are represented projected on a plane normal to the axis $Y_3$. Green arrows represent a bi-vector (e.g. a Stokes Green function) with origin at the point ${\pmb \xi}$ and ${\pmb x}$.}
\label{fig1}
\end{figure}

This section focuses on the specific case of cylindrical boundaries.
Before starting the evaluation of the regular part of the Green function bounded by cylindrical walls, it is useful to provide the bitensorial representation of the Stokeslet in cylindrical coordinates at both the field and pole points (hence, the cylindrical-cylindrical entries of the Stokeslet for brevity). As shown in Appendix \ref{App_Stokeslet}, the bi-invariant Stokeslet reads
\begin{equation}
S^b_{\, \, \beta}({\pmb x},{\pmb \xi})
= \dfrac{g^b_{\, \, \beta}({\pmb x},{\pmb \xi})}{r}- \dfrac{r_b\, r_\beta}{r^3}
\label{eq16}
\end{equation}
and the associated pressure
\begin{equation}
{\rm P}_\beta({\pmb x},{\pmb \xi})
=-2 \,   
\dfrac{r_\beta}{r^3} 
\label{eq17}
\end{equation}
where $r=\sqrt{({\pmb x} - {\pmb \xi})\cdot ({\pmb x} - {\pmb \xi})}$ is the distance between the points ${\pmb x}$ and ${\pmb \xi}$, and
\begin{equation}
r_a =\dfrac{\partial\, r^2/2 }{\partial x^a}; \quad r_\alpha = \dfrac{\partial\, r^2/2 }{\partial \xi^\alpha}
\label{eq18}
\end{equation} 
$r^2/2$ corresponding to the distance function (or Synge world function \cite{synge_relativity_1960,poisson_motion_2011}) of the Euclidean space.
A Cartesian coordinate system $(Y_1,Y_2,Y_3)$ is defined with origin on the symmetrical axis of the cylindrical boundaries corresponding to the coordinate $Y_3$ as represented in Figure \ref{fig1}.
Furthermore, a cylindrical coordinate system $(R,\Phi,Y_3)$ is introduced such that
\begin{equation}
Y_1 = R\, \cos{(\Phi)}; \quad Y_2 = R \sin{(\Phi)}
\label{eq19}
\end{equation} 
In what follows, the coordinates $(x^1,x^2,x^3)$ indicate the entries of the field point ${\pmb x}$ in the cylindrical coordinate 
system $(R,\Phi,Z)$, while $(\xi^1,\xi^2,\xi^3)$ the entries of the pole point ${\pmb \xi}$ in the same cylindrical coordinate system. The entries of the field point ${\pmb x}$ and of the pole point ${\pmb \xi}$ in the Cartesian coordinate system $(Y_1,Y_2,Y_3)$ are indicated by $(\overline{x}_1,\overline{x}_2,\overline{x}_3)$ and $(\overline{\xi}_1,\overline{\xi}_2,\overline{\xi}_3)$ respectively. Therefore, the following relations hold
\begin{equation}
\overline{x}_1 = x^1 \cos{(x^2)}; \quad \overline{x}_2 = x^1 \sin{(x^2)}; \quad \overline{x}_3=x^3
\label{eq20}
\end{equation}
and
\begin{equation}
\overline{\xi}_1 = \xi^1 \cos{(\xi^2)}; \quad \overline{\xi}_2 = \xi^1 \sin{(\xi^2)}; \quad \overline{\xi}_3=\xi^3
\label{eq21}
\end{equation}
The inverse relations of eqs. (\ref{eq20}) and (\ref{eq21}) are
\begin{equation}
\begin{array}{l}
{x}^1 = \sqrt{ (\overline{ x}_1)^2 + (\overline{ x}_2)^2 }; \quad {x}^2 = \arctan{ \left( \dfrac{\overline{x}_2}{\overline{x}_1} \right)} + \text{c}_1; \quad x^3 = \overline{x}_3
\\
\begin{cases}
\text{c}_1= 0\quad \text{for}\quad \overline{x}_1 > 0
\\
\text{c}_1= \pi \quad \text{for}\quad \overline{x}_1 < 0 \quad \text{and}\quad  \overline{x}_2 \geq 0
\\
\text{c}_1= -\pi \quad \text{for}\quad \overline{x}_1  < 0 \quad \text{and}\quad  \overline{x}_2 < 0
\end{cases}
\end{array}
\label{eq22}
\end{equation}
and
\begin{equation}
\begin{array}{l}
{\xi}^1 = \sqrt{ (\overline{ \xi}_1)^2 + (\overline{ \xi}_2)^2 }; \quad {\xi}^2 = \arctan{ \left( \dfrac{\overline{\xi}_2}{\overline{\xi}_1} \right)}  + \text{c}_2; \quad \xi^3 = \overline{\xi}_3
\\
\begin{cases}
\text{c}_2= 0\quad \text{for}\quad \overline{\xi}_1 > 0
\\
\text{c}_2= \pi \quad \text{for}\quad \overline{\xi}_1 < 0 \quad \text{and}\quad \overline{\xi}_2 \geq 0
\\
\text{c}_2= -\pi \quad \text{for}\quad \overline{\xi}_1  < 0 \quad \text{and}\quad  \overline{\xi}_2 < 0
\end{cases}
\end{array}
\label{eq23}
\end{equation}

In the remainder, the entries of the bitensors $S^b_{\, \, \beta}({\pmb x},{\pmb \xi})$, $W^b_{\, \, \beta}({\pmb x},{\pmb \xi})$, 
$P_{\, \, \beta}({\pmb x},{\pmb \xi})$, ${\rm P}_{\, \, \beta}({\pmb x},{\pmb \xi})$, etc. are considered in the cylindrical-cylindrical (at the field and pole point) coordinate systems unless otherwise specified. Therefore, for instance, $S^b_{\, \, \beta}({\pmb x},{\pmb \xi})$ correspond to the entries of a {\em bi-vector} expressed in the local coordinate system
\begin{equation}
\left(
\dfrac{\partial {\pmb \xi}}{\partial \xi^1},
\dfrac{\partial {\pmb \xi}}{\partial \xi^2},
\dfrac{\partial {\pmb \xi}}{\partial \xi^3}
\right)
\nonumber
\end{equation}
at the pole point and
\begin{equation}
\left(
\dfrac{\partial {\pmb x}}{\partial x^1},
\dfrac{\partial {\pmb x}}{\partial x^2},
\dfrac{\partial {\pmb x}}{\partial x^3},
\right)
\nonumber
\end{equation}
at the field point. See the schematic representation Figure \ref{fig1} panel (b), where a bi-vector (such as a Stokeslet) is represented by two green arrows with origin at two different points.

In order to obtain the explicit expression for 
$S^b_{\, \, \beta}({\pmb x},{\pmb \xi})$, it is convenient to employ
the following representation which is equivalent to eq. (\ref{eq16})
\begin{equation}
S^b_{\, \, \beta}({\pmb x},{\pmb \xi})
=
\left(\,
g^b_{\, \, \beta}({\pmb x},{\pmb \xi})
+
r^b\, \dfrac{\partial}{\partial{\xi}^{\beta}} \, 
\right) \dfrac{1}{r}
\label{eq24}
\end{equation}
The expression for $r^{-1}$ 
in terms of the cylindrical coordinates is reported in \cite[p. 361]{watson_treatise_1944} (see also \cite{happel_1983} or \cite{liron_stokes_1978}), according to which
\begin{equation}   
\dfrac{1}{r}\, =\,
\left\{
\begin{array}{l}
\displaystyle 
\dfrac{2}{\pi} \sum_{n=-\infty}^\infty \cos{ (n\, \phi)} \int^\infty_0 K_n(\lambda\, x) I_n(\lambda\, \xi) \, \cos{(\lambda\, z)} \, d\lambda \quad \text{for}\quad x > \xi
\\
[20pt]
\displaystyle 
\dfrac{2}{\pi} \sum_{n=-\infty}^\infty \cos{ (n\, \phi)} \int^\infty_0 I_n(\lambda\, x) K_n(\lambda\, \xi) \, \cos{(\lambda\, z)} \, d\lambda \quad \text{for}\quad x < \xi
\end{array}
\right.
\label{eq25}
\end{equation}
where $I_{n}(z)$ and
$K_{n}(z)$ are the modified Bessel functions of first and second kind, and where $x \equiv x^1$, $\xi \equiv \xi^1$,
 $\phi=x^2 -\xi^2$ and $z=x^3-\xi^3$. Where it does not cause confusion, the symbols $x$ and $\xi$ will be used in place of $x^1$ and $\xi^1$ for the sake of compactness. Moreover, for the sake of clarity, parentheses will always be used to indicate the powers of a quantity in the following. For example, the square of $x$ will be always written as $(x)^2 = x x$ (or equivalently as $(x^1)^2$ ) in order to avoid confusion with the second contravariant component $x^2$.

The expression of the distance function in terms of the cylindrical coordinates of the points ${\pmb x}$ and ${\pmb \xi}$ is obtained by substituting the relations eqs. (\ref{eq20}) and (\ref{eq21}) into the definition
\begin{equation}
\dfrac{r^2}{2} = \delta_{i\, j}	(\overline{x}_i - \overline{\xi}_i ) 	\,		 (\overline{x}_j - \overline{\xi}_j ) 
\label{eq26}
\end{equation}
which provides
\begin{equation}
\dfrac{(r)^2}{2} = \dfrac{(x)^2 +(\xi)^2 + (z)^2 - 2\, x\, \xi \cos{(\phi)} }{2}
\label{eq27}
\end{equation}
From eq. (\ref{eq27}), it is possible to obtain the position vector $r^b=g^{b a}({\pmb x}) \, r_a$, where $r_a$ is defined in eqs. (\ref{eq18}) and the metric tensor  $g^{b a}({\pmb x})$  of the cylindrical coordinate systems at the point ${\pmb x}$ is reported in Appendix \ref{app_useful_geo} eq. (\ref{eqA3}).
Specifically, then position vector $r^b = \nabla^b (r)^2/2$ is
\begin{equation}
\{r^b\} \equiv (x-\xi \cos{(\phi)}\, , \, \dfrac{\xi}{x}\sin{(\phi)}\, ,\,  z)
\label{eq28}
\end{equation}
Consider, now, the parallel propagator in the cylindrical-cylindrical coordinate system.
From its definition eq. (\ref{eq3}), and considering that $Y_j \equiv \overline{x}_j$ at the point ${\pmb x}$ and 
$Y_j \equiv \overline{\xi}_j$ at the point ${\pmb \xi}$
\begin{equation}
g^b_{\, \, \beta}({\pmb x},{\pmb \xi})
=
\dfrac{\partial x^b}{\partial  \overline{x}_i}\,  \dfrac{\partial \overline{\xi}_i}{\partial \xi^\beta}
\label{eq29}
\end{equation}
Using eqs. (\ref{eq22}) and (\ref{eq21}), the cylindrical-cylindrical entries of the parallel propagator read
\begin{equation}
\begin{array}{l}
g^1_{\,\, 1}({\pmb x},{\pmb \xi}) = \cos{(\phi)}; \quad g^2_{\,\, 1}({\pmb x},{\pmb \xi})=-\dfrac{\sin{(\phi)}}{x};\quad g^3_{\,\, 1}({\pmb x},{\pmb \xi}) = 0
\\
[10pt]
g^1_{\,\, 2}({\pmb x},{\pmb \xi})=\xi \sin{(\phi)}; \quad g^2_{\,\, 2}({\pmb x},{\pmb \xi})=\dfrac{\xi }{x} \cos{(\phi)};\quad g^3_{\,\, 2}({\pmb x},{\pmb \xi}) = 0 
\\
[10pt]
g^1_{\,\, 3}({\pmb x},{\pmb \xi}) = 0; \quad g^2_{\,\, 3}({\pmb x},{\pmb \xi})=0 ;\quad g^3_{\,\, 3}({\pmb x},{\pmb \xi}) = 1
\end{array}
\label{eq30}
\end{equation}
Finally, let us consider the cylindrical-cylindrical entries of the Stokeslet.
Substituting eqs. (\ref{eq25}), (\ref{eq28}) and (\ref{eq30}) into eq. (\ref{eq24}), and using also the relation \cite{happel_1983}
\begin{equation}   
\dfrac{z}{r}\, =\,
\left\{
\begin{array}{l}
\displaystyle 
-\dfrac{2}{\pi} \sum_{n=-\infty}^\infty \cos{ (n\, \phi)} \int^\infty_0
\left[ x K'_n(\lambda\, x) I_n(\lambda\, \xi) 
+\xi K_n(\lambda\, x) I'_n(\lambda\, \xi) \right]
 \, \sin{(\lambda\, z)} \, d\lambda \quad \text{for}\quad x > \xi
\\
[20pt]
\displaystyle -
\dfrac{2}{\pi} \sum_{n=-\infty}^\infty \cos{ (n\, \phi)} \int^\infty_0 \left[ x K_n(\lambda\, \xi) I'_n(\lambda\, x) 
+\xi K'_n(\lambda\, \xi) I_n(\lambda\, x) \right]
\, \sin{(\lambda\, z)} \, d\lambda \quad \text{for}\quad x < \xi
\end{array}
\right.
\label{eq31}
\end{equation}
where
$K'(z)$ and $I'(z)$ are the derivatives of
$K(z)$ and $I(z)$ with respect to their argument,
the entries of the Stokeslet $S^b_{\,\, \beta}({\pmb x},{\pmb \xi})$ in cylindrical-cylindrical coordinates read
\begin{equation}
\begin{array}{l}
\displaystyle
S^1_{\,\, 1}({\pmb x},{\pmb \xi})=\sum^{*}\cos(n \phi)\cos(\lambda z)\, {\rm A}_{(11)} (\lambda\, x,\lambda\, \xi, \lambda, n)
\\
[10pt]
\displaystyle
S^1_{\,\, 2}({\pmb x},{\pmb \xi})=\sum^{*}\sin(n \phi)\cos(\lambda z)\, {\rm A}_{(12)} (\lambda\, x,\lambda\, \xi, \lambda, n)
\\
[10pt]
\displaystyle
S^1_{\,\, 3}({\pmb x},{\pmb \xi})=\sum^{*}\cos(n \phi)\sin(\lambda z)\, {\rm A}_{(13)} (\lambda\, x,\lambda\, \xi, \lambda, n)
\\
[10pt]
\displaystyle
S^2_{\,\, 1}({\pmb x},{\pmb \xi})=\sum^{*}\sin(n \phi)\cos(\lambda z)\, {\rm A}_{(21)} (\lambda\, x,\lambda\, \xi, \lambda, n)
\\
[10pt]
\displaystyle
S^2_{\,\, 2}({\pmb x},{\pmb \xi})=-\sum^{*}\cos(n \phi)\cos(\lambda z)\, {\rm A}_{(22)} (\lambda\, x,\lambda\, \xi, \lambda, n)
\\
[10pt]
\displaystyle
S^2_{\,\, 3}({\pmb x},{\pmb \xi})=\sum^{*}\sin(n \phi)\sin(\lambda z)\, {\rm A}_{(23)} (\lambda\, x,\lambda\, \xi, \lambda, n)
\\
[10pt]
\displaystyle
S^3_{\,\, 1}({\pmb x},{\pmb \xi})=\sum^{*}\cos(n \phi)\sin(\lambda z)\, {\rm A}_{(31)} (\lambda\, x,\lambda\, \xi, \lambda, n)
\\
[10pt]
\displaystyle
S^3_{\,\, 2}({\pmb x},{\pmb \xi})=\sum^{*}\sin(n \phi)\sin(\lambda z)\, {\rm A}_{(32)} (\lambda\, x,\lambda\, \xi, \lambda, n)
\\
[10pt]
\displaystyle
S^3_{\,\, 3}({\pmb x},{\pmb \xi})=-\sum^{*}\cos(n \phi)\cos(\lambda z)\, {\rm A}_{(33)} (\lambda\, x,\lambda\, \xi, \lambda, n)
\end{array}
\label{eq32}
\end{equation}
where  $\sum^{*}$ is the operator defined by the relation
\begin{equation}
\sum^{*} f_n (\lambda) =\dfrac{2}{\pi} \, \sum_{n=-\infty}^{\infty} \int_0^\infty f_n (\lambda) \,d\lambda
\label{eq33}
\end{equation}
for any $f_n(\lambda)$.
Moreover, in eqs. (\ref{eq32})
\begin{equation}
{\rm A}_{( b\, \beta)} (\lambda\, x,\lambda\, \xi, \lambda, n) =
\left\{
\begin{array}{l}
A^{>}_{( b\, \beta)} (\lambda\, x,\lambda\, \xi, \lambda, n) \quad \text{for}\quad x > \xi
\\
[5pt]
 A^{<}_{( b\, \beta)} (\lambda\, x,\lambda\, \xi, \lambda, n) \quad \text{for}\quad x < \xi
\end{array}
\right.
\label{eq34}
\end{equation}
Explicit expressions for $A^{>}_{( b\, \beta)} (\lambda\, x,\lambda\, \xi, \lambda, n) $ and $A^{<}_{( b\, \beta)} (\lambda\, x,\lambda\, \xi, \lambda, n)$
are reported in the Appendix \ref{app_entries_A>}
 and \ref{app_entries_A<} respectively.

From eqs. (\ref{eq32}), it can be noted that the Stokeslet can be compactly written as
\begin{equation}
S^b_{\,\, \beta}({\pmb x},{\pmb \xi})=\sum^{*}\cos(n \phi + p_{(b)}+p_{(\beta)})\cos(\lambda z  + q_{(b)}+q_{(\beta)})\, {\rm A}_{(b \,\beta)} (\lambda\, x,\lambda\, \xi, \lambda, n)
\label{eq35}
\end{equation}
where
\begin{equation}
\{ p_{(i)} \} = (0,-\pi/2,0); \quad 
\{ q_{(i)} \} = (0,0,-\pi/2)
\label{eq36}
\end{equation}

In eq. (\ref{eq36}) and in the following, the Einstein summation convention is not adopted for indexes between parenthesis.

\section{The Green function}
\label{sec_green_function}

\subsection{General expression for two cylinders}
In order to obtain the regular part of the Green function in the cylindrical annulus (and, as consequence, the associated Green function), let us consider the Brenner and Happel solution of the Stokes flow in cylindrical
harmonics \cite{brenner_slow_1958} (see \cite[pp. 71-78]{happel_1983} for the proof), according to which a solution of the Stokes flow in the cylindrical coordinate system $(v^a({\pmb x}), p({\pmb x}))$ can be expressed as \footnote{The Brenner expressions reported in \cite{brenner_slow_1958,happel_1983} are represented in terms of the physical components of the velocity, while in this work the contravariant components are used. If $v^{\theta}({\pmb x})$ is the physical component along the angular direction used in \cite{brenner_slow_1958,happel_1983}, the relation $v^2({\pmb x})=v^{\theta}({\pmb x})/R$ holds. See \cite[pp. 142-149]{synge_tensor_1978} for a detailed discussion.}
\begin{equation}
\begin{array}{l}
v^1 ({\pmb x}) = x^1 \dfrac{\partial}{ \partial x^1} \dfrac{\partial\, \Pi}{ \partial x^1}
+ \dfrac{\partial \, \Psi}{\partial x^1} 
+ \dfrac{1}{x^1} \dfrac{\partial\, \Omega}{ \partial x^2}
\\
[10 pt]
v^2 ({\pmb x}) =
 \dfrac{\partial}{ \partial x^1} \dfrac{1}{x^1} \dfrac{\partial\, \Pi}{ \partial x^2}
+ \dfrac{1}{(x^1)^2} \dfrac{\partial \, \Psi}{\partial x^2} 
-\dfrac{1}{x^1}  \dfrac{\partial\, \Omega}{ \partial x^1}
\\
[10 pt]
v^3 ({\pmb x})=
x^1 \dfrac{\partial}{ \partial x^1} \dfrac{\partial\, \Pi}{ \partial x^3}
+
 \dfrac{\partial\, \Pi}{ \partial x^3}
+
\dfrac{\partial \, \Psi}{\partial x^3} 
\\
[10 pt]
p({\pmb x})= -2 \mu
\dfrac{\partial }{\partial x^3} \dfrac{\partial \,\Pi}{\partial x^3} 
\end{array}
\label{eq37}
\end{equation}
where  $\Phi$, $\Psi$ and $\Omega$ are harmonic functions, i.e. $\Delta_x \{\Pi, \Psi, \Omega\}=0$.
The velocity field expression in eqs. (\ref{eq37}) in the tensorial notation reads
\begin{equation}
v^a({\pmb x})=
x^c \delta^1_c \, \delta^b_1 \nabla_b \nabla^a \Pi
+
\delta^a_3 \delta^b_3 \nabla_b \Pi
+
\nabla^a \Psi 
+
\dfrac{\varepsilon^{a b c}}{\sqrt{g({\pmb x})}}
\nabla_b \, \delta_c^3 \, \Omega
\label{eq38}
\end{equation}
where $\varepsilon^{a b c}/\sqrt{g({\pmb x})} $
is the absolute contravariant Ricci--Levi-Civita tensor. Since $ v^a({\pmb x})$ are the entries in the cylindrical coordinate system, $ \sqrt{g({\pmb x})} = x^1$

It is possible to use eq. (\ref{eq38}) 
for obtaining the entries of the regular part of the Green function $W^b_{\, \, \beta}({\pmb x},{\pmb \xi})$, which represent three distinct Stokes problems at ${\pmb x}$, one for each direction $\beta=1,2,3$ of the Stokeslet at the pole ${\pmb \xi}$.
Therefore, eq. (\ref{eq38}) for the regular part of the Green function can be expressed as
\begin{equation}
W^b_{\,\, \beta}({\pmb x}, {\pmb \xi})=
x^c \delta^1_c \, \delta^a_1 \nabla_a \nabla^b \Pi_{\beta}
+
\delta^b_3 \delta^a_3 \nabla_a \Pi_\beta
+
\nabla^b \Psi_\beta 
+
\dfrac{\varepsilon^{b c d}}{x^1}
\nabla_c \, \delta_d^3 \, \Omega_\beta
\label{eq39}
\end{equation}
with the associated pressure equal to
\begin{equation}
Q_\beta({\pmb x},{\pmb \xi})= -2
\dfrac{\partial }{\partial x^3} \dfrac{\partial \,\Pi_\beta}{\partial x^3}
\label{eq40}
\end{equation}
{\color{black} Determining the disturbance flow of the Green function due to the cylindrical walls thus requires the determination of nine harmonic fields (three for each direction of the Stokeslet). From a more general perspective, it is always possible to express the disturbance flow in terms of harmonic functions, as done for the flow inside a cylinder in \cite{hasimoto_slow_1976} using the Papkovich–Neuber representation \cite{papkovich,neuber}. Howevere, this approach requires an higher number of harmonic fields consisting in four harmonic fields per direction (twelve total). The choice of Happel and Brenner's harmonic functions is thus motivated by both the reduced number of harmonics required and the straightforward implementation of boundary conditions.
Specifically, in order to enforce boundary conditions, it is possible to express
the harmonic functions  $\Pi_\beta$, $\Psi_\beta$ and $\Omega_\beta$  as sums of cylindrical harmonic functions according the method reported in \cite{happel_1983}.} Considering the form assumed by the Stokeslet entries eq. (\ref{eq35}), let us choose the harmonic functions $\Pi_\beta$, $\Psi_\beta$ and $\Omega_\beta$ as follows
\begin{equation}
\begin{array}{l}
\displaystyle
\Pi_\beta=  \sum^{*}\cos(n \phi + p_{(\beta)})\cos(\lambda z + q_{(\beta)})\, \left[ {B}_{(1,\beta)} (\lambda\, \xi, \lambda, n)\, I_n(\lambda x)
+
{B}_{(4,\beta)} (\lambda\, \xi, \lambda, n)\, K_n(\lambda x)
\right] 
\\
[10pt]
\displaystyle
\Psi_\beta= \sum^{*}\cos(n \phi + p_{(\beta)})\cos(\lambda z + q_{(\beta)})\, \left[ {B}_{(2,\beta)} (\lambda\, \xi, \lambda, n)\, I_n(\lambda x)
+
{B}_{(5,\beta)} (\lambda\, \xi, \lambda, n)\, K_n(\lambda x)
\right] 
\\
[10pt]
\displaystyle
\Omega_\beta= \sum^{*}\sin(n \phi + p_{(\beta)})\cos(\lambda z + q_{(\beta)})\, \left[ {B}_{(3,\beta)} (\lambda\, \xi, \lambda, n)\, I_n(\lambda x)
+
{B}_{(6,\beta)} (\lambda\, \xi, \lambda, n)\, K_n(\lambda x)
\right] 
\end{array}
\label{eq41}
\end{equation}
where $ {B}_{(s,\beta)} $, with the index $s=1,2, ..., 6$, are constant with respect to the coordinates of the field point to be determined as function radial position of the pole point.

By substituting the expressions eqs. (\ref{eq41}) into eq. (\ref{eq39}), the entries of the regular part of the Green function read
\begin{equation}
\begin{array}{l}
W^1_{\,\, \beta}({\pmb x},{\pmb \xi})=
\displaystyle \sum^{*}
 \sum_{s=1}^6 
  \cos(n \phi + p_{(\beta)})\cos(\lambda z + q_{(\beta)}) \,  {B}_{(s,\beta)} (\lambda\, \xi, \lambda, n)\,
{C}_{(1,s)} (\lambda\, x, \lambda, n)
\\
[10pt]
W^2_{\,\, \beta}({\pmb x},{\pmb \xi})=
\displaystyle
 \sum^{*}
 \sum_{s=1}^6 
  \sin(n \phi + p_{(\beta)})\cos(\lambda z + q_{(\beta)})\,   {B}_{(s,\beta)} (\lambda\, \xi, \lambda, n)\,
{C}_{(2,s)} (\lambda\, x, \lambda, n)
\\
[10pt]
W^3_{\,\, \beta}({\pmb x},{\pmb \xi})=
\displaystyle
 \sum^{*}
 \sum_{s=1}^6 
  \cos(n \phi + p_{(\beta)})\sin(\lambda z + q_{(\beta)}) \,  {B}_{(s,\beta)} (\lambda\, \xi, \lambda, n)\,
{C}_{(3,s)} (\lambda\, x, \lambda, n)
\end{array}
\label{eq42}
\end{equation}
with the coefficients $C_{(b,s)}(\lambda x, \lambda, n )$
reported in Appendix \ref{app_entries_C}.
By eq. (\ref{eq40}), the associated pressure reads
\begin{equation}
Q_{\,\, \beta}({\pmb x},{\pmb \xi})= 2
\displaystyle
 \sum^{*} 
  \cos(n \phi + p_{(\beta)})\cos(\lambda z + q_{(\beta)}) \,  \lambda^2 \left[
   {B}_{(1,\beta)} (\lambda\, \xi, \lambda, n)\, I_n(\lambda x)
+
{B}_{(4,\beta)} (\lambda\, \xi, \lambda, n)\, K_n(\lambda x)
  \right]
\label{eq43}
\end{equation}

Observing eqs. (\ref{eq42}), it is possible to deduce the compact form for the entries of the regular part of the Green function, which is
\begin{equation}
W^b_{\,\, \beta}({\pmb x},{\pmb \xi})=
\displaystyle \sum^{*}
 \sum_{s=1}^6 
  \cos(n \phi+p_{(b)} + p_{(\beta)})\cos(\lambda z + q_{(b)} + q_{(\beta)}) \,  {B}_{(s,\beta)} (\lambda\, \xi, \lambda, n)\,
{C}_{(b,s)} (\lambda\, x, \lambda, n)
\label{eq44}
\end{equation}
The regular part of the Green function in eq. (\ref{eq44}) is determined by the $18$ coefficients ${B}_{(s,\beta)} (\lambda\, \xi, \lambda, n)$, where $\beta=1,2,3$ and $s=1,2, ... ,6$. To determine ${B}_{(s,\beta)} (\lambda\, \xi, \lambda, n)$, the boundary conditions eq. (\ref{eq7}) need to be imposed. 
Considering the schematic representation of the geometry of the system Fig \ref{fig1}, the boundaries of the fluid domain $\partial V_f$  are
\begin{equation}
\partial V_f=\left \{ (x^1,x^2,x^3) : x^1 = R_o \right \} \cup \left \{ (x^1,x^2,x^3) : x^1 = R_i \right \} 
\label{eq45}
\end{equation}
with $R_o$ and $R_i$ being the radius of the external and internal cylinder, respectively (see Figure \ref{fig1}).
Therefore, the boundary conditions at the external surface, where $x=R_o > \xi$, read
\begin{equation}
\begin{array}{l}
\displaystyle \sum^{*}
  \cos(n \phi + p_{(b)} + p_{(\beta)})\cos(\lambda z  + q_{(b)}+ q_{(\beta)}) \times 
\\
[10pt]
\times
\displaystyle\left( 
 \sum_{s=1}^6  {B}_{(s,\beta)} (\lambda\, \xi, \lambda, n)\,
{C}_{(b,s)} (\lambda\, R_o, \lambda, n)+
{A}^>_{(b \,\beta)} (\lambda\, R_o,\lambda\, \xi, \lambda, n)
\right)
=0
\end{array}
\label{eq45}
\end{equation}
which are fulfilled if
\begin{equation}
 \sum_{s=1}^6   {B}_{(s,\beta)} (\lambda\, \xi, \lambda, n)\,
{C}_{(b,s)} (\lambda\, R_o, \lambda, n)=-
{A}^>_{(b \,\beta)} (\lambda\, R_o,\lambda\, \xi, \lambda, n)
\label{eq47}
\end{equation}
Similarly, at the internal surface, where $x=R_i < \xi$, the boundary conditions require
\begin{equation}
 \sum_{s=1}^6   {B}_{(s,\beta)} (\lambda\, \xi, \lambda, n)\,
{C}_{(b,s)} (\lambda\, R_i, \lambda, n)=-
{A}^<_{(b \,\beta)} (\lambda\, R_i,\lambda\, \xi, \lambda, n)
\label{eq48}
\end{equation}
Eqs. (\ref{eq47}) and (\ref{eq48}) represent a linear system of $18$ 
equations from which it is possible to obtain the coefficients 
$ {B}_{(s,\beta)} (\lambda\, \xi, \lambda, n) $. In order to solve the system, it is useful to introduce the matrix notation.
The entries  
 $ {B}_{(s,\beta)} (\lambda\, \xi, \lambda, n) $ are collected
in $3$ vectors ${\bf B}_\beta$ having $6$ rows
\begin{equation}
{\bf B}_\beta (\xi,R_o,R_i)= \left( {B}_{(s,\beta)} (\lambda\, \xi, \lambda, n)  \right)_{s=1,2...,6}
\label{eq49}
\end{equation}
where the dependence on $n$ and $\lambda$ have been omitted for the sake of simplicity.

Similarly, the entries of ${A}^>_{(b \,\beta)} (\lambda\, R_o,\lambda\, \xi, \lambda, n)$ and 
$ {A}^<_{(b \,\beta)} (\lambda\, R_i,\lambda\, \xi, \lambda, n)$
are collected in $3$ vectors $\overline{\bf A}_\beta(\xi,R_o,R_i)$ with $6$ rows as follows
\begin{equation}
{\pmb A}_\beta^{>}(\xi,R_o)=
\left( 
{A}^>_{(b \,\beta)} (\lambda\, R_o,\lambda\, \xi, \lambda, n)
\right)_{b=1,2,3} 
, \quad
{\pmb A}_\beta^{<}(\xi,R_i)=
\left( 
{A}^<_{(b \,\beta)} (\lambda\, R_i,\lambda\, \xi, \lambda, n)
\right)_{b=1,2,3}
\label{eq50}
\end{equation} 
and
\begin{equation}
\overline{\bf A}_\beta(\xi,R_o,R_i) =
\left(
\begin{array}{c}
{\pmb A}_\beta^{>}(\xi,R_o)
\\
{\pmb A}_\beta^{<}(\xi,R_i)
\end{array}
\right)
\label{eq51}
\end{equation}
{\color{black}
Next, we define the $3 \times 3$ matrices
\begin{equation}
{\pmb C}^o (x)
=
\left( 
{C}_{(b,s)} (\lambda\, x, \lambda, n)
\right)_{s=1,2,3} 
, \quad
{\pmb C}^i (x)
=
\left( 
{C}_{(b,s)} (\lambda\, x, \lambda, n)
\right)_{s=4,5,6}
\label{eq52}
\end{equation} 
so that the entries ${C}_{(b,s)} (\lambda\, x, \lambda, n)$
are collected in the $3 \times 6$ matrix}
\begin{equation}
{\bf C} (x) =
\left(
\begin{array}{cc}
{\pmb C}^o (x)
&
{\pmb C}^i (x)
\end{array}
\right)
\label{eq53}
\end{equation}
Finally, the entries  ${C}_{(b,s)} (\lambda\, x, \lambda, n)$,
evaluated for $x=R_o$ and $x=R_i$,
are collected in the $6 \times 6$ block matrix
\begin{equation}
\overline{\bf C}(R_o,R_i) = 
\left(
\begin{array}{c}
{\bf C} (R_o)
\\
 {\bf C} (R_i)
\end{array}
\right)
=
\left(
\begin{array}{cc}
{\pmb C}^o (R_o) & {\pmb C}^i (R_o) 
\\
{\pmb C}^o (R_i) & {\pmb C}^i (R_i)
\end{array}
\right)
\label{eq54}
\end{equation}
By using the notation introduced in eqs. (\ref{eq49})-(\ref{eq54}), the linear system eqs. (\ref{eq47}) and (\ref{eq48}) read
\begin{equation}
\overline{\bf C}(R_o,R_i)  \, {\bf B}_\beta (\xi,R_o,R_i) \, = \, -\overline{\bf A}_\beta(\xi,R_o,R_i)
\label{eq55}
\end{equation}
And, finally, the regular part of the Green function reads
\begin{equation}
\begin{array}{l}
W^b_{\,\, \beta}({\pmb x},{\pmb \xi})=-
\displaystyle \sum^{*}
  \cos(n \phi+p_{(b)} + p_{(\beta)})\cos(\lambda z + q_{(b)} + q_{(\beta)}) \times 
  \\  
  [10pt]
  \hspace{2.2cm}
 \times  \left(\, {\bf C}(x) \left( \overline{\bf C}(R_o,R_i) \right)^{-1} \overline{\bf A}_\beta (\xi,R_o,R_i)
\, \right)_{(b)}
\end{array}
\label{eq56}
\end{equation}
from which the Green function in the cylindrical-cylindrical coordinate system is obtained considering eq. (\ref{eq6}) and eq. (\ref{eq35}). {\color{black}The entries of the matrix $ \left( \overline{\bf C}(R_o,R_i) \right)^{-1} $ are explicitly reported in Appendix \ref{app_entries_B4}.}

Cartesian entries of the Green function eq. (\ref{eq6}) can be obtained enforcing bitensorial transformations. In the Cartesian-Cartesian coordinate system, the entries of the Green function are
\begin{equation}
{\overline G}_{i\, j}({\pmb x},{\pmb \xi})\,=\,
\dfrac{\partial \overline{x}_i}{\partial x^b}\, \dfrac{\partial \xi^\beta}{\partial{\overline \xi}_j}
\, { G}^{b}_{\,\, \beta}({\pmb x},{\pmb \xi})
\end{equation}
with $ \dfrac{\partial \overline{x}_i}{\partial x^b}$ and $\dfrac{\partial\xi^\beta}{\partial\overline{ \xi}_j} $ given by eq. (\ref{eqA8}) and (\ref{eqA11}), respectively.

\begin{figure}
\centering
\includegraphics[scale=0.40]{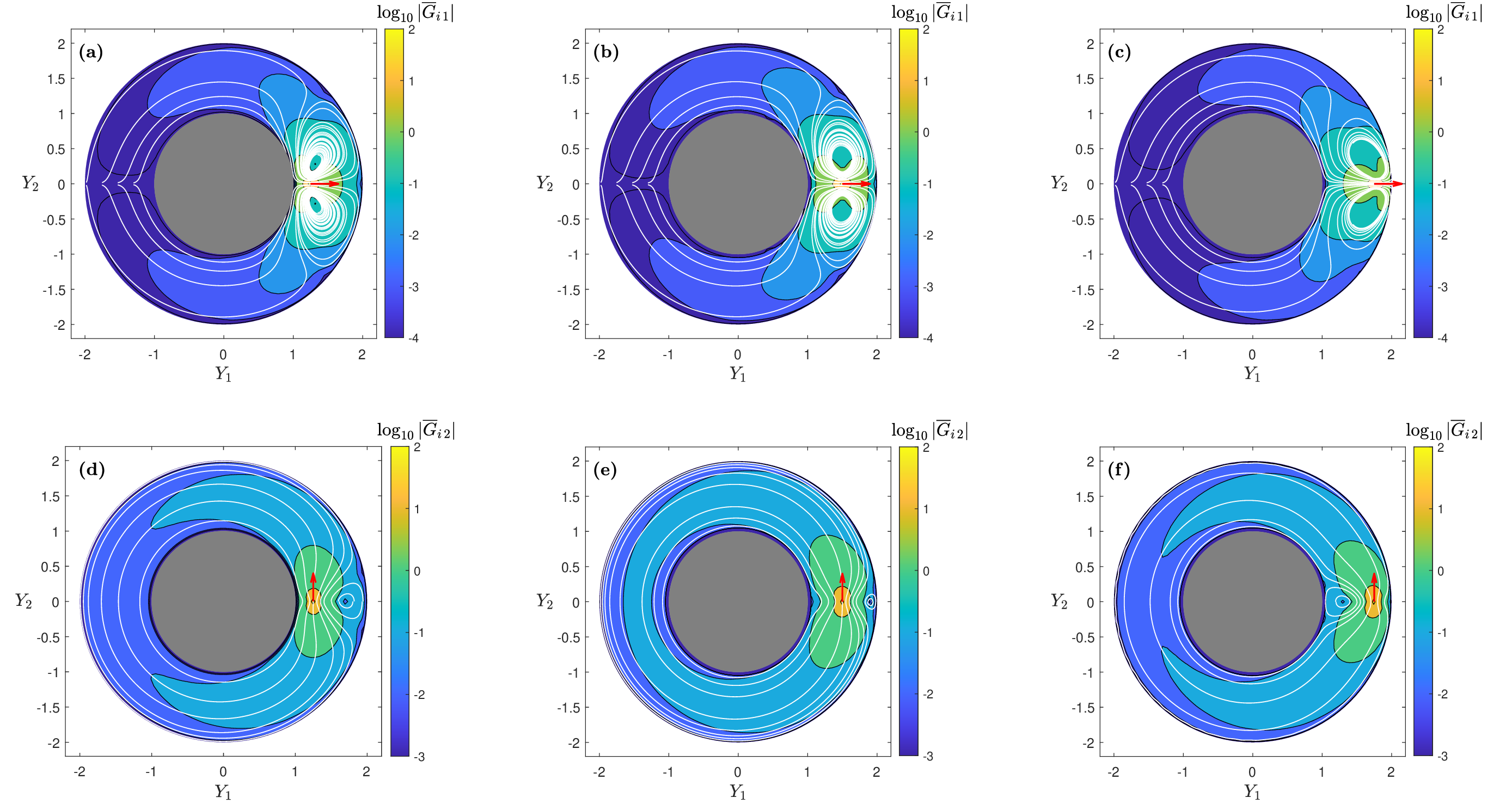}
\caption{ Streamlines (white solid lines) and intensity of the velocity field in the annular region between two concentric cylinder with $R_i=1$ and $R_o=2$ due to a Stokeslet (red arrows in the figure) in the radial direction (panels (a)-(c))  and in the angular direction (panel (d)-(f)). All the data are reported for $x^3=\xi^3$ and $\xi^2=0$. The radial position of the Stokeslet is $\xi^1=1.25$ in panels (a) and (d), $\xi^1=1.5$ in panels (b) and (e), and $\xi^1=1.75$ in panels (c) and (f). {\color{black} Henceforth, we adopt $|\overline{G}_{i j}| = \sqrt{|\overline{G}_{1 j}|^2 + |\overline{G}_{2 j}|^2 + |\overline{G}_{3 j}|^2}$ to indicate the norm of the vector with component $ \overline{G}_{i j}$  at the field point}.}
\label{fig2}
\end{figure}

{\color{black}By expressing the entries of the unbounded Stokeslet in the Cartesian coordinate system (i.e., the so-called Oseen tensor), the entries of the bounded Stokeslet read}
\begin{equation}
{\overline G}_{i\, j}({\pmb x},{\pmb \xi})\,=
\,\dfrac{\delta_{i\, j}}{r}
+
\dfrac{(\overline{x}_i -\overline{\xi}_i)\, (\overline{x}_j - \overline{\xi}_j)}{r^3}\,
+\,
\dfrac{\partial \overline{x}_i}{\partial x^b}\, \dfrac{\partial \xi^\beta}{\partial {\overline \xi}_j}
\, { W}^{b}_{\,\, \beta}({\pmb x},{\pmb \xi})
\end{equation}
Figure \ref{fig2} shows the velocity field in the annular space between two concentric cylinders with radii $R_i=1$ and $R_o=2$ due to a radially oriented Stokeslet (panels (a)-(c)) and an angularly oriented Stokeslet (panels (d)-(f)) at different distances from the center.
Specifically, 
the streamlines and the intensity of the velocity fields in Cartesian-Cartesian coordinates $\overline{G}_{i \, 1}({\pmb x},{\pmb \xi})$ and $\overline{G}_{i \, 2}({\pmb x},{\pmb \xi})$ are reported for $x^3=\xi^3$ and are represented (without loss of generality) for $\xi^2=0$.

The solution was obtained by approximating the summation eq. (\ref{eq33}) as 
\begin{equation}
\sum^{*}\approx \dfrac{2}{\pi} \, \sum_{n=-N_{max}}^{N_{max}} \int_0^\infty d\lambda
\label{eq65}
\end{equation}
with $N_{max}=50$. Figure \ref{fig2} and all the figure in the remainder are obtained by using $N_{max}$ such that the error committed in satisfying the boundary conditions is less that $10^{-6}$, which correspond to $N_{max} \approx 25 \text{--} 100 $.

Due to the symmetry of the problem, the axial component $\overline{G}_{3 \, 1}({\pmb x},{\pmb \xi})$ is vanishing everywhere along the plane $x^3=\xi^3$. Specifically, it changes sign  
for $x^3<0$ (for which $\overline{G}_{3 \, 1}({\pmb x},{\pmb \xi})>0$) and $x^3>0$ (for which $\overline{G}_{3 \, 1}({\pmb x},{\pmb \xi})<0$). 
As shown in panels (a)-(c) of Figure \ref{fig2}, for $x^2=\pi$ also the angular component is vanishing and the only non-vanishing entry is the radial component, possessing a slight intensity $\lesssim 10^{-4} $.

On the contrary, as shown in  panels (d)-(f) of Figure \ref{fig2}, the intensity of the flow does not decay along to the annular region when the singularity is angularly oriented (especially in the case where the pole of the singularity is at the mean distance between the internal and external cylinder), possessing an intensity almost unitary in the entire annular region. 
Moreover, in this case, the confinement of the two cylinders leads to the formation of a vortex near the angularly oriented Stokeslet. This vortex appears near the outer wall when the Stokeslet is closer to the inner wall, tends to disappear as the Stokeslet approaches the midway distance, and reappears near the inner wall when the Stokeslet is located closer to the outer wall.
Also in this case, the axial component $ \overline{G}_{3 \, 2}({\pmb x},{\pmb \xi})=0$ at $x^3=\xi^3$ everywhere due to the symmetry of the problem.

\begin{figure}
\centering
\includegraphics[scale=0.40]{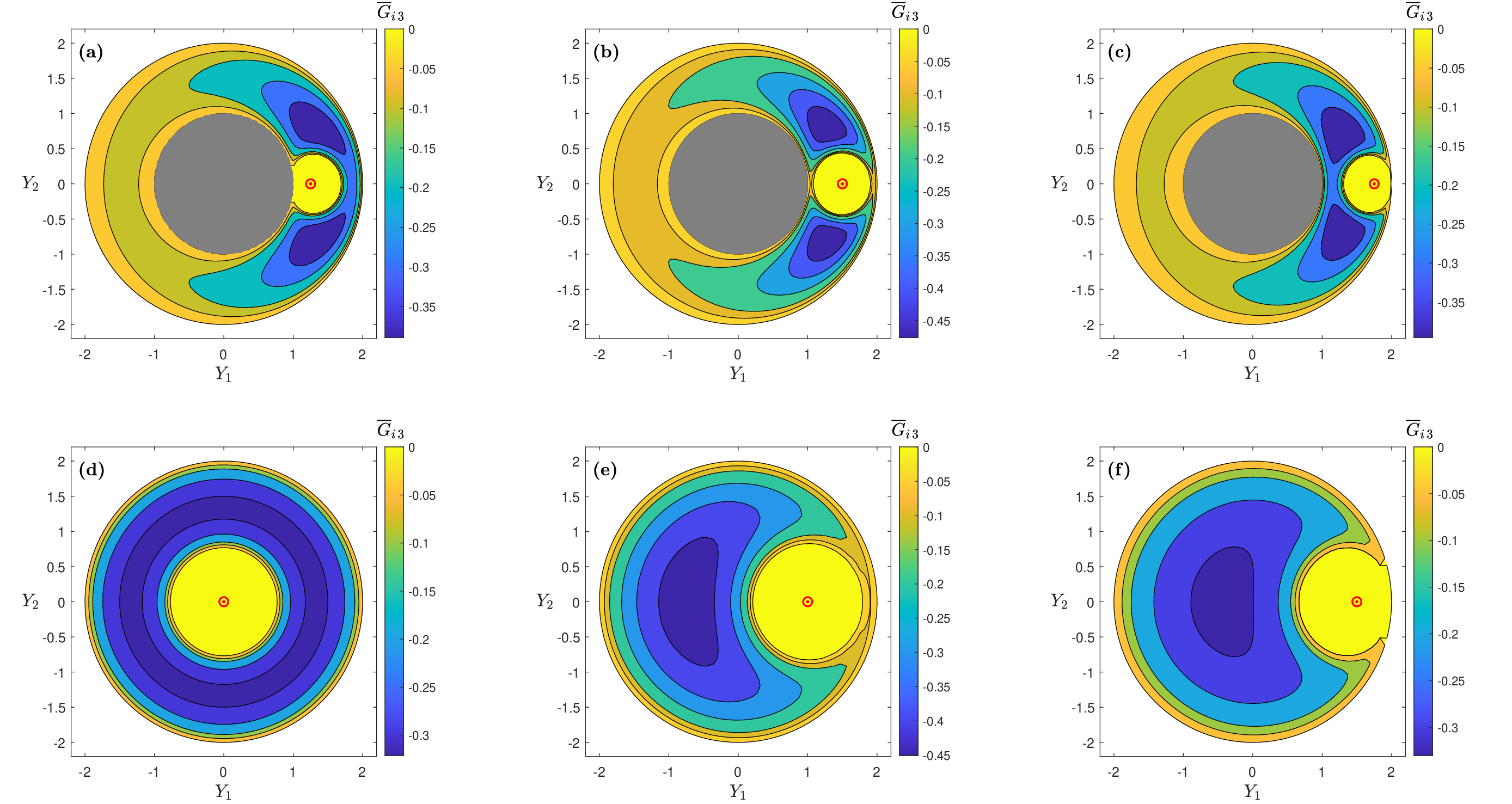}
\caption{Contour plot of the velocity field induced by a Stokeslet (red circled dots in the figure) in the fluid bounded by cylindrical walls for $x^3=\xi^3$.
 In panel (a)-(c) the fluid is confined between two concentric cylinders with internal radius $R_i=1$ and external radius $R_o=2$ and the radial position of the Stokeslet are $\xi^1=1.25$ (panel (a)), $\xi^1=1.5$ (panel (b)), $\xi^1=1.75$ (panel (c)).
 In panel (d)-(f) the fluid is confined an external cylinder with  radius $R_o=2$, and the radial position of the Stokeslet are $\xi^1=0$ (panel (d)), $\xi^1=1$ (panel (e)), $\xi^1=1.5$ (panel (f)).}
 
\label{fig3}
\end{figure}

In the case the Stokeslet is oriented along the symmetry axis of the cylinders $Y_3$, the only non-vanishing entry at the cylinder section $x^3=\xi^3$ is $\overline{G}_{3 \, 3}({\pmb x},{\pmb \xi})$. As shown in Figure \ref{fig3} panels (a)-(c), a backflow, which keeps the total flux vanishing in the section, occupies most of the section, while the flow in the same direction as the Stokeslet is restricted to a region (represented in yellow in the figure) with a size of the same order as the distance between the cylinders and representing about $8 \%$ of the total available section for the fluid. This backflow is equivalent to that reported in \cite{liron_stokes_1978} for the case of a Stokeslet in a single cylindrical channel and comes from the incompressibility constraint and the symmetry of the problem, which require that the total flow on the section of the cylinder is equal to zero.
\newpage
However, as shown in Figure \ref{fig3} panels (d)-(f), in the case where the internal cylinder is not present (the expressions for this case are reported and discussed below in the article), the zone in which the fluid velocity has the same direction of the Stokeslet is larger (it can be considered as a circle with radius about $0.4 \, R_o$ representing about $\approx 14 \%$ of the section of the cylinder)
and it slightly restricts when the singularity approaches the cylindrical wall at $\xi^1 = 3/4\,R_o$ (Figure \ref{fig3} panel (f)).

The large backflow region
in case the fluid is between two cylinders affects the sedimentation of particles. In fact, by considering two equal sedimenting spheres (small enough to make valid the singular approximation) and placed at the same radial distance from the center of the cylinder (so that we ensure they moves with the same velocity), it is possible to conclude that the spheres will sediment faster than the single sphere only if they are very close to each other. Despite the screen effect of the internal cylinder, two spheres will slow down each other, even if they are placed at opposite points in the annular region.
 
\subsection{Cases with a single cylinder ($R_i=0$ or $R_o \rightarrow \infty$).}
\label{sec_green_func_1_cil}

By using the properties of block matrices
\cite{bernstein_scalar_2018}, it is possible to express the solution of the system eq. (\ref{eq55}) as
\begin{equation}
{\bf B}_\beta (\xi) \, =- 
\left(
\begin{array}{c}
 ( {\pmb C}^o (R_o) )^{-1}
{\pmb \Delta}^o (R_o,R_i)\,
\left[
{\pmb A}_\beta^{>}(\xi, R_o)
-  {\pmb D}^i(R_o,R_i)\,
{\pmb A}_\beta^{<}(\xi, R_i) 
\right]
\\
[5pt]
 ( {\pmb C}^i (R_i) )^{-1}
{\pmb \Delta}^i (R_o,R_i)
\,
\left[
{\pmb A}_\beta^{<}(\xi,R_i)
-  {\pmb D}^o(R_o,R_i)\,
{\pmb A}_\beta^{>}(\xi,R_o) 
\right]
\end{array}
\right)
\label{eq57}
\end{equation}
where
\begin{equation}
 {\pmb D}^o(R_o,R_i)
={\pmb C}^o\, (R_i)\, 
 ( {\pmb C}^o (R_o) )^{-1},
 \quad
  {\pmb D}^i(R_o,Ri)
={\pmb C}^i\, (R_o)\, 
 ( {\pmb C}^i (R_i) )^{-1}
\label{eq58}
\end{equation}
and
\begin{equation}
\begin{array}{l}
{\pmb \Delta}^o (R_o,R_i)=
\left[
I-
  {\pmb D}^i(R_o,R_i)
 \, 
{\pmb D}^o(R_o,R_i)
 \right]^{-1},
 \\
 [5pt]
 {\pmb \Delta}^i (R_o,R_i)=
 \left[
I -
{\pmb D}^o(R_o,R_i)
\, 
 {\pmb D}^i(R_o,R_i)
 \right]^{-1}
\end{array}
\label{eq59}
\end{equation}
Since the modified Bessel function of the second kind  (and its derivatives) decay exponentially to zeros for any $n$ as their argument tends to infinity,
the matrix $ {\pmb C}^i (R_o)$ and ${\pmb A}^{>}_\beta(\xi,R_o)$  vanish for $R_o \rightarrow \infty$. Therefore, in this limit, the regular part of the Green function is equal to the regular part of the Green function in the domain external to a infinite long pillar
\begin{equation}
\begin{array}{l}
W^b_{\,\, \beta}({\pmb x},{\pmb \xi})=-
\displaystyle \sum^{*}
  \cos(n \phi+p_{(b)} + p_{(\beta)})\cos(\lambda z + q_{(b)} + q_{(\beta)}) \times 
  \\  
  [10pt]
  \hspace{2.2cm}
 \times  \left(\, {\pmb C}^i(x) \left( {\pmb C}^i(R_i) \right)^{-1} {\pmb A}^{<}_\beta (\xi,R_i)
\, \right)_{(b)}
\end{array}
\label{eq60}
\end{equation}
obtained imposing $B_{s,\beta}(\lambda \xi, \lambda, n)=0$ in eqs. (\ref{eq41}) for $s=1,2,3$ and without considering boundary condition at the external surface. {\color{black} The explicit expression for the inverse matrix of ${\pmb C}^i(x)$ entering eq. (\ref{eq60}) is reported in the Appendix \ref{app_entries_C}.} 
Figure \ref{fig4} depicts the streamlines of the Green function external to a cylinder with radius $R_i=1$ for $x^3=\xi^3=0$. The streamlines of the fluid around the cylinder are compared with the streamlines around a sphere, evaluated by the Oseen's expression \cite{oseen_neuere_1927} reported in \cite[pp. 246-247]{kim_1991}, with the same radius $R_i=1$ and with the center at the origin of the coordinate system $(Y_1,Y_2,Y_3)$.
Although the cylindrical obstacle might be approximated by a sphere for the sake of simplicity and the availability of solutions, the figure shows that the difference between the two flows is remarkable. In particular, the vortices occurring in the wake of the cylindrical obstacle are completely absent when the obstacle is spherical. Therefore, the approximation of the cylindrical obstacle with a sphere could be misleading, especially in the case where the interaction between more particles is under investigation.
\begin{figure}
\centering
\includegraphics[scale=0.5]{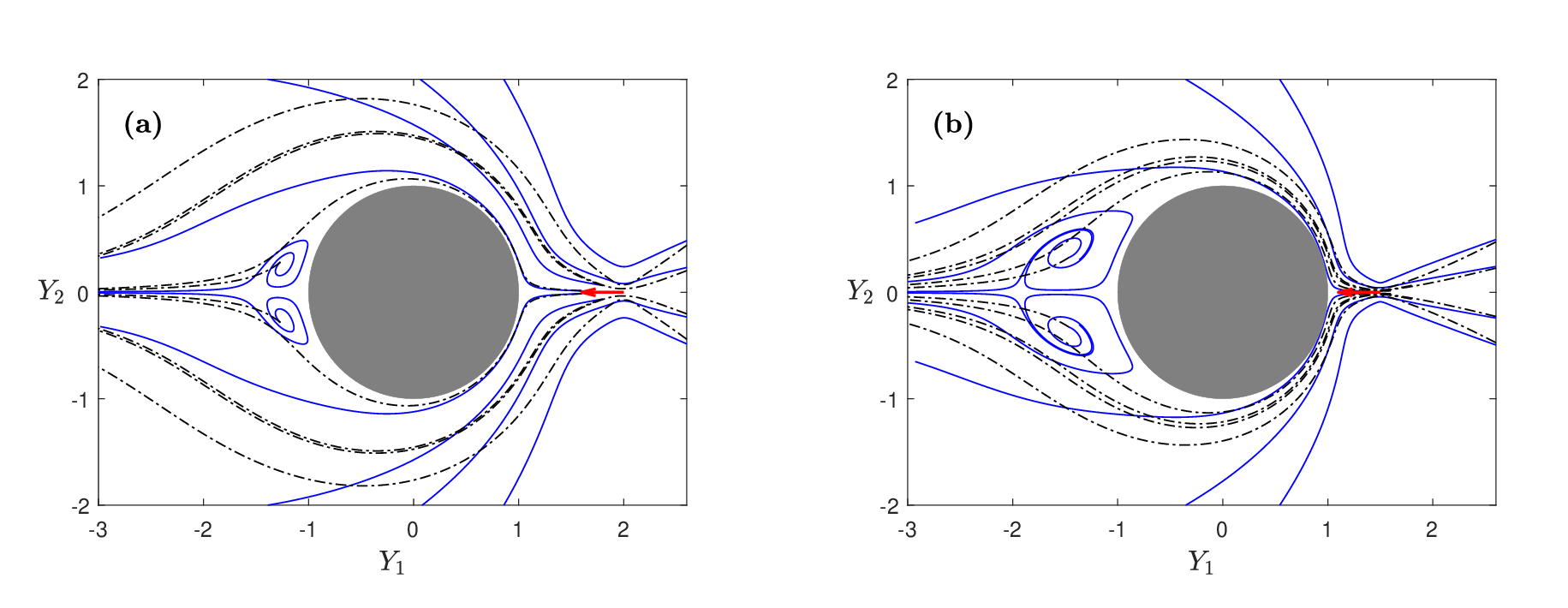}
\caption{Streamlines at the plane $x^3=0$ of a Stokeslet (red arrows) with pole at $\xi^3=0$, $\xi^2=0$ and $\xi^1=2$ (panel (a)) or $\xi^1=1.5$ (panel (b)) near a cylindrical or spherical obstacle. Solid blue lines represent the streamlines in the case the obstacle is a cylinder with radius $R_i=1$ and black dashed-dotted lines represents streamlines in the case the obstacle is a sphere with the same radius.}
\label{fig4}
\end{figure}

\begin{figure}
\centering
\includegraphics[scale=0.5]{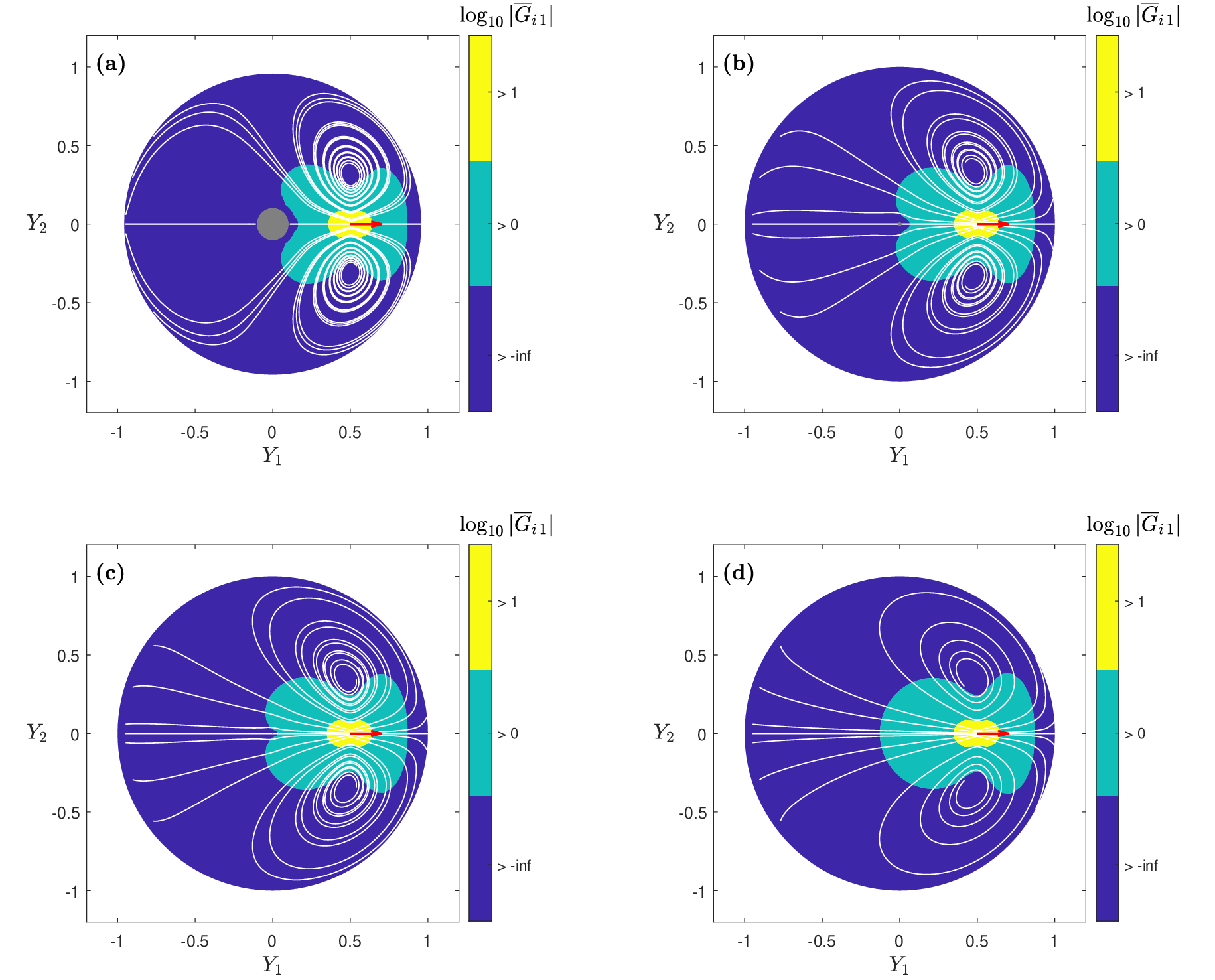}
\caption{Streamlines (withe solid lines) and intensity of the norm of the velocity field at $x^3=\xi^3$ for a Stokeslet placed at $\xi^1=1.5$ in a fluid domain bounded by two concentric cylinders with an external radius $R_o=1$ and an internal radius $R_i=0.1$ (panel (a)), $R_i=0.01$ (panel (b)), $R_i=0.001$ (panel (c)). In panel (d), the solution for the fluid bounded by a single external cylinder with radius $R_o=1$ is depicted. The intensity of the norm of the velocity field is subdivided in three zone: (i) the yellow zone corresponds to an intense flow where $ |\overline{G}_{a\, 1}|\, > \, 10 $, (ii) the turquoise zone corresponds to a moderate flow where $ 1\, < \,|\overline{G}_{a\, 1}|\, \leq \, 10 $, (iii) the blue zone corresponds to a weak flow where $ |\overline{G}_{a\, 1}|\, < \, 1 $.}
\label{fig5}
\end{figure}

On the other hand, for $R_i \rightarrow 0$, the solution corresponding to a Stokeslet bounded by a single cylinder is not recovered. In this limit, in fact, the entries $ {\pmb A}^{<}_\beta(\xi,R_i)$, which depend on $R_i$ through Bessel functions of the first kind, do not vanish. Physically, the limit for $R_i \rightarrow 0$ represents the case in which a thin cord is fixed at the center of the cylinder and, hence, hydrodynamically interacts with the singularity.
This can be seen in Figure \ref{fig5} panels (a)-(c), the streamlines and the norm of the velocity field at $x^3=\xi^3$ are depicted for the flow induced by a Stokeslet radially oriented at the point $\xi^1=0.5$ in a fluid bounded by an external cylinder with radius $R_o=1$ and an internal cylinder with radii $R_i=0.1$ (panel (a)), $R_i=0.01$ (panel (b)) and $R_i=0.001$ (panel (c)). To catch the influence of the internal cylinder in the limit for $R_i\rightarrow 0$, the domain is subdivided in three zone: (i) a zone of intense flow where $ |\overline{G}_{a\, 1}|\, > \, 10 $ (yellow in the figure), (ii) a zone of moderate flow where $1 <|\overline{G}_{a\, 1}|\, \leq \, 10$ (turquoise in the figure), (iii) a zone of weak flow where  $ |\overline{G}_{a\, 1}|\, < \, 1 $ (blue in the figure). The screen effect of the internal cylinder is evident in panel (a) where $R_i=0.1$. However, also in panel (b) where $R_i=0.01$ the moderate zone is "hooked" at the internal cylinder. Even for $R_i=0.001$ the moderate zone cannot overcome the obstacle. On the other hand, by considering the problem without the internal cylinder, the moderate zone occupies all the space around the center of the cylinder.
Specifically, for obtaining the regular part of the Green function in the domain internal to a single cylinder, the constants $B_{s,\beta}(\lambda \xi, \lambda, n)$ in eqs. (\ref{eq41}) for $s=4,5,6$ should be set equal to zero, since they would provide a singularity for $x = \xi \in [0,R_o]$.
Without considering the boundary conditions at the internal surface in eq. (\ref{eq55}), the regular part of the Green function in the cylindrical domain is
\begin{equation}
\begin{array}{l}
W^b_{\,\, \beta}({\pmb x},{\pmb \xi})=-
\displaystyle \sum^{*}
  \cos(n \phi+p_{(b)} + p_{(\beta)})\cos(\lambda z + q_{(b)} + q_{(\beta)}) \times 
  \\  
  [10pt]
  \hspace{2.2cm}
 \times  \left(\, {\pmb C}^o(x) \left( {\pmb C}^o(R_o) \right)^{-1} {\pmb A}^{>}_\beta (\xi,R_o)
\, \right)_{(b)}
\end{array}
\label{eq61}
\end{equation}
 {\color{black} The explicit expression for the inverse matrix of ${\pmb C}^o(x)$ entering eq. (\ref{eq61}) is reported in the Appendix \ref{app_entries_C}. The Green function solution in the domain internal to a cylinder has been addressed in \cite{liron_stokes_1978}
by enforcing the boundary conditions on the Cartesian-Cartesian entries of the Green function. Where the boundary conditions are enforced in the Cartesian-Cartesian coordinate system, the nine unknown coefficients $B_{(i,j)}(\lambda \xi,\lambda,n)$ providing the solution internal to a cylinder are solution of the system
\begin{equation}
\lambda
\left[
\begin{array}{ccc}
\lambda x I'_{n+1}(\lambda x), & I_{n+1}(\lambda x), &- I_{n+1}(\lambda x)
\\
\lambda x I'_{n-1}(\lambda x), & I_{n-1}(\lambda x), & I_{n-1}(\lambda x)
\\
-\lambda x I'_{n-1}(\lambda x) - I_{n}(\lambda x), &- I_{n}(\lambda x), & 0
\end{array}
\right]
\, 
\left[
\begin{array}{c}
{B}_{(1,j)} 
\\
{B}_{(2,j)}
\\
{B}_{(3,j)}
\\
\end{array} \right]
=-
\left[
\begin{array}{c}
{A}^>_{(1 \,j)}
+ x
{A}^>_{(2 \,j)} 
\\
{A}^>_{(1 \,j)}
- x
{A}^>_{(2 \,j)} 
\\
{A}^>_{(3\,j)}
\end{array} \right]
\label{eqS11}
\end{equation}
The coefficient matrix in eq. (\ref{eqS11}) is equivalent to that provided in \cite{liron_stokes_1978}. However, the solution of eq. (\ref{eqS11}) does not fulfill the boundary conditions when employed in the expressions reported in \cite{liron_stokes_1978}. This reveals an evident inconsistency in the solution provided in \cite{liron_stokes_1978}, arising from an inconsistency in the choice of phase constants. In fact, a simple change in the choice of the phase constant and in the sign in eq. (4.16) of \cite{liron_stokes_1978} makes the Liron and Shahar expressions the correct Green function internal to a cylinder. Specifically, in expressions (4.15) and (4.16) of \cite{liron_stokes_1978}, the phase constants must be $\alpha _k^1 = \alpha _k^2 =
  \pi /2$ and the terms $H_k^2,G_k^2,L_k^2$ should have reversed sign.
}

\section{Stokeslet Dipole}
\label{sec_dipole}
In order to obtain {\color{black} the bounded Stokeslet dipole (and, more generally, any higher order multipole of the bounded Green function)}, it is possible to differentiate the solution eq. (\ref{eq56}) obtained in Section \ref{sec_green_function} at the pole point ${\pmb \xi}$. 
From eq. (\ref{eq8}) and from the expression of the Stokeslet dipole eq. (\ref{eqB16}), the Stokes dipole can be expressed as
\begin{equation}
\begin{array}{l}
 \nabla_{\alpha} \,G^b_{\,\, \beta} ({\pmb x},{\pmb \xi}) =
-\dfrac{r^b\, g_{\alpha\, \beta}({\pmb \xi})}{r^3}
+
3\dfrac{r^b\, r_\alpha\, r_\beta}{r^5}
+
\dfrac{r_\beta\, g^b_{\,\, \alpha} ({\pmb x}.{\pmb \xi}) \, - \, r_\alpha \,g^b_{\,\, \beta}({\pmb x},{\pmb \xi})}{r^3}
\\
[10pt]
\hspace{2.2cm}+
\dfrac{\partial \, W^b_{\,\, \beta} ({\pmb x},{\pmb \xi}) }{\partial \xi^{\alpha} }
- \Gamma^\gamma_{\alpha\, \beta} ({\pmb \xi})\,
  W^b_{\,\, \gamma} ({\pmb x},{\pmb \xi})
\end{array}
\label{eq64_8}
\end{equation}
where the $\Gamma^\gamma_{\alpha\, \beta} ({\pmb \xi}) $
are the Christoffel symbols of the second kind reported in eqs. (\ref{eqA7ch}). Focusing on the regular part, we have that
\begin{equation}
\Gamma^\gamma_{\alpha\, \beta} ({\pmb \xi})\,
  W^b_{\,\, \gamma} ({\pmb x},{\pmb \xi})
  =
  - \delta^2_\alpha \, \delta^2_\beta \, \xi \,W^b_{\, 1}({\pmb x},{\pmb \xi})
  \\
  + \left( \delta^1_\alpha \, \delta^2_\beta \, + \, 
  \delta^2_\alpha \, \delta^1_\beta 
    \right) \dfrac{W^b_{\, 2}({\pmb x},{\pmb \xi}) }{\xi^1} 
\label{eq66_}
\end{equation}
and
\begin{equation}
\begin{array}{l}
\dfrac{\partial\, W^b_{\,\, \beta}({\pmb x},{\pmb \xi})}{\partial \xi^1} =-
\displaystyle \sum^{*}
  \cos(n \phi+p_{(b)} + p_{(\beta)})\cos(\lambda z + q_{(b)} + q_{(\beta)}) \times 
  \\  
  [10pt]
  \hspace{2.2cm}
 \times  \left(\,{\bf C}(x) \left( \overline{\bf C}(R_o,R_i) \right)^{-1}  \dfrac{\partial \overline{\bf A}_\beta(\xi,R_o,R_i) }{\partial\, \xi}
\, \right)_{(b)}
\\
  [10pt]
\dfrac{\partial\, W^b_{\,\, \beta}({\pmb x},{\pmb \xi})}{\partial \xi^2} =-
\displaystyle \sum^{*}
  \sin(n \phi+p_{(b)} + p_{(\beta)})\cos(\lambda z + q_{(b)} + q_{(\beta)}) \times 
  \\  
  [10pt]
  \hspace{2.2cm}
 \times  \left( \,{\bf C}(x) \left( \overline{\bf C}(R_o,R_i) \right)^{-1}\,n\, \overline{\bf A}_\beta (\xi,R_o,R_i)
\, \right)_{(b)}
\\
  [10pt]
\dfrac{\partial\, W^b_{\,\, \beta}({\pmb x},{\pmb \xi})}{\partial \xi^3} =-
\displaystyle \sum^{*}
  \cos(n \phi+p_{(b)} + p_{(\beta)})\sin(\lambda z + q_{(b)} + q_{(\beta)}) \times 
  \\  
  [10pt]
  \hspace{2.2cm}
 \times  \left(\,{\bf C}(x) \left( \overline{\bf C}(R_o,R_i) \right)^{-1}\, \lambda \, \overline{\bf A}_\beta (\xi,R_o,R_i)
\, \right)_{(b)}
\end{array}
\label{eq566_}
\end{equation}
In a more compact form, eqs. (\ref{eq566_}) read
\begin{equation}
\begin{array}{l}
\dfrac{\partial\, W^b_{\,\, \beta}({\pmb x},{\pmb \xi})}{\partial \xi^\gamma} =-
\displaystyle \sum^{*}
  \cos(n \phi+p_{(b)} + p_{(\beta)}+ p_{(\gamma)})\cos(\lambda z + q_{(b)} + q_{(\beta)}+ q_{(\gamma)}) \times 
  \\  
  [10pt]
  \hspace{2.2cm}
 \times  \left(\,{\bf C}(x) \left( \overline{\bf C}(R_o,R_i) \right)^{-1}  \overline{\bf A'}_{(\beta,\, \gamma)} (\xi,R_o,R_i)
\, \right)_{(b)}
\end{array}
\label{eq57_}
\end{equation}
where $ \overline{\bf A'}_{(\beta,\, \gamma)}  (\xi,R_o,R_i)$ is given by
\begin{equation}
\begin{array}{l}
\overline{\bf A'}_{(\beta,\, 1)} (\xi,R_o,R_i) = \dfrac{\partial \overline{\bf A}_\beta(\xi,R_o,R_i) }{\partial\, \xi},\quad 
\overline{\bf A'}_{(\beta,\, 2)} (\xi,R_o,R_i) =\,n\,\overline{\bf A}_\beta (\xi,R_o,R_i) 
  \\  
  [10pt]
  \overline{\bf A'}_{(\beta,\, 3)} (\xi,R_o,R_i) =\,\lambda\,\overline{\bf A}_\beta (\xi,R_o,R_i) 
  \end{array}
\label{eq51_}
\end{equation}
In the Cartesian-Cartesian coordinate system, the confined Stokes dipole reads
\begin{equation}
\begin{array}{l}
 \dfrac{\partial \overline{G}_{i \, j}({\pmb x},{\pmb \xi}) }{\partial \overline{\xi}_k} 
= 
-\dfrac{(\overline{x}_i-\overline{\xi}_i)\, \delta_{j\, k}}{r^3}
+
3\dfrac{(\overline{x}_i-\overline{\xi}_i)\, (\overline{x}_j-\overline{\xi}_j)\,(\overline{x}_k-\overline{\xi}_k)}{r^5}
+
\dfrac{(\overline{x}_k-\overline{\xi}_k)\, \delta_{i\, j}  \, - \, (\overline{x}_j-\overline{\xi}_j)\, \delta_{i\, k} }{r^3}
\\
[15pt]
+\, \dfrac{\partial \overline{x}_i}{\partial x^b}
\left(
\dfrac{\partial \xi^\alpha}{\partial \overline{\xi}_k}
\dfrac{\partial \xi^\beta}{\partial \overline{\xi}_j}
\dfrac{\partial\, W^b_{\,\, \beta}({\pmb x},{\pmb \xi}) }{\partial\, \xi^\alpha}\,+ \, \xi^1 \, 
\dfrac{\partial \xi^2}{\partial \overline{\xi}_j}
 \dfrac{\partial \xi^2}{\partial \overline{\xi}_k}
 W^b_{\,\, 1}({\pmb x},{\pmb \xi})
 - \dfrac{1}{\xi^1} \left(
 \dfrac{\partial \xi^1}{\partial \overline{\xi}_j}
 \dfrac{\partial \xi^2}{\partial \overline{\xi}_k}
 +
  \dfrac{\partial \xi^2}{\partial \overline{\xi}_j}
 \dfrac{\partial \xi^1}{\partial \overline{\xi}_k}
  \right) 
  W^b_{\,\, 2}({\pmb x},{\pmb \xi})
 \right)
\end{array}
\label{eq69_}
\end{equation}
By defining the symmetry operator
\begin{equation}
\eta^{\gamma \, \delta}_{\alpha \, \beta}=
\delta^{\gamma}_{\alpha}\, \delta^{\delta}_{\beta}\,
+\,
\delta^{\delta}_{\alpha}\, \delta^{\gamma}_{\beta}
\label{eqB17}
\end{equation}
the confined Stokes dipole can be expressed in term of its symmetric (the first and second terms at the r.h.s of eq. (\ref{eqB16})) and antisymmetric (the third term at the r.h.s of eq. (\ref{eqB16})) parts as
\begin{equation}
\nabla_\alpha\, G^b_{\, \, \beta}({\pmb x},{\pmb \xi})
\,=\,
\dfrac{ \eta^{\gamma \, \delta}_{\alpha \, \beta} }{2}
\,
\nabla_\gamma\, G^b_{\, \, \delta}({\pmb x},{\pmb \xi})
\,+\,
\dfrac{ \varepsilon_{\zeta \alpha \beta}\, \varepsilon^{\zeta \gamma \delta} }{2}
\,
\nabla_\gamma\, G^b_{\, \, \delta}({\pmb x},{\pmb \xi})
\label{eq71_}
\end{equation}
where $\varepsilon_{\zeta \alpha \beta} \sqrt{g({\pmb \xi})} $
and $ \varepsilon^{\zeta \gamma \delta} / \sqrt{g({\pmb \xi})} $
are the absolute covariant and contravariant Ricci--Levi-Civita tensors respectively.

The term $\varepsilon^{\zeta \gamma \delta}\, \nabla_\gamma\, G^b_{\, \, \delta}({\pmb x},{\pmb \xi})\, /\,(2 \sqrt{g({\pmb \xi})}) $ in eq. (\ref{eq71_}), with $ \sqrt{g({\pmb \xi})}) = \xi^1$ in the cylindrical coordinate system,
is referred to as the confined {Couplet} (or also {Rotlet} \cite{blake_fundamental_1974}) and represents the far field 
velocity field of a rigid body rotating without translating at
the point ${\pmb \xi}$.

The first term at the l.h.s of eq. (\ref{eq71_}),
$  \eta^{\gamma \, \delta}_{\alpha \, \beta} 
\,
\nabla_\gamma\, G^b_{\, \, \delta}({\pmb x},{\pmb \xi})/2$, represents the far field flow due to a deforming body, such as a swimming microorganism, and it is referred to as the {Stresslet} (alternatively we could refer to it also as the {Strainlet} \footnotemark[1]).
Given the importance of these two singularities in the study of fluid-particle interaction in the Stokes regime, let us focus on each of them separately.

\subsection{Couplet}

\begin{figure}
\centering
\includegraphics[scale=0.55]{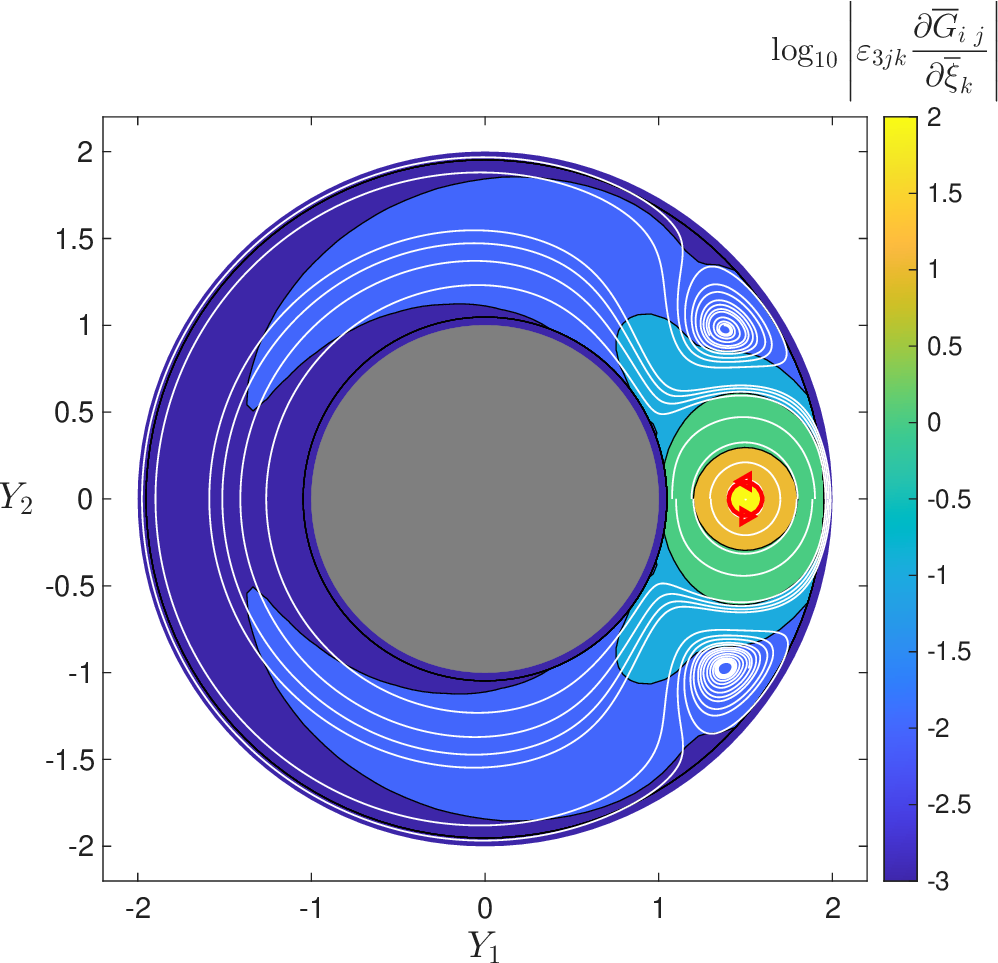}
\caption{Streamlines (white solid lines) and intensity of the norm of the velocity field at $x^3=\xi^3$ induced by an anticlockwise couplet (represented by the red arrows) placed at $\xi^1=1.5$ in a fluid bounded by an internal cylinder with radius $R_i=1$ and by an external cylinder with radius $R_o=2$.}
\label{fig6}
\end{figure}

Since the Christoffel symbol is symmetric, by applying the Ricci--Levi-Civita symbol, the contribute of the last term at the l.h.s of eq. (\ref{eq64_8}) is vanishing. Therefore, the confined Couplet in the cylindrical system reads
\begin{equation}
\dfrac{\varepsilon^{\beta \gamma \delta} }{2\, \xi}
\,
\nabla_\gamma\, G^b_{\, \, \delta}({\pmb x},{\pmb \xi})
=
\dfrac{\varepsilon^{\beta \gamma \delta}\, g^b_{\, \, \gamma}({\pmb x},{\pmb \xi}) }{2 \, \xi }
\, \dfrac{r_\delta}{r^3}
+
\dfrac{\varepsilon^{\beta \gamma \delta} }{2 \, \xi}
\,
\dfrac{\partial \, W^b_{\,\, \delta} ({\pmb x},{\pmb \xi}) }{\partial \xi^{\gamma} }
\label{eq72_}
\end{equation} 
In the Cartesian-Cartesian coordinate system, it can be expressed as
\begin{equation}
\dfrac{\varepsilon_{j k h} }{2}
\,
\dfrac{\partial \,  \overline{G}_{i \, h}({\pmb x},{\pmb \xi})}{\partial \, \overline{\xi}_k}
=
\dfrac{\varepsilon_{i j k}}{2 }
\, \dfrac{(x_k-\xi_k) }{r^3}
+ \dfrac{\varepsilon_{j h k}}{2}\,
\dfrac{\partial \overline{x}_i}{\partial x^b}\,
\dfrac{\partial \xi^\gamma}{\partial \overline{\xi}_h}\,
\dfrac{\partial \xi^\delta}{\partial \overline{\xi}_k}\,
\dfrac{\partial \, W^b_{\,\, \delta} ({\pmb x},{\pmb \xi})}{\partial\, \xi^\gamma}
\label{eq72_}
\end{equation} 
where the relation \cite{synge_tensor_1978}
\begin{equation}
\dfrac{\varepsilon^{\beta \gamma \delta} }{\sqrt{g({\pmb \xi})}}
=
\dfrac{\varepsilon^{\beta \gamma \delta} }{\xi}
=
\varepsilon_{i j k}
\dfrac{\partial \xi^\beta}{\partial \overline{\xi}_i}\,
\dfrac{\partial \xi^\gamma}{\partial \overline{\xi}_j}\,
\dfrac{\partial \xi^\delta}{\partial \overline{\xi}_k}\,
\end{equation}
has been used.
Figure \ref{fig6} shows, at the section of the cylinder $x^3=\xi^3$, the streamlines and the norm of the velocity field due to a Couplet oriented parallel to the symmetry axis of the cylinders $Y_3$ and placed at $\xi^1=(R_i+R_o)/2$. An effect of the Couplet in the annular region is the generation of two connected vortices in the confined fluid with their centers at around $x^1 \approx 1.7$ and $x^2 \approx \xi^2 \pm \pi/5 $. Specifically, for the case of an anticlockwise couplet, the fluid elements are repulsed from the vortex with center at $ x^2 \approx \xi^2 + \pi/5 $ and attracted
by the vortex centred at $ x^2 \approx \xi^2 - \pi/5$. The opposite occurs in the clockwise arrangement.

\subsection{Stresslet}
The confined Stresslet is obtained by applying the operator $\eta^{\gamma \, \delta}_{\alpha \, \beta}/2 $ to the Stokes dipole eq. (\ref{eq64_8}). By using eq. (\ref{eqB18}), we have
\begin{equation}
\begin{array}{l}
\dfrac{ \eta^{\gamma \, \delta}_{\alpha \, \beta} }{2}
\,
\nabla_\gamma\, G^b_{\, \, \delta}({\pmb x},{\pmb \xi})=
\left( 
- \dfrac{g^{\gamma \, \delta}({\pmb \xi}) \, g_{\alpha \,\beta}({\pmb \xi}) }{3}
+
\dfrac{ \eta^{\gamma \, \delta}_{\alpha \, \beta}  }{2}
\right)
3\dfrac{r^b\, r_\gamma\, r_\delta}{r^5}
\\
[10pt]
\hspace{2.2cm}+
\dfrac{ \eta^{\gamma \, \delta}_{\alpha \, \beta} }{2}
\,
\dfrac{\partial \, W^b_{\,\, \delta} ({\pmb x},{\pmb \xi}) }{\partial \xi^{\gamma} }
- \Gamma^\gamma_{\alpha\, \beta} ({\pmb \xi})\,
  W^b_{\,\, \gamma} ({\pmb x},{\pmb \xi})
\end{array}
\label{eq75_}
\end{equation}
In the Cartesian-Cartesian coordinate system, the latter expression attains the form
\begin{equation}
\begin{array}{l}
\dfrac{ \eta_{j k h l} }{2}
\,
 \dfrac{\partial \overline{G}_{i \, l}({\pmb x},{\pmb \xi}) }{\partial \overline{\xi}_h} 
= 
3\left( -\dfrac{\delta_{j\, k} \delta_{h\, l}}{3}+ \dfrac{\eta_{jkhl}}{2} \right)
\dfrac{(\overline{x}_i-\overline{\xi}_i)\, (\overline{x}_h-\overline{\xi}_h)\,(\overline{x}_l-\overline{\xi}_l)}{r^5}
\\
[15pt]
+\, \dfrac{\partial \overline{x}_i}{\partial x^b}
\left(\dfrac{\eta_{jkhl}}{2}
\dfrac{\partial \xi^\alpha}{\partial \overline{\xi}_k}
\dfrac{\partial \xi^\beta}{\partial \overline{\xi}_j}
\dfrac{\partial\, W^b_{\,\, \beta}({\pmb x},{\pmb \xi}) }{\partial\, \xi^\alpha}\,+ \, \xi^1 \, 
\dfrac{\partial \xi^2}{\partial \overline{\xi}_j}
 \dfrac{\partial \xi^2}{\partial \overline{\xi}_k}
 W^b_{\,\, 1}({\pmb x},{\pmb \xi})
 - \dfrac{1}{\xi^1} \left(
 \dfrac{\partial \xi^1}{\partial \overline{\xi}_j}
 \dfrac{\partial \xi^2}{\partial \overline{\xi}_k}
 +
  \dfrac{\partial \xi^2}{\partial \overline{\xi}_j}
 \dfrac{\partial \xi^1}{\partial \overline{\xi}_k}
  \right) 
  W^b_{\,\, 2}({\pmb x},{\pmb \xi})
 \right)
\end{array}
\label{eq76_}
\end{equation}
with 
\begin{equation}
\eta_{jkhl}=\delta_{j\, h}\delta_{k \, l}+\delta_{k\, h}\delta_{j \, l}
\end{equation}
The case where the indexes $j$ and $k$ in eq. (\ref{eq76_}) are equal is of particular interest since it provides a simple model flow generated by force-free microswimmers \cite{berke_hydrodynamic_2008,drescher_fluid_2011,lauga_2020}. Considering a microswimmer with symmetry axis along the direction $j$, the induced flow is 
\begin{equation}
\begin{array}{l}
 \dfrac{\partial \overline{G}_{i \, (j)}({\pmb x},{\pmb \xi}) }{\partial \overline{\xi}_{(j)}} 
= 
-\dfrac{(\overline{x}_i-\overline{\xi}_i) }{r^3}
+3
\dfrac{(\overline{x}_i-\overline{\xi}_i)\, (\overline{x}_{(j)}-\overline{\xi}_{(j)})^2}{r^5}
\\
[15pt]
+\, \dfrac{\partial \overline{x}_i}{\partial x^b}
\left(
\dfrac{\partial \xi^\alpha}{\partial \overline{\xi}_{(j)}}
\dfrac{\partial \xi^\beta}{\partial \overline{\xi}_{(j)}}
\dfrac{\partial\, W^b_{\,\, \beta}({\pmb x},{\pmb \xi}) }{\partial\, \xi^\alpha}\,+ \, \xi^1 \, 
\left(\dfrac{\partial \xi^2}{\partial \overline{\xi}_{(j)}}
\right)^2
 W^b_{\,\, 1}({\pmb x},{\pmb \xi})
 - \dfrac{2}{\xi^1} \left(
 \dfrac{\partial \xi^1}{\partial \overline{\xi}_{(j)}}
 \dfrac{\partial \xi^2}{\partial \overline{\xi}_{(j)}}
  \right) 
  W^b_{\,\, 2}({\pmb x},{\pmb \xi})
 \right)
\end{array}
\label{eq79_}
\end{equation}
where the parenthesis for the index $(j)$ highlight that the Einstein summation is not applied. In fact, the field in eq. (\ref{eq79_}) is not a bitensor since it does not transform as a vector at the pole point ${\pmb \xi}$.

\section{Sourcelet and Sourcelet Multipoles}
\label{sec_sourcelet_sourcedipole}
\subsection{Sourcelet}
By eqs. (\ref{eq14}) and (\ref{eq15}) it is possible to obtain the  Sourcelet confined in the annular region from the expression of the pressure of the Green function eq. (\ref{eq43}).
More specifically, the solution of eq. (\ref{eq9}) can be expressed
as
\begin{equation}
 M^a ({\pmb x},{\pmb \xi}) =
 \dfrac{r^a}{r^3}-\dfrac{g^{a\, b}({\pmb x})}{2}
\, Q_b ({\pmb \xi},{\pmb x}) 
\label{eq66}
\end{equation}
where, by exchanging ${\pmb x} \leftrightarrow {\pmb \xi}$ in eq. (\ref{eq43}), the covariant vector (at the point ${\pmb x}$) $Q_{\,\, b}({\pmb x},{\pmb \xi}) $ reads
\begin{equation}
Q_{\,\, b}({\pmb x},{\pmb \xi})= 2
\displaystyle
 \sum^{*} 
  \cos(n \phi\, -\, p_{(b)})\cos(\lambda z\, -\, q_{(b)}) \,  \lambda^2 \left[
   {B}_{(1,b)} (\lambda\, x, \lambda, n)\, I_n(\lambda \xi)
+
{B}_{(4,b)} (\lambda\, x, \lambda, n)\, K_n(\lambda \xi)
  \right]
\label{eq67}
\end{equation}
with
\begin{equation}
 {B}_{(1,b)} (\lambda\, x, \lambda, n)
 = -\left(\, \left(
 \overline{\bf C}(R_o,R_i)
 \right)^{-1} \overline{\bf A}_b(x,\, R_o,\,R_i) \right)_1
 \label{eq68}
\end{equation}
and
\begin{equation}
 {B}_{(4,b)} (\lambda\, x, \lambda, n)
 = -\left(\, \left(
 \overline{\bf C}(R_o,R_i)
 \right)^{-1} \overline{\bf A}_b(x,\, R_o,\,R_i) \right)_4
 \label{eq69}
\end{equation}
The Cartesian entries $\overline{M}_i({\pmb x},{\pmb \xi})$ of the bounded Sourcelet are
\begin{equation}
\overline{M}_i({\pmb x},{\pmb \xi})=
\dfrac{\overline{x}_i-\overline{\xi}_i}{r^3}
-
\dfrac{g^{a\, b}({\pmb x})}{2}
\dfrac{\partial \overline{x}_i }{\partial x^a} 
\, Q_b ({\pmb \xi},{\pmb x})
 \label{eq70}
\end{equation}

\begin{figure}
\centering
\includegraphics[scale=0.40]{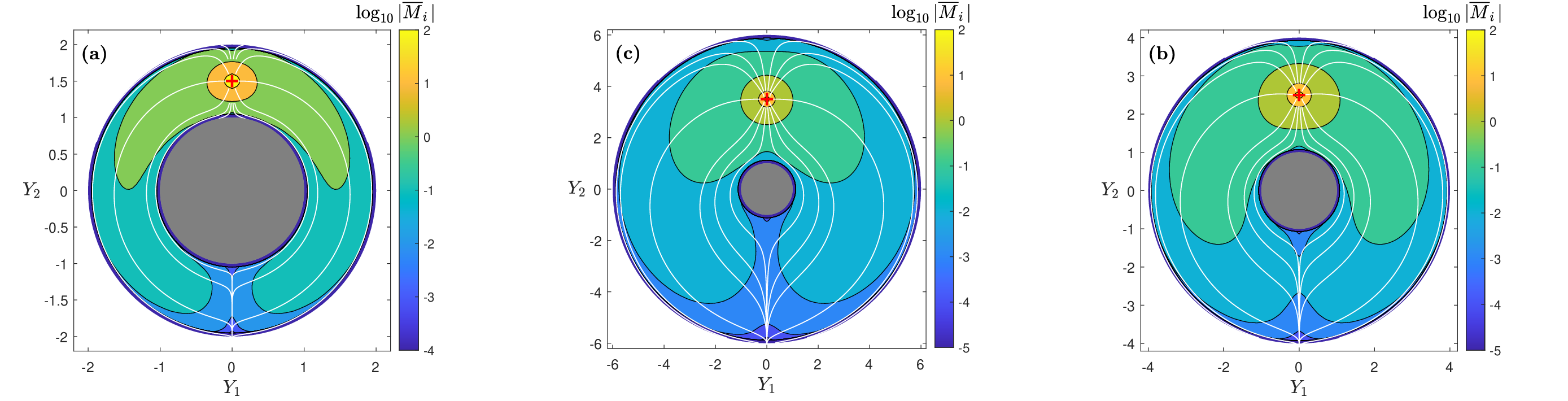}
\caption{Streamlines (white solid lines) and intensity of the velocity field  generated by a Sourcelet at the middle point between an internal cylinder with radius $R_i=1$ and an external cylinder with radius $R_o=2$ (panel (a)), $R_o=4$ (panel (b)) and $R_o=6$ (panel (c))}
\label{fig10}
\end{figure}

Figure \ref{fig10} shows the velocity field, obtained by eq. (\ref{eq70}), generated by a Sourcelet at the middle point between an internal cylinder  with radius $R_i=1$ and an external cylinder with different radii: $R_o=2,4,6$.

\subsection{Sourcelet dipole}
In order to obtain the $n$th order Source singularities
it is possible to differentiate $n$ times eq. (\ref{eq66})
at the pole point ${\pmb \xi}$. Since the Sourcelet is a scalar field at the pole point ${\pmb \xi}$, the covariant derivative corresponds to the partial derivative with respect to the components $\xi^\beta$.
Therefore, the cylindrical-cylindrical coordinates of the Source dipole are
\begin{equation}
\nabla_\beta\, {M}^b({\pmb x},{\pmb \xi})\,=\,
-
\left(
\dfrac{g^b_{\, \, \beta}({\pmb x},{\pmb \xi})}{r^3}
+
3
\dfrac{r^b\,r_\beta}{r^5}
\right) 
-\dfrac{g^{a\, b}({\pmb x})}{2}
\, \dfrac{\partial \,  Q_b ({\pmb \xi},{\pmb x}) }{\partial\, \xi^\beta}
\label{eq88}
\end{equation}
where
\begin{equation}
\begin{array}{l}
\dfrac{\partial \,  Q_b ({\pmb \xi},{\pmb x}) }{\partial\, \xi^1}=
2
\displaystyle
 \sum^{*} 
  \cos(n \phi\, -\, p_{(b)})\cos(\lambda z\, -\, q_{(b)}) \,  \lambda^3 \left[
   {B}_{(1,b)} (\lambda\, x, \lambda, n)\, I_n'(\lambda \xi)
+
{B}_{(4,b)} (\lambda\, x, \lambda, n)\, K_n'(\lambda \xi)
  \right]
  \\
  [15pt]
  \dfrac{\partial \,  Q_b ({\pmb \xi},{\pmb x}) }{\partial\, \xi^2}=
2
\displaystyle
 \sum^{*} 
  \sin(n \phi\, -\, p_{(b)})\cos(\lambda z\, -\, q_{(b)}) \,  \lambda^2\, n\, \left[
   {B}_{(1,b)} (\lambda\, x, \lambda, n)\, I_n(\lambda \xi)
+
{B}_{(4,b)} (\lambda\, x, \lambda, n)\, K_n(\lambda \xi)
  \right]
    \\
  [15pt]
  \dfrac{\partial \,  Q_b ({\pmb \xi},{\pmb x}) }{\partial\, \xi^3}=
2
\displaystyle
 \sum^{*} 
  \cos(n \phi\, -\, p_{(b)})\sin(\lambda z\, -\, q_{(b)}) \,  \lambda^3\, \left[
   {B}_{(1,b)} (\lambda\, x, \lambda, n)\, I_n(\lambda \xi)
+
{B}_{(4,b)} (\lambda\, x, \lambda, n)\, K_n(\lambda \xi)
  \right]
\end{array}
\label{eq89}
\end{equation}
In the Cartesian-Cartesian coordinate system
\begin{equation}
\dfrac{\partial\, \overline{M}_i({\pmb x},{\pmb \xi})}{\partial \, \overline{\xi}_j}
=
\left(
\dfrac{\delta_{i\, j}}{r^3}
-
3
\dfrac{(\overline{x}_i-\overline{\xi}_i)\,(\overline{x}_j-\overline{\xi}_j)}{r^5}
\right) 
-
\dfrac{g^{a\, b}({\pmb x})}{2}
\dfrac{d \overline{x}_i }{d x^a} \dfrac{\partial \xi^\beta}{\partial \overline{\xi}_j}
\, \dfrac{\partial \,  Q_b ({\pmb \xi},{\pmb x}) }{\partial\, \xi^\beta}
 \label{eq90}
\end{equation}
Further derivation at the pole, provides the higher order source singularities.


{\color{black}
\section{Interaction between particles and cylinders}
\label{sec_applications}
As no ambiguity arises between powers and covariant indices in this section, we adopt the standard power notation without parentheses ($\zeta \zeta = \zeta^2$, $\zeta \zeta \zeta= \zeta^3$, etc.).

\subsection{Sedimenting particles}
Deriving the hydrodynamic interaction between colloids and the wall from the Green function of the confinement requires evaluating the regular part of the singularities at their pole, i.e., for ${\pmb x}={\pmb \xi}$.  Beyond the mathematical developments in the literature \cite{happel_1983,kim_1991,procopio_theory_2024}, from a physical point of view, the regular part of the Green function at its pole represents the intensity of the wall-induced advective flow that modifies the hydrodynamic field around the particle.
In order to obtain convenient expressions for the value of the regular part of the Green function at its pole, it is useful to introduce a normalized coordinate with the origin at the mid distance between the cylinders surfaces, namely 
\begin{equation}
{\zeta}= \dfrac{\xi}{L_c}-\hat{R}_i-1
\label{eqzeta}
\end{equation}
where 
\begin{equation}
L_C=(R_o-R_i)/2;\quad \hat{R}_i=R_i/L_c
\label{eq91v2}
\end{equation}
We denote by $w_{i\, j}( \zeta)$ the dimensionless value of the regular part of the Green function at the pole, hence
\begin{equation}
w_{i\, j}( \zeta)
=L_C\,\dfrac{\partial \overline{x}_i}{\partial x^b}\, \dfrac{\partial \xi^\beta}{\partial {\overline \xi}_j}
\, 
W^b_{\,\, \beta}({\pmb x},{\pmb \xi})\bigg|_{{\pmb x}={\pmb \xi}=(\xi_0,0,0)}
\label{eq94}
\end{equation}
where 
\begin{equation}
\xi_0=L_C(\zeta+ 1+\hat{R}_i)
\label{eq93v2}
\end{equation}
Specifically, the quantity $w_{i\, j}(\zeta)$ provides the leading-order term of the additional hydrodynamic force acting on a body due to the presence of the walls. For instance, the drag acting on a sphere with radius $r_S\ll L_C$, sedimenting with velocity $U_j$ and with its center at ${\pmb \xi}$, is given by the relation \cite{kim_1991,procopio_theory_2024}
\begin{equation}
F_i=-
6 \pi \mu r_S\, U_j \left(1 - \dfrac{3}{4}\dfrac{r_S}{L_C} w_{i\, j}(\zeta) \right)
\label{eq95}
\end{equation}
A power series expansion of $w_{i\, j}(\zeta)$ around $\zeta=0$
can be obtained by expanding the Bessel functions with argument $1+\hat{R}_i + \zeta$ that enter the matrices ${\pmb C}(x)$ and $\overline{\bf A}_\beta (\xi,R_o,R_i)$ in eq. (\ref{eq56}), and then numerically integrating the resulting coefficients. 
Considering that $w_{i\, j}(\zeta)=0$ for $i\neq j$, the power series expansion reads
\begin{equation}
w_{i\, i}(\zeta) = -\sum_{n=0}^{\infty} a_{(i,n)}\, \zeta^n
\label{eq64v2}
\end{equation} 
The lower-order terms in the power series in eq. (\ref{eq64v2}) account for the value of $w_{ii}(\zeta)$ far from the walls, while the higher-order terms, which must be truncated to a cut-off order $N_C$ in applications, contribute to $w_{ii}(\zeta)$ for $\zeta \rightarrow \pm 1$.
However, in order to obtain a valid expression in the entire range of $\zeta$, we can express $ w_{i\, i}(\zeta) $
to explicitly incorporate the asymptotic behavior as $\zeta$ approaches the cylindrical walls. In this limit, $w_{i\, i}(\zeta)$ approaches the values corresponding to a domain confined by a plane.
Specifically, for $\zeta \rightarrow 1$, \cite{kim_1991,procopio_theory_2024}
\begin{equation}
-w_{1\, 1}(\zeta) \rightarrow \dfrac{3}{2(1-\zeta)};\quad   -w_{2\, 2}(\zeta)=-w_{3\, 3}(\zeta) \rightarrow \dfrac{3}{4(1-\zeta)}
\label{eq66v2}
\end{equation}
Which, for $\zeta \to 1$, can also expressed as 
\begin{equation}
-w_{1\, 1}(\zeta) \rightarrow \dfrac{3}{(1-\zeta^2)};\quad   -w_{2\, 2}(\zeta)=-w_{3\, 3}(\zeta) \rightarrow \dfrac{3}{2(1-\zeta^2)}
\label{limit_w}
\end{equation}
Considering the 
series expansion of the limit expressions in eqs. (\ref{limit_w}), the series in eqs. (\ref{eq64v2}) can be expressed as follows
\begin{equation}
\begin{array}{l}
\displaystyle
-w_{1\, 1}(\zeta) =\dfrac{3}{1-\zeta^2}
-3\sum_{n=0}^{N_C/2} \, \zeta^{2n}+
\sum_{n=0}^{N_C} a_{(1,n)} \, \zeta^n
+O(\zeta^{N_C+1})
;
\\[20pt]
\displaystyle
-w_{2\, 2}(\zeta) =\dfrac{3}{2(1-\zeta^2)}
-\dfrac{3}{2}\sum_{n=0}^{N_C/2} \, \zeta^{2n}+
\sum_{n=0}^{N_C} a_{(2,n)} \, \zeta^n
+O(\zeta^{N_C+1});
\\[20pt]
\displaystyle
-w_{3\, 3}(\zeta) =\dfrac{3}{2(1-\zeta^2)}
-\dfrac{3}{2}\sum_{n=0}^{N_C/2} \, \zeta^{2n}+
\sum_{n=0}^{N_C} a_{(3,n)} \, \zeta^n
+O(\zeta^{N_C+1});
\end{array}
\label{eq97v2}
\end{equation} 
The values of the coefficients $a_{(i,n)}$ in the power series eq. (\ref{eq64v2}) for different values of the dimensionless radius $\hat{R}_i$ up to order $\zeta^4$ are reported in Table \ref{tab1}.\\

\begin{table}[htbp]
\centering
\begin{tabular}{c|ccccc|ccccc|ccccc}
\hline
$\hat{R}_i$ & $a_{(1,0)}$ & $a_{(1,1)}$ & $a_{(1,2)}$ & $a_{(1,3)}$ & $a_{(1,4)}$
& $a_{(2,0)}$ & $a_{(2,1)}$ & $a_{(2,2)}$ & $a_{(2,3)}$ & $a_{(2,4)}$ 
& $a_{(3,0)}$ & $a_{(3,1)}$ & $a_{(3,2)}$ & $a_{(3,3)}$ & $a_{(3,4)}$
 \\
\hline
$1$ & $1.913$ & $0.406$ & $3.24$ & $-0.194$ & $2.999$ 
& $1.347$ & $0.399$ & $0.916$ & $0.201$ & $1.514$
& $1.366 $ & $ 0.102$ & $ 0.958$ & $ 0.139$ & $ 1.628$
\\
$10$ & $1.935$ & $0.069$ & $3.45$ & $-0.054$ & $2.861$ 
& $1.339$ & $ 0.076$ & $ 1.057$ & $0.024$ & $1.657$
& $1.34 $ & $ 0.02$ & $ 1.059$ & $ 0.018$ & $1.661$
\\
$100$ & $1.935$ & $0.008$ & $3.457$ & $-0.006$ & $2.854$ 
& $1.339 $ & $ 0.008$ & $ 1.063$ & $ 0.003$ & $1.661 $
& $1.339$ & $ 0.002$ & $ 1.063$ & $ 0.002$ & $1.661 $
\\
\hline
\end{tabular}
\caption{\color{black} Values of the coefficients $a_{(i,n)}$ for different values of the dimensionless radius $\hat{R}_i$ up to order $\zeta^4$.}
\label{tab1}
\end{table}

As can be seen from Table \ref{tab1}, the coefficients associated with even powers tend to a constant that satisfies the planar limit. Specifically, $a_{(1,2n)} \to 3$ and $a_{(2,2n)} \to a_{(3,2n)} \to 3/2$.
Furthermore, the coefficients associated with odd powers, $a_{(i,2n+1)}$, approach zero and $a_{(2,n)} \rightarrow a_{(3,n)}$ as $\hat{R}_i \rightarrow \infty$. In this limit, while keeping $L_c$ and $\zeta$ constant, the solution in eq. (\ref{eq56}) approaches the Green function confined by two parallel walls. The vanishing coefficients $a_{(i,2 n+1)}$ are therefore a consequence of the symmetry of a Stokeslet confined by two plane walls, for which $w_{i\, i}(\zeta)=w_{i\, i}(-\zeta)$ and $w_{2\, 2}(\zeta)=w_{3\, 3}(\zeta)$.
We can therefore exploit this limit to obtain the expressions for $w_{ii}$ for two parallel planes, which have been numerically calculated and tabulated in \cite{swan2010particle} (see the Supplementary Material of \cite{swan2010particle} where the factors $3 w_{i\, i}(\zeta)/2$ are tabulated as $f^{(UF)}_1$ and $g^{(UF)}_1$).
In the limit where the two cylinders become parallel planes (hence, for $\hat{R}_i\rightarrow \infty$ with $L_c$ and $\zeta$ kept constant), we obtain
\begin{equation}
\begin{array}{l}
-w_{1\, 1}(\zeta)= 1.935+3.457\,\zeta^2+2.854\, \zeta^4 +O(\zeta^6);
\\
-w_{2\, 2}(\zeta)=-w_{3\, 3}(\zeta)=1.339+1.063\, \zeta^2+1.661\, \zeta^4+O(\zeta^6); 
\end{array}
\label{eq65v2}
\end{equation}
The expressions in eq. (\ref{eq65v2}) can be further improved by using eqs. (\ref{eq97v2}) to take into account the limit $\zeta \rightarrow \pm 1$. 
According to eqs. (\ref{eq97v2}), the expressions for $w_{i\, i}(\zeta)$ valid for any $\zeta$ between two parallel walls read
\begin{equation}
\begin{array}{ll}
-w_{1\, 1}(\zeta)= -1.065+0.457\,\zeta^2-0.146\, \zeta^4 + \dfrac{3}{1-\zeta^2};
\\
[10pt]
-w_{2\, 2}(\zeta)=-w_{3\, 3}(\zeta)=-0.161-0.437\, \zeta^2+0.161\, \zeta^4+\dfrac{3}{2(1-\zeta^2)} 
\end{array}
\label{eq67v2}
\end{equation}
Both the truncated relations in eq. (\ref{eq65v2}) and the matched relations in eq. (\ref{eq67v2}) are depicted in Figure \ref{fig11} and compared with the numerical results obtained in \cite{swan2010particle}. As can be observed, the results obtained from eq. (\ref{eq67v2}) show perfect agreement with the numerical results from \cite{swan2010particle} over the entire range of $\zeta$.

\begin{figure}
\centering
\includegraphics[scale=0.75]{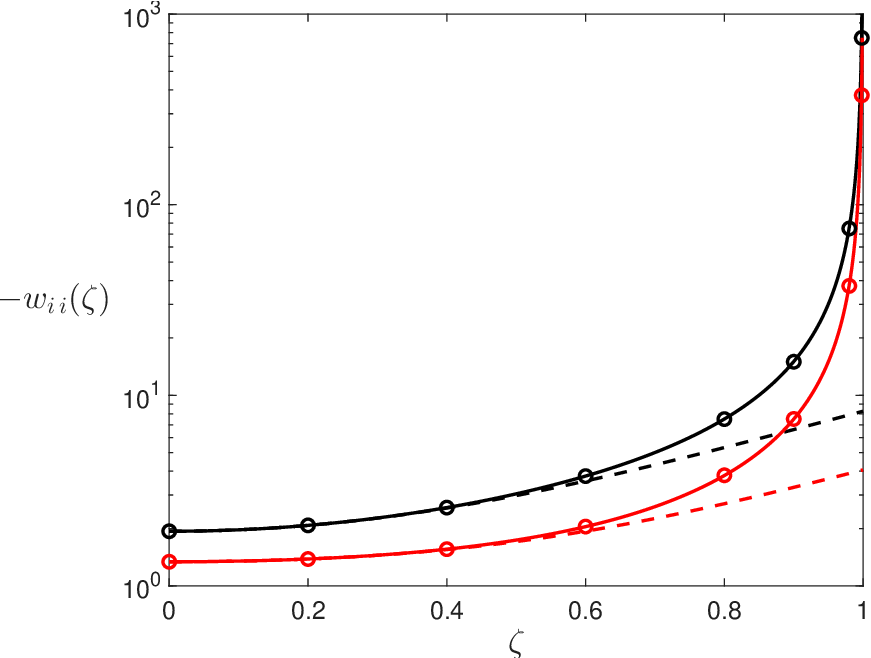}
\caption{ \color{black} Dimensionless regular part of the Green function at its pole, $w_{1\, 1}(\zeta)$ (black) and $w_{2\, 2}(\zeta) = w_{3\, 3}(\zeta)$ (red) in the domain between two parallel walls as a function of the dimensionless distance $\zeta$ from the midpoint between two parallel walls. Dashed lines are obtained using the truncated relation in eq. (\ref{eq65v2}), solid lines using the relation in eq. (\ref{eq67v2}), and symbols represent the numerical results obtained in \cite{swan2010particle} (see the Supplementary Material to \cite{swan2010particle}).}
\label{fig11}
\end{figure}

Similar expressions can be obtained for the Green functions bounded by a single cylinder. In these cases, the definition eq. (\ref{eq94}) and the expression in eq. (\ref{eq95}) for the drag acting on a small sphere ($r_S \ll R_o$ or $r_S\ll R_i$) holds once it is assumed that 
\begin{equation}
L_C=R_o;\quad \xi_0= \zeta\, R_o
\label{def1_LC}
\end{equation}
for the case where only the external cylinder is present, and 
\begin{equation}
L_C=R_i;\quad \xi_0= \zeta \,R_i
\label{def2_LC}
\end{equation}
when only the internal cylinder is present. Also in both these cases, $w_{i\, j}(\zeta)$ vanishes for $i \neq j$.

For the Green function internal to a cylindrical domain, the dimensionless regular part at the pole reads
\begin{equation}
\begin{array}{ll}
-w_{1\, 1}(\zeta)= 2.406+3.334\,\zeta^2+2.879\, \zeta^4 + O(\zeta^6);
\\
[10pt]
-w_{2\, 2}(\zeta)=2.406+1.2\, \zeta^2+1.458\, \zeta^4+ O(\zeta^6); 
\\
[10pt]
-w_{3\, 3}(\zeta)=2.805-0.931\, \zeta^2+2.365 \zeta^4+ O(\zeta^6) 
\end{array}
\label{eq100}
\end{equation}
Eqs. (\ref{eq100}) agrees with the expressions provided, up to second order, in \cite{hasimoto_slow_1976}. Considering that as $\zeta \to 1$, the regular part at the pole of the Green function in a cylindrical domain becomes equal to that bounded by a plane wall. From the matched expressions eqs. (\ref{eq97v2}), we obtain
\begin{equation}
\begin{array}{ll}
-w_{1\, 1}(\zeta)= -0.594+0.334\,\zeta^2-0.121\, \zeta^4 +\dfrac{3}{1-\zeta^2}+ O(\zeta^6);
\\
[10pt]
-w_{2\, 2}(\zeta)=0.906-0.38\, \zeta^2-0.042\, \zeta^4+\dfrac{3}{2(1-\zeta^2)}+ O(\zeta^6); 
\\
[10pt]
-w_{3\, 3}(\zeta)=1.306-2.431\, \zeta^2+0.865, \zeta^4
+\dfrac{3}{2(1-\zeta^2)}+ O(\zeta^6)
\end{array}
\label{eq101}
\end{equation}
\begin{figure}
\includegraphics[scale=0.75]{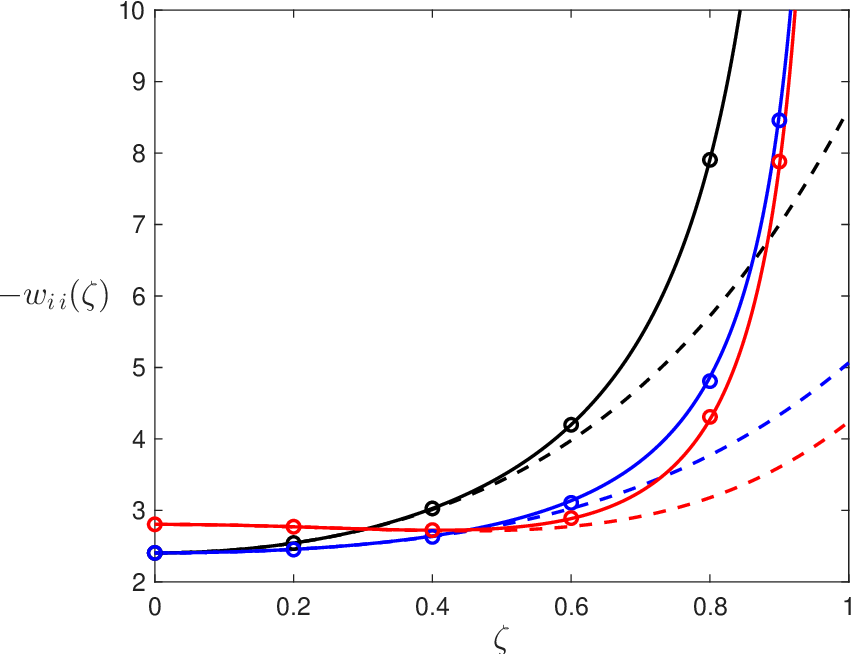}
\caption{ \color{black} Dimensionless regular part of the Green function at its pole, $w_{1\, 1}(\zeta)$ (black), $w_{2\, 2}(\zeta)$ (blue), and $w_{3\, 3}(\zeta)$ (red), in the domain internal to a cylinder as a function of the dimensionless distance $\zeta$ from the cylinder axis. Dashed lines are obtained using the truncated relation in eq. (\ref{eq100}), solid lines using the relation in eq. (\ref{eq101}), and symbols represent the numerical results obtained from the direct calculation of $w_{i\, i}(\zeta)$ at each point using eq. (\ref{eq61}).}
\label{fig12}
\end{figure}
As shown in \cite{md_shamsul_alam_slow_1980} and \cite{tanasijevic_hydrodynamic_2021} the regular part of the Green function at the pole $w_{i\, i}(\zeta)$ in the domain external to a cylinder decay as $1/(\zeta \,\log(\zeta))$ in the limit $\zeta = \xi/R_i \rightarrow \infty$. This behavior is due to the modified Bessel functions of the second kind entering the matrix $\left( {\pmb C}^i(R_i) \right)^{-1}$, for which $K_0(x) \sim -\log(x)$ as $x \rightarrow 0$.
In this limit, the force in eq. (\ref{eq95}) (with $L_C=R_i$) represents the hydrodynamic interaction between a translating small sphere and a slender fiber, since it is valid when both the cylinder radius and the sphere radius are much smaller than the distance between the sphere and the cylinder, hence for $r_S\ll \xi $ and $R_i \ll \xi$. Expanding eq. 
eq. (\ref{eq60}) in logarithmic powers around $R_i=0$ and using the definition in eq. (\ref{eq94}) (enforcing $x=\xi=R_i \, \zeta$ and $L_c=R_i$), we obtain
\begin{equation}
\begin{array}{ll}
-w_{1\, 1}(\zeta)=
\\
[10pt]
 \dfrac{\pi}{\zeta}\left(
 \dfrac{39}{32 \log(2\, \zeta)} -\dfrac{  117 \log(4)-22 }{64\left( \log\left(2\, \zeta\right) \right)^2} +
 \dfrac{\pi\, ( 52\, \pi^2-9 + (351 \log (4) - 132) \log (4) }{128 \left( \log(2\, \zeta) \right)^3}+
 O\left(\dfrac{1}{\log(\zeta)}\right)^4 \right);
\\
[10pt]
-w_{2\, 2}(\zeta)=
\\[10pt]
 \dfrac{\pi}{\zeta}\left(
\dfrac{1}{2 \log (2\,\zeta)}
-
\dfrac{(1 + 3\log (4) )}{4 \left( \log(2\,\zeta)  \right)^2}+
\dfrac{4\pi^2 + 3(1 + 3\log(4))^2}{24 \left( \log(2\,\zeta)  \right)^3}
+
 O\left(\dfrac{1}{\log(\zeta)}\right)^4 \right)
 ; 
\\
[10pt]
-w_{3\, 3}(\zeta)= 
\\[10pt]
 \dfrac{\pi}{\zeta}\left(
 \dfrac{21}{32 \log (2\,\zeta)}
 -
 \dfrac{3 (9\log (4) + \log (8)-1) }{32 \left( \log (2\, \zeta) \right)^2}
 +
 \dfrac{ (5 + 28\pi^2 - 36\log 4 + 189(\log 4)^2)}{128 \left( \log(2\,\zeta) \right)^3}+
 O\left(\dfrac{1}{\log(\zeta)}\right)^4 \right)
 ; 
\end{array}
\label{eq102}
\end{equation}

\begin{figure}
\includegraphics[scale=0.75]{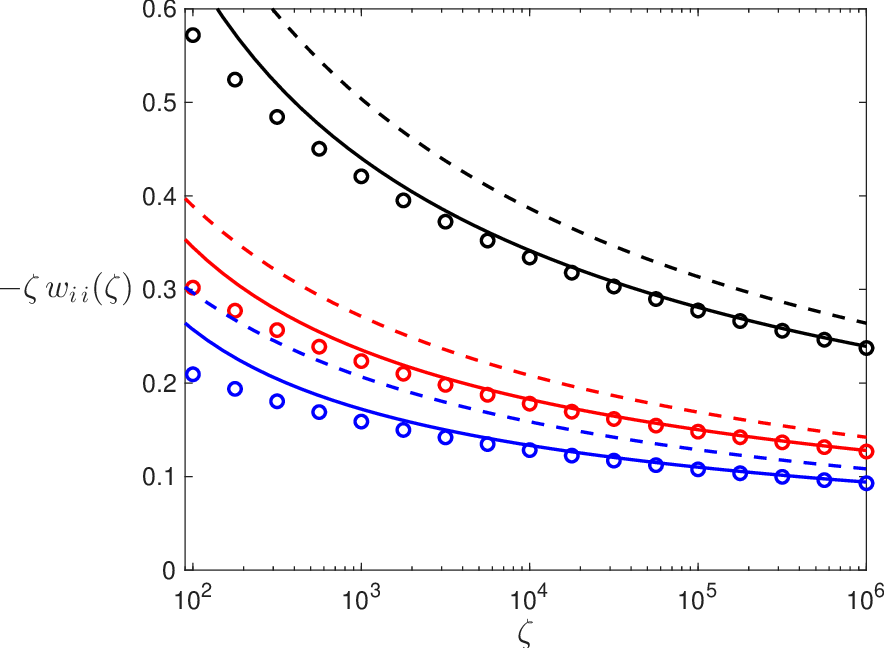}
\caption{\color{black} Dimensionless regular part of the Green function at its pole in the domain external to a cylinder multiplied by $\zeta$: $\zeta\, w_{1\, 1}(\zeta)$ (black), $\zeta\, w_{2\, 2}(\zeta)$ (blue), and $\zeta\, w_{3\, 3}(\zeta)$ (red), as a function of the dimensionless distance $\zeta$ from the cylinder axis. Dashed lines are obtained using the
limit trend obtained in \cite{md_shamsul_alam_slow_1980}, solid lines using the relation in eq.~(\ref{eq102}), and symbols represent the numerical results obtained from the direct calculation of $w_{i\, i}(\zeta)$ at each point using eq.~(\ref{eq61}).}
\label{fig13}
\end{figure}
As shown in Figure~\ref{fig13}, where $-\zeta\, w_{i\,i}(\zeta)$ is plotted for large values of $\zeta$, the asymptotic limit obtained in \cite{md_shamsul_alam_slow_1980} is reached very slowly as the distance from the cylindrical wall increases. The relations in eqs.~(\ref{eq102}) improve the convergence to the asymptotic limit, which is reached at approximately $\zeta \approx 10^3$ -- $10^4$.

\subsection{Microswimmers}

Considering, without loss of generality, a microswimmer being at $\xi^2=0$ and  $\xi^3=0$ between two concentric cylinders (with characteristic size much smaller than $L_C$), the attractive (or repulsive) leading order hydrodynamic force through the walls $F_1$ is proportional by a constant $c_{(j)}>0$ to the intensity of the strain $\mathcal{S}$ introduced by the microswimmer into the fluid and to the regular part term in eq. (\ref{eq79_}) evaluated at ${\pmb x}={\pmb \xi}$, where $c_{(j)}$ depends on the specific geometry of the microswimmer \cite{lauga_2020,procopio_theory_2024}. Namely,
\begin{equation}
F_1=\dfrac{ \mu \, \mathcal{S}\, c_{(j)}\, u_{(j)}(\zeta)}{L_C^2}
\label{eq105}
\end{equation}
where $u_{(j)}(\zeta)$ is the regular part of the 
confined stresslet at its pole.
In the case of a microswimmer oriented along the radial direction, hence for $j=1$, we have
\begin{equation}
u_{(1)}(\zeta)
=L_C^2\,\dfrac{\partial\, W^1_{\,\, 1}({\pmb x},{\pmb \xi}) }{\partial\, \xi^1}\bigg|_{{\pmb x}={\pmb \xi}=(\xi_0,0,0)}
\label{eq80}
\end{equation}
where $L_C$ and $\xi_0$ are defined in eqs. (\ref{eq91v2}) and (\ref{eq93v2}).
In the case the microswimmer is oriented parallel to the walls along the angular direction ($j=2$), the radial force is
\begin{equation}
u_{(2)}(\zeta)
=
\dfrac{L_C^2}{(\xi^1)^2}
\dfrac{\partial\, W^1_{\,\, 2}({\pmb x},{\pmb \xi}) }{\partial\, \xi^2}\bigg|_{{\pmb x}={\pmb \xi}}\,+\, \dfrac{L_C^2}{\xi^1} 
 W^1_{\,\, 1}({\pmb x},{\pmb \xi})\bigg|_{{\pmb x}={\pmb \xi}=(\xi_0,0,0)}
\label{eq81}
\end{equation}
Finally, for a microswimmer oriented parallel to the symmetry axis of the cylinders ($j=3$), we have
\begin{equation}
u_{(3)}(\zeta)
=L_C^2
\dfrac{\partial\, W^1_{\,\, 3}({\pmb x},{\pmb \xi}) }{\partial\, \xi^3}\bigg|_{{\pmb x}={\pmb \xi}=(\xi_0,0,0)}
\label{eq82}
\end{equation}
The regular part of the stresslet at its pole
$u_{(i)}(\zeta)$ can be expressed in a power series around $\zeta=0$, hence
\begin{equation}
u_{(i)}(\zeta)= \sum_{n=0}^{\infty} b_{(i,n)}\, \zeta^n
\label{eq109}
\end{equation}
Similarly to eq. (\ref{eq97v2}) for $w_{ii}(\zeta)$, $u_{(i)}(\zeta)$ can be expressed to explicitly incorporate the asymptotic behavior as $\zeta \rightarrow \pm 1$, corresponding to the solution near a plane wall, reading
\begin{equation}
u_{(1)}(\zeta) \rightarrow \dfrac{3}{4(1-\zeta)^2};
\quad
u_{(2)}(\zeta)= u_{(3)}(\zeta)\rightarrow -\dfrac{3}{8(1-\zeta)^2}
\label{eq110}
\end{equation}
that, for $\zeta\rightarrow 1$, can be expressed also as
\begin{equation}
u_{(1)}(\zeta) \rightarrow \dfrac{3}{(1-\zeta^2)^2};
\quad
u_{(2)}(\zeta)= u_{(3)}(\zeta)\rightarrow -\dfrac{3}{2(1-\zeta^2)^2}
\label{limit_u}
\end{equation}
Taking into account eq. (\ref{limit_u}), the expressions for  $u_{(i)}(\zeta)$ can be developed as follows
\begin{equation}
\begin{array}{l}
\displaystyle
u_{(1)}(\zeta) = -\dfrac{3 \,\zeta}{(1-\zeta^2)^2}
+3\sum_{n=0}^{(N_C-1)/2} \binom{2+n-1}{n}\, \zeta^{2 n+1} 
 +\sum_{n=0}^{N_C} b_{(1,n)} \, \zeta^n
 +O(\zeta^{N_C+1});
\\[20pt]
\displaystyle
u_{(2)}(\zeta) = \dfrac{3 \,\zeta}{2(1-\zeta^2)^2}
-\dfrac{3}{2}\sum_{n=0}^{(N_C-1)/2} \binom{2+n-1}{n}\, \zeta^{2 n+1} 
 +\sum_{n=0}^{N_C} b_{(2,n)} \, \zeta^n
 +O(\zeta^{N_C+1});
\\[20pt]
\displaystyle
u_{(3)}(\zeta) =\dfrac{3 \,\zeta}{2(1-\zeta^2)^2}
-\dfrac{3}{2}\sum_{n=0}^{(N_C-1)/2} \binom{2+n-1}{n}\, \zeta^{2 n+1} 
 +\sum_{n=0}^{N_C} b_{(3,n)} \, \zeta^n
 +O(\zeta^{N_C+1})
\end{array}
\label{eq111}
\end{equation} 
The coefficients $b_{(i,n)}$ for different values of $\hat{R}_i$ have been computed using eqs. (\ref{eq80})--(\ref{eq82}) and (\ref{eq56}) and are reported in Table \ref{tab2} up to order $\zeta^4$.\\

\begin{table}[htbp]
\centering
\begin{tabular}{c|ccccc|ccccc|ccccc}
\hline
$\hat{R}_i$ & $b_{(1,0)}$ & $b_{(1,1)}$ & $b_{(1,2)}$ & $b_{(1,3)}$ & $b_{(1,4)}$
& $b_{(2,0)}$ & $b_{(2,1)}$ & $b_{(2,2)}$ & $b_{(2,3)}$ & $b_{(2,4)}$ 
& $b_{(3,0)}$ & $b_{(3,1)}$ & $b_{(3,2)}$ & $b_{(3,3)}$ & $b_{(3,4)}$
 \\
\hline
$1/75$ & $-0.54$ & $-2.175$ & $-1.116$ & $-4.538$ & $-1.714$
&$0.155$ & $1.007$ & $1.275$ & $2.523$ & $0.326$
&$0.385$ & $1.169$ & $0.833$ & $2.016$ & $1.388$
\\
$2/15$ & $-0.421$ & $-2.612$ & $-0.366$ & $-5.539$ & $-0.441$ 
& $0.103$ & $1.269$ & $-0.2$ & $3.177$ & $-0.451$
& $0.318$ & $1.343$ & $0.565$ & $2.362$ & $0.892$
\\
$2$ & $-0.131$ & $-3.366$ & $0.257$ & $-5.866$ & $-0.096$ 
& $0.017$ & $1.69$ & $-0.345$ & $3.03$ & $-0.1$
& 
$0.113$ & $1.679$ & $0.087$ & $2.830$ & $0.171$
\\
$100$ & $-0.004$ & $-3.457$ & $0.009$ & $-5.708$ & $-0.005$
& $0$ & $1.728$ & $0$ & $2.853$ & $0$
& $0.003$ & $1.728$ & $0.002$ & $2.854$ & $0.005$

\\
\hline
\end{tabular}
\caption{\color{black}
Values of the coefficients $b_{(i,n)}$ for different values of the dimensionless radius $\hat{R}_i$ up to order $\zeta^4$.}
\label{tab2}
\end{table}

In Figure \ref{fig7}, the expressions reported in eq. (\ref{eq111}) with $N_C=4$ and coefficients $b_{(i,n)}$ from Table \ref{tab2} are compared with exact expressions (\ref{eq80})--(\ref{eq82}) and (\ref{eq56}).
As can be observed in Figure \ref{fig7} panel (a), where $u_{(i)}(\zeta)$ is reported as a function of the normalized position $\zeta$, the cylindrical walls induce a repulsive hydrodynamic force on the microswimmer when it is radially oriented (note that, based on the dipolar description of the flow due to a microswimmer, it does not depend on the direction of motion). Specifically, 
within a region, starting at the inner cylinder's surface and ending slightly before the halfway point between the two surfaces, the microswimmer is repelled by the inner surface. Beyond this point, it is repelled by the outer surface.
As the internal radius decreases (or the distance between the cylinders increases), the point at which the force reverses direction moves closer to the inner surface (see the inset).
Contrarily, as shown in panels (b) and (c) of Figure \ref{fig7}, when the microswimmer is oriented parallel to the surfaces (hence angularly and axially), the microswimmer is attracted by the inner surface in a zone starting from the inner cylinder's surface and ending slightly before the halfway point between the two surfaces. In the remainder zone, it s attracted by the outer surface.
\begin{figure}
\centering
\includegraphics[scale=0.50]{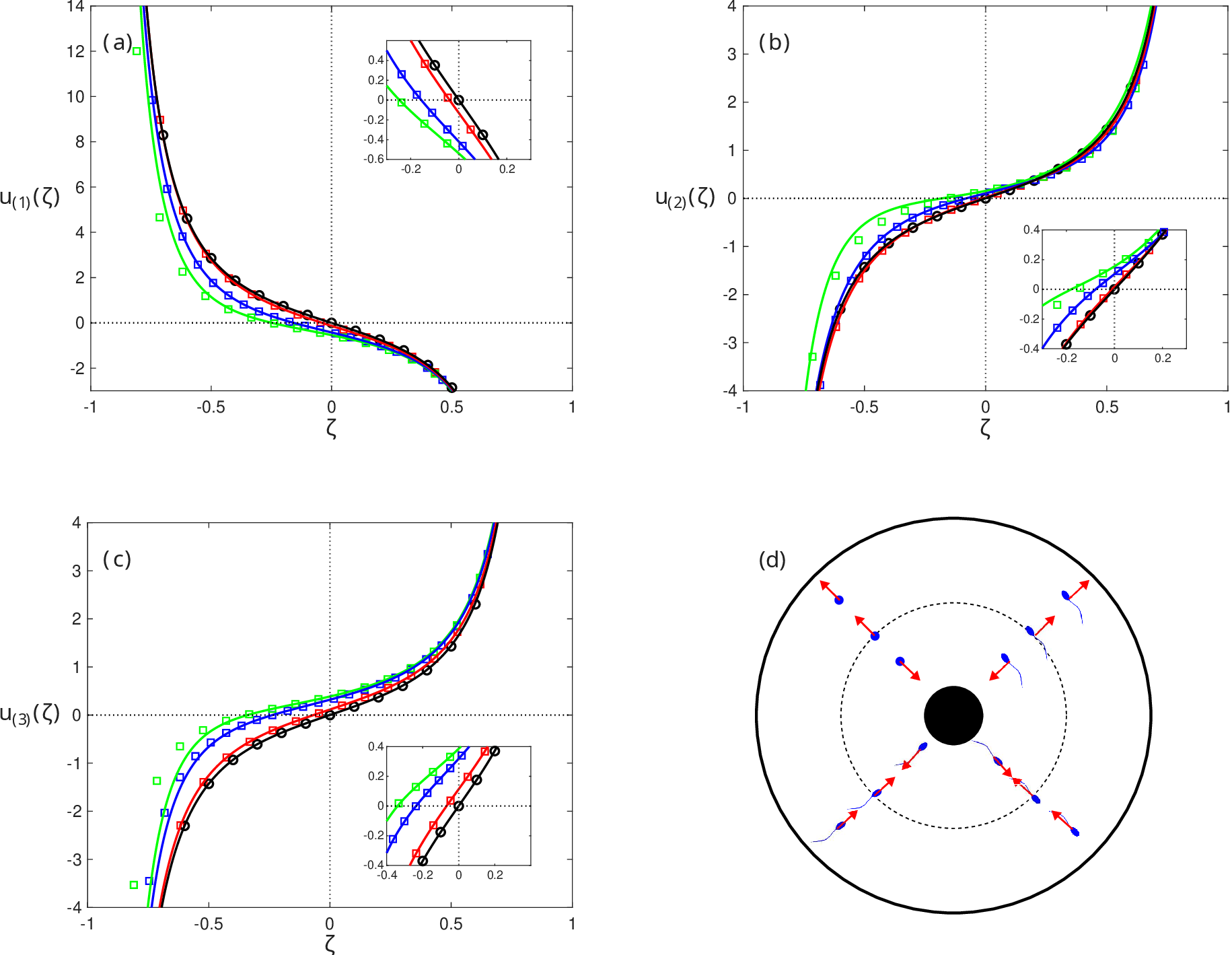}
\caption{ \color{black}
Dimensionless stresslet experienced by a microswimmer in the annular region between two concentric cylinders as a function of the dimensionless distance $\zeta$ from the midpoint between the cylinders for different orientations: (a) radial, (b) angular, and (c) axial. Black curves and symbols represent the limit of two parallel planes ($\hat{R}_i\rightarrow \infty$), red corresponds to $\hat{R}_i = 2$, blue to $\hat{R}_i = 2/15$, and green to $\hat{R}_i = 1/75$. Solid lines are obtained from eq. (\ref{eq111}) using the coefficients in Table \ref{tab2}, circles represent the numerical results reported in \cite{swan2010particle} for parallel planes, and squares are exact expressions computed point-by-point using eqs. (\ref{eq80})--(\ref{eq82}) and (\ref{eq56}). Insets in panels (a), (b), and (c) show zoom-in views of the values near $\zeta = 0$. Panel (d) shows a schematic representation of pusher microswimmers, where the orientation is defined by the head-tail structure, the dashed circle represents the midpoint between the cylinders, and red arrows indicate the direction of the hydrodynamic attraction/repulsion force induced by the cylindrical walls (for pullers, the arrows would be reversed).}
\label{fig7}
\end{figure}

Carrying out 
the limit for $\hat{R_i} \rightarrow \infty$, keeping 
$L_c$ and $\zeta$ constant, we obtain
the coefficients for $u_{(i)}(\zeta)$ associated with a microswimmer between two parallel walls. In this limit, $ u_{(i)}(\zeta)$ reads
\begin{equation}
\begin{array}{l}
u_{(1)}(\zeta)=-3.457\, \zeta - 5.708\, \zeta^3 + O\left(\zeta^5\right)
\\[10pt]
u_{(2)}(\zeta)=u_{(3)}(\zeta)=1.728\, \zeta + 2.854\, \zeta^3 + O\left(\zeta^5\right)
\end{array}
\end{equation} 
Using the representation eqs. (\ref{eq111}), the expression for $u_{(i)}(\zeta)$ between two parallel plane walls, valid over the entire range of $\zeta$, read
\begin{equation}
\begin{array}{l}
u_{(1)}(\zeta)=-0.457\, \zeta + 0.292\, \zeta^3 - \dfrac{3	\, \zeta}{(1-\zeta^2)^2}+ O\left(\zeta^5\right)
\\[20pt]
u_{(2)}(\zeta)=u_{(3)}(\zeta)=0.228\, \zeta - 0.146\, \zeta^3 +
\dfrac{3	\, \zeta}{2(1-\zeta^2)^2}+
O\left(\zeta^5\right)
\end{array}
\label{eq113}
\end{equation} 
The values of the stresslet at its pole computed with eqs. (\ref{eq113}) are reported in Figure \ref{fig7} panels (a)--(c) and compared with the numerical results obtained  in \cite{swan2010particle}. In the supplementary material of \cite{swan2010particle}, $3 u_{(1)}(\zeta)/2$ is tabulated as $g_2^{\text{(US)}}$ and $3 u_{(3)}(\zeta)$ correspond to $-2 g_2^{\text{(US)}}$ (the minus sign does not appear in eq. (37) of the main text, which is likely a typographical error, as the single-plane limit reported in \cite{berke_hydrodynamic_2008} would otherwise be violated). As can be observed, eqs. (\ref{eq113}) fit perfectly the numerical results over the entire range of $\zeta$.
Due to the symmetry of the problem, in this limit $u_{(i)}(\zeta)$ changes sign at the midpoint between the two planes.

In order to compare the intensity of the hydrodynamic interaction induced by the walls, Figure \ref{fig8} depicts the modulus $|u_{(i)}(\zeta)|$ for the three different principal orientations $j=1,2,3$ of the microswimmer.
The repulsive flow with respect to the closest surface, generated by the radially oriented microswimmer, 
is almost everywhere (apart from a small region around the point where it changes sign) more intense than the attractive flows generated by the microswimmer oriented parallel to the surfaces. The internal surface results more attractive when the microswimmer is angularly oriented than when it is axially oriented, possessing a wider attractive region and, in this region, higher intensity. Conversely, the outer surface is more attractive for axially oriented microswimmers.

In the case that the fluid is bounded by a single cylinder, eqs. (\ref{eq105})--(\ref{eq82}) hold by considering the definitions for $L_C$ and $\xi_0$ given in eq. (\ref{def1_LC}) when the fluid is bounded externally and in eq. (\ref{def2_LC}) when the fluid is bounded by an internal cylinder.
In the case where only the external cylinder is present, the power expansion of $u_{(i)}(\zeta)$ around $\zeta=0$ reads
\begin{equation}
\begin{array}{l}
u_{(1)}=-3.334\, \zeta-5.758\, \zeta^3+O(\zeta^5);
\\[10pt]
u_{(2)}=0.808\, \zeta+2.749\, \zeta^3+O(\zeta^5);
\\[10pt]
u_{(3)}=2.526\, \zeta+3.009\, \zeta^3+O(\zeta^5);
\end{array}
\end{equation}
Including the plane wall limit as $\zeta \rightarrow 1$ by eqs. (\ref{eq111}), the expressions valid over the entire range of $\zeta$ read
\begin{equation}
\begin{array}{l}
u_{(1)}=-0.334\, \zeta+0.242\, \zeta^3-
\dfrac{3\, \zeta}{(1-\zeta^2)^2}
+
O(\zeta^5);
\\[10pt]
u_{(2)}=-0.692\, \zeta-0.251\, \zeta^3
+\dfrac{3\, \zeta}{2(1-\zeta^2)^2}
+O(\zeta^5);
\\[10pt]
u_{(3)}=1.026\, \zeta+0.009\, \zeta^3
+\dfrac{3\, \zeta}{2(1-\zeta^2)^2}
+O(\zeta^5);
\end{array}
\label{eq115}
\end{equation}
In the case of a single internal cylinder bounding the fluid, the stresslet at its pole $u_{(i)}(\zeta)$, in the limit $\zeta\rightarrow \infty$ reads
\begin{equation}
\begin{array}{ll}
u_{(1)}(\zeta)=
 \dfrac{\pi}{\zeta^2}\left(
 \dfrac{39}{64 \log(2\, \zeta)} -\dfrac{  117 \log(4)-100}{128\left( \log\left(2\, \zeta\right) \right)^2} + O\left(\dfrac{1}{\log(\zeta)}\right)^3 \right);
\\
[20pt]
 u_{(2)}(\zeta)=-
 \dfrac{15\, \pi}{32\, \zeta^2}\left(
 \dfrac{1}{ \log(2\, \zeta)} -\dfrac{   \log(8)-1}{\left( \log\left(2\, \zeta\right) \right)^2} + O\left(\dfrac{1}{\log(\zeta)}\right)^3 \right);
 \\
[20pt]
 u_{(3)}(\zeta)=-
 \dfrac{\pi}{\zeta^2}\left(
 \dfrac{9}{ 64\, \log(2\, \zeta)} -\dfrac{  27 \log(4)-40}{128\left( \log\left(2\, \zeta\right) \right)^2} + O\left(\dfrac{1}{\log(\zeta)}\right)^3 \right)
\end{array}
\label{eq102}
\end{equation}

In order to understand the effect of curvature on the hydrodynamic repulsion and attraction experienced by a microswimmer, let us consider a microswimmer at a dimensionless distance $h$ from the cylindrical surface, where $h=1+|\zeta|$ for the internal cylinder and $h=1-|\zeta|$ for the external cylinder.
 Figure \ref{fig9} shows $h\, u_{(i)}(h)$ 
 the case the microswimmer is near a single cylinder for different radii of the cylinder itself, as a function of the distance $h$ between the singularity and the cylinder surface. 
 The reported results were obtained by computing eqs. (\ref{eq80})-(\ref{eq82}) and (\ref{eq56}); in the case of the external cylinder, the values obtained using eqs. (\ref{eq115}) are also reported.
  For infinite radii, i.e., when the singularity is close to a plane, $h\, u_{(i)}(h)$ is constant with respect to $h$ \cite{berke_hydrodynamic_2008}.
As shown in panel (a) of Figure \ref{fig9}, at larger distances, curvature reduces the repulsive force on a radially oriented microswimmer in both the external and internal cases. In the internal case, the forces vanish at the center of the cylinder due to the symmetry of the system.
Similarly, as shown in panel (b) of Figure \ref{fig9}, the attractive force on an angularly oriented microswimmer is reduced when it is inside a cylinder. On the other hand, if the microswimmer is angularly oriented outside a cylinder, the attractive force increases compared to the planar case when the singularity is close to the cylinder, and decreases when it is sufficiently far away (or the cylinder radius is sufficiently small, see the case $L_C=R_i=1$).
Finally, panel (c) of Figure \ref{fig9} shows that the attractive force on an axially oriented microswimmer increases in the internal case (provided the microswimmer is not too close to the center of the cylinder, where the force vanish) and decreases when the microswimmer is outside the cylinder.

\begin{figure}
\includegraphics[scale=0.75]{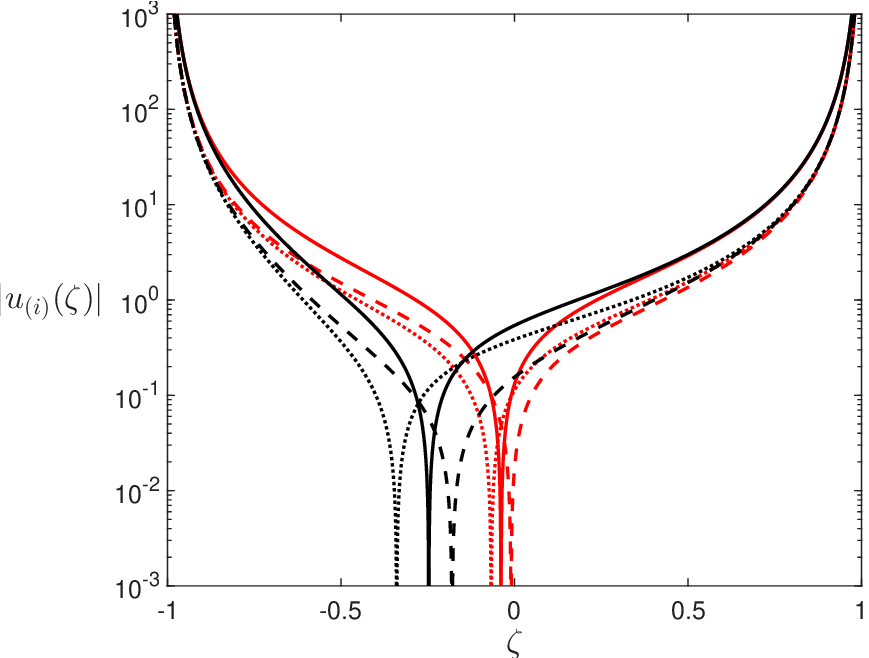}
\caption{\color{black} Absolute value of $u_{(i)}(\zeta)$ for different inner cylinder radii and orientations. Red curves: $\hat{R}_i = 2$; black curves: $\hat{R}_i = 1/75$. Solid lines represent radial orientation, dashed lines angular orientation, and symbols axial orientation. }
\label{fig8}
\end{figure}

\begin{figure}
\includegraphics[scale=0.4]{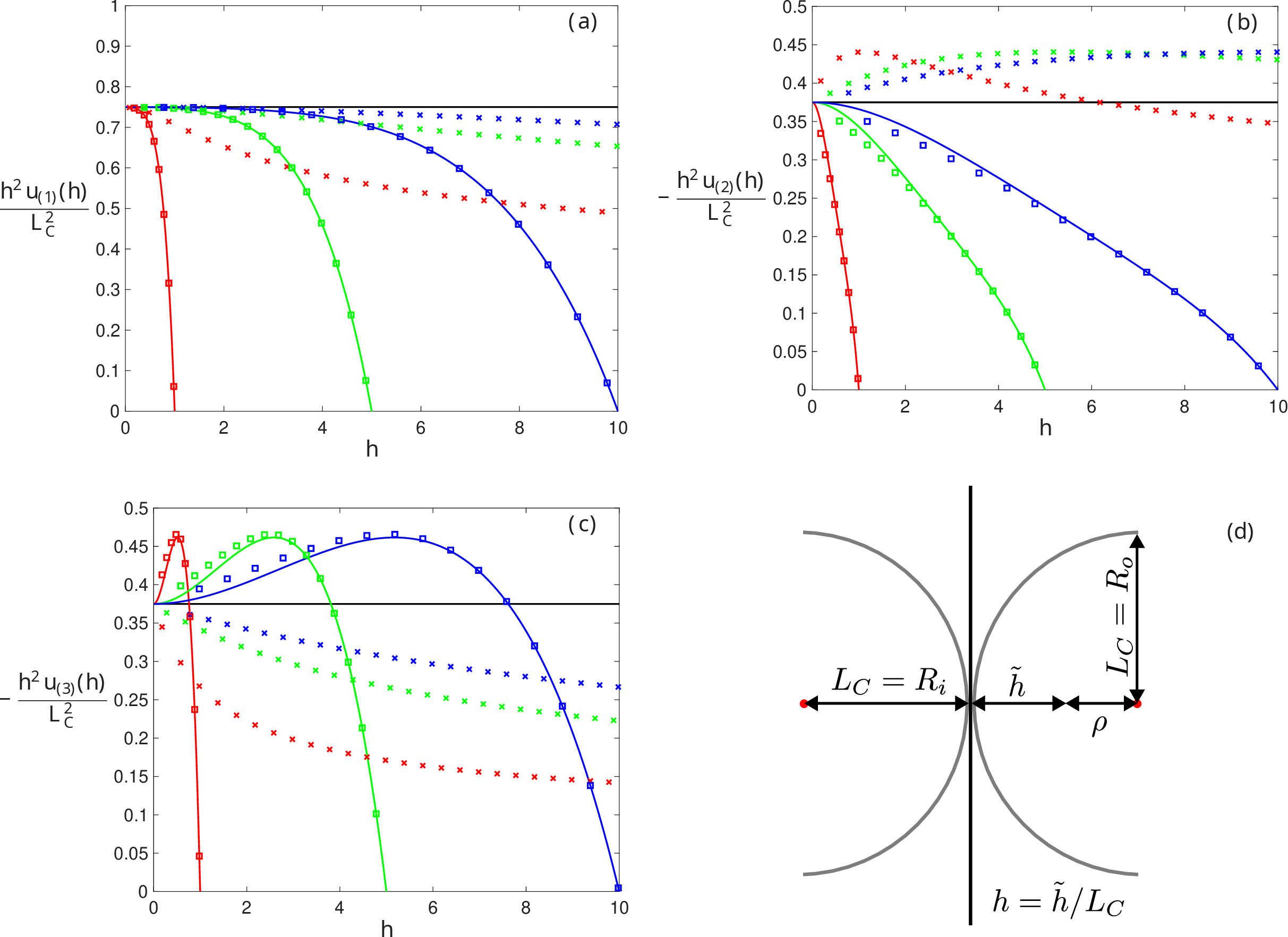}
\caption{\color{black} Dimensionless group $h^2 u_{(i)}(h)/L_C^2$ as a function of the distance $h$ from the cylindrical wall for different orientations: (a) radial, (b) angular, and (c) axial. The solid black line represents the constant value for a plane wall ($3/4$ for normal orientation, $-3/8$ for parallel orientation). Squares denote results for the external cylinder, crosses for the internal cylinder, and solid lines show the internal cylinder predictions from eq.~(\ref{eq115}). Colors indicate $L_C = 1$ (red), $L_C = 5$ (green), and $L_C = 10$ (blue). Panel (d) shows a schematic representation of the distance $h$ from the cylindrical/planar surface.}
\label{fig9}
\end{figure}

}

\section{Conclusions}
In a general perspective, the method applied here shows how, by employing the bitensorial formalism introduced in \cite{procopio_bitensorial_2022}, it is possible to obtain all the Stokes singularities in a confined domain once the Green function is known. More specifically, it is shown that{\color{black}, assuming no-slip boundary conditions,} all singularities can be derived either by differentiating the Green function at its pole (in the case of "momentum" singularities such as the Couplet, Stresslet, and higher-order Stokes multipoles), or by using the reciprocal relations of the Green function (in the case of "mass" singularities such as the Sourcelet, Source Dipole, and higher-order Source multipoles). Therefore, this represents an efficient method for evaluating the singularities associated with a given confinement in a unified manner. {\color{black} While this work focuses on no-slip boundary conditions, the method for obtaining higher-order singularities via Green function differentiation is general and extends to other boundary conditions, provided that the Green functions for both concentrated momentum and mass sources are available, since differentiation at the pole is independent of that at the field point where boundary conditions are enforced.}

The bitensorial representation — which takes into account the tensorial nature of the singular fields at both the field and pole points — paves the way for describing the hydrodynamics of many singularities in confined domains, and thus for investigating the mutual interactions between multiple colloids and, simultaneously, their interactions with the confining walls.
Moreover, as shown in \cite{procopio_theory_2024}, knowledge of all derivatives of the Green function can provide the full hydromechanics of colloids in confined flows, even in cases involving complex colloid shapes and characteristic sizes comparable to that of the confinement. Therefore, the results obtained in this article might represent the starting point for characterizing the complex hydrodynamic interactions that can arise in other systems of scientific and technological interest.

More specifically, the study of hydromechanics and colloid transport near cylindrical walls reveals peculiar characteristics that might differ considerably from those occurring in other well-studied geometries such as planar or spherical walls. The importance in applications, the widespread adoption and the simplicity of this geometry call for a thorough understanding of its microscale hydrodynamics, to which this work aims to contribute. In fact, Stokes singularities provide a valid model for describing the hydromechanics of colloids, and this article addresses examples of colloidal hydrodynamic interactions with cylindrical walls — including interactions with sedimenting particles and microswimmers — though only at a preliminary level, as a comprehensive analysis of these phenomena falls outside the scope of this study.



{\color{black}
\section*{Nomenclature}

\begin{tabbing}
  \hspace{4.5cm} \= \hspace{10cm} \= \\
  
    \em{Indices} \\
  $a, b, c=1,2,3$ \> Indices for tensorial entries at the field point\\
  $\alpha, \beta, \gamma, ...=1,2,3$ \> Indices for tensorial entries at the pole point \\
  $h, i, j, k, l=1,2,3$ \> Indices for Cartesian entries \\
 $n=1, \ldots, \infty $ \>  Order of cylindrical harmonic functions \\
 $s=1, \ldots, 6 $ \> Index for harmonic function coefficients, see eqs. (\ref{eq41}) \\
 
  \em{Points and Coordinates} \\
    $(Y_1, Y_2, Y_3)$ \> Cartesian coordinate system, see Fig. \ref{fig1}\\
    $(R, \Phi, Y_3)$ \> Cylindrical coordinate system, see Fig. \ref{fig1}\\ 
  ${\pmb x}$ \> Field point \\
  ${\pmb \xi}$ \> Pole point \\
  $x^b$ \> Contravariant coordinates of the field point in the
  \\ \> cylindrical (generic in Sec. \ref{setting}) coordinate system \\
  $\xi^\beta$ \> Contravariant coordinates of the pole point in the
  \\ \> cylindrical (generic in Sec. \ref{setting}) coordinate system \\
    $\overline{x}_j$ \> Contravariant coordinates of the field point in the
  \\ \> Cartesian coordinate system \\
  $\overline{\xi}_j$ \> Contravariant coordinates of the pole point in the
  \\ \> Cartesian coordinate system  \\
  
  \em{Geometric Quantities} \\
  $r$ \> Distance between field and pole points \\
  $r^2/2$ \> Distance function (Synge world function) \\
  $r_b$, $r^b$ \> Covariant and contravariant position vector at field
  \\
  \> point, see eqs. (\ref{eq18}) and (\ref{eq28}) \\
  $r_\beta$, $r^\beta$ \> Covariant and contravariant position vector at pole \\ \> point, see eqs. (\ref{eq18})  \\
  $x = x^1$ \> Radial position of the field point \\
  $\xi = \xi^1$ \> Radial position of the pole point\\
  $\phi=x^2 - \xi^2$ \> Angular difference between field and pole point \\
  $z = x^3 - \xi^3$ \> Axial difference between field and pole point\\
  $g_{ab}({\pmb x}), g^{ab}({\pmb x})$ \> Covariant and contravariant metric tensor at field point \\
  $g_{\alpha\beta}({\pmb \xi}),g^{\alpha\beta}({\pmb \xi})$ \> Covariant and contravariant metric tensor at pole point \\
  $g({\pmb x})$ \> Determinant of metric tensor at field point \\
  $g({\pmb \xi})$ \> Determinant of metric tensor at pole point \\
  $g^b_{\,\, \beta}({\pmb x}, {\pmb \xi})$ \> Parallel propagator from pole to field point \\
  $R_i$ \> Radius of internal cylinder \\
  $R_o$ \> Radius of external cylinder \\
  $V_f$ \> Fluid domain \\
  $\partial V_f$ \> Boundary of fluid domain \\
  
   \em{Tensorial Quantities} \\
     $\varepsilon_{ijk}$ \> Ricci--Levi-Civita symbol (Cartesian coordinates) \\
  $\eta^{\gamma\delta}_{\alpha\beta}= \delta^\gamma_\alpha \delta^\delta_\beta + \delta^\delta_\alpha \delta^\gamma_\beta$ \> Symmetry operator \\
  $\sqrt{g({\pmb x})}\, \varepsilon_{abc}\,$, $\varepsilon^{abc}/\sqrt{g({\pmb x})}$ \> Covariant and contravariant Ricci--Levi-Civita tensors  \\
  $\delta({\pmb x}, {\pmb \xi})  $
   \> Three-dimensional Dirac delta function \\
   
  \em{Fluid Dynamics} \\
  $\mu$ \> Dynamic viscosity of the fluid\\
  $v^b({\pmb x})$ \> Contravariant entries of the velocity field \\
  $\overline{v}_i({\pmb x})$ \> Cartesian entries of the velocity field \\
  $p({\pmb x})$ \> Pressure field \\
  $f^\beta({\pmb \xi})$ \> External force field entries at the pole point
  \\
 $f^a({\pmb x})$ \> Entries of the external force field parallel\\ \> transported at the field point, see eq.  (\ref{eq2}) \\
 
  \em{Operators} \\
  $\nabla^b$, $\nabla_b$ \> Contravariant and covariant derivative at field point \\
  $\nabla^\beta$, $\nabla_\beta$ \> Contravariant and covariant derivative at pole point \\
  $\Delta_x$, $\Delta_\xi$ \> Laplacian operator at field and at the pole point \\
  $\Gamma^a_{bc}({\pmb x}),\Gamma^\alpha_{\beta\gamma}({\pmb \xi})$ \> Christoffel symbols at field and at the pole point \\
    $\displaystyle \sum^*$ \> Summation-integration operator, see eq. (\ref{eq33})\\
      $\lambda$ \> Integration variable, see eq. (\ref{eq33}) \\
  $N_{\max}$ \> Maximum order for numerical summation \\
  
  \em{Singularities (Bitensorial Entries)} \\
  $G^b_{\,\, \beta}({\pmb x}, {\pmb \xi})$ \> Bounded Stokeslet (or Green function), see eq. (\ref{eq5})\\
  $\overline{G}_{ij}({\pmb x}, {\pmb \xi})$ \> Bounded Stokeslet in the Cartesian coordinate system \\
    $P_\beta({\pmb x}, {\pmb \xi})$ \> Pressure associated with the Stokeslet $G^b_{\,\, \beta}({\pmb x}, {\pmb \xi}) $ \\
  $S^b_{\,\, \beta}({\pmb x}, {\pmb \xi})$ \> Unbounded Stokeslet (or  unbounded Green function) \\
    ${\rm P}_\beta({\pmb x}, {\pmb \xi})$ \> Pressure associated with $S^b_{\,\, \beta}({\pmb x}, {\pmb \xi})$  \\
  $W^b_{\,\, \beta}({\pmb x}, {\pmb \xi})$ \> Regular part of the Green function (see eq. (\ref{eq7}))\\
  $Q_\beta({\pmb x}, {\pmb \xi})$ \> Pressure associated with  $ W^b_{\,\, \beta}({\pmb x}, {\pmb \xi})$ \\
  $M^b({\pmb x}, {\pmb \xi})$ \> Bounded Sourcelet (see eq. (\ref{eq9}))\\
  $\overline{M}_i({\pmb x}, {\pmb \xi})$ \> Bounded Sourcelet in the Cartesian coordinate system \\
  $\phi({\pmb x}, {\pmb \xi})$ \> Pressure associated with the Sourcelet $ M^b({\pmb x}, {\pmb \xi})$ \\
  $\nabla_\alpha G^b_{\,\, \beta}({\pmb x}, {\pmb \xi})$ \> Bounded Stokeslet dipole \\
  $\dfrac{\varepsilon^{\beta\gamma\delta}}{2\sqrt{g({\pmb \xi})}} \nabla_\gamma G^b_{\,\, \delta}({\pmb x}, {\pmb \xi})$ \> Bounded Couplet (or Rotlet) \\
  $\dfrac{\eta^{\gamma\delta}_{\alpha\beta}}{2} \nabla_\gamma G^b_{\,\, \delta}({\pmb x}, {\pmb \xi})$ \> Bounded Stresslet (or Strainlet) \\
  $\nabla_\beta M^b({\pmb x}, {\pmb \xi})$ \> Bounded Sourcelet dipole \\
  
  \em{Cylindrical Harmonics Representation of the Solution} 
   \\
  $I_n(y)$,$K_n(y)$ \> Modified Bessel function of first and second kind \\
  $I'_n(y)$, $K'_n(y)$ \> Derivatives of Bessel functions with respect to $y$ 
  \\
  $\Pi_\beta({\pmb x}, {\pmb \xi})$,$\Psi_\beta({\pmb x}, {\pmb \xi})$,\> \\  $\Omega_\beta({\pmb x}, {\pmb \xi})$ \> Harmonic functions  \\
    ${\rm A}_{(b\beta)}(\lambda x, \lambda\xi, \lambda, n)$ \> Functions, depending on the radial coordinates, \\
    \> providing   $S^b_{\,\, \beta}({\pmb x}, {\pmb \xi})$, see eqs. (\ref{eq32}) \\
    $A^>_{(b\beta)}(\lambda x, \lambda\xi, \lambda, n)$ \> Expressions of ${\rm A}_{(b\beta)}(\lambda x, \lambda\xi, \lambda, n)$ for $x > \xi$, see Appendix \ref{app_entries_A>} \\
  $A^<_{(b\beta)}(\lambda x, \lambda\xi, \lambda, n)$ \> Expressions of ${\rm A}_{(b\beta)}(\lambda x, \lambda\xi, \lambda, n)$ for $x < \xi$, see Appendix \ref{app_entries_A<} \\
    $p_{(i)}$ \> Angular phase constant, see eqs. (\ref{eq36}) \\
  $q_{(i)}$ \> Axial phase constant, see eqs. (\ref{eq36}) \\
  $B_{(s,\beta)}(\lambda\xi, \lambda, n)$ \> Functions, depending on the pole point radial coordinate, \\
    \> providing   $W^b_{\,\, \beta}({\pmb x}, {\pmb \xi})$, see eqs. (\ref{eq44}) \\
  $C_{(b,s)}(\lambda x, \lambda, n)$ \> Functions, depending on the field point radial coordinate, \\
    \> providing   $W^b_{\,\, \beta}({\pmb x}, {\pmb \xi})$, see eqs. (\ref{eq44}) and Appendix
    \ref{app_entries_C}\\
    
  \em{Matrix Notation} \\
  ${\bf B}_\beta(\xi, R_o, R_i)$ \> See eq. (\ref{eq49}) \\
  ${\pmb A}^>_\beta(\xi, R_o)$, \> \\
  ${\pmb A}^<_\beta(\xi, R_i)$ \> See eqs. (\ref{eq50}) \\
  $\overline{\bf A}_\beta(\xi, R_o, R_i)$ \> See eq. (\ref{eq51}) \\
  ${\pmb C}^o(x)$, \>  \\
  ${\pmb C}^i(x)$ \> See eqs. (\ref{eq52}) \\
  ${\bf C}(x)$ \>  See eq. (\ref{eq53}) \\
  $\overline{\bf C}(R_o, R_i)$ \> See eq. (\ref{eq54})  \\
  ${\pmb D}^o(R_o, R_i)$, \> \\
  ${\pmb D}^i(R_o, R_i)$ \> See eqs. (\ref{eq58}) \\
  ${\pmb \Delta}^o(R_o, R_i)$, \>  \\
  ${\pmb \Delta}^i(R_o, R_i)$ \>See eqs. (\ref{eq59})  \\
  
  \em{Physical Applications} \\
  $\zeta$ \>  Dimensionless radial distance, see eqs. (\ref{eqzeta}), (\ref{def1_LC}),(\ref{def2_LC})  \\
  $L_c$ \> Characteristic length of the system, see eqs. (\ref{eq91v2}), (\ref{def1_LC}),(\ref{def2_LC})\\
  $U_j$ \> Sedimentation velocity of a particle\\
  $r_S$ \> Radius of a spherical particle\\
  $w_{i\, j} (\zeta)$ \> Dimensionless regular part of the Green function at its pole,
  \\  \> see eq. (\ref{eq94})\\
  $F_1$ \> Radial force on microswimmer \\
  $\mathcal{S}$ \> Strain intensity of microswimmer \\
  $c_{(j)}$ \> Friction constants for microswimmer\\
  $h$ \> Dimensionless distance between surface and microswimmer,\\ \>  see Figure \ref{fig9} panel (d) \\
    $u_{(i)} (\zeta)$ \> Dimensionless regular part of the axialsymmetric stresslet at its pole,
  \\  \> see eqs. (\ref{eq80})--(\ref{eq82})\\
  $\xi_0$ \> Auxiliary parameter for defining  $w_{i\, j} (\zeta)$ and $u_{(i)} (\zeta)$ in different
  \\  \> 
   geometries, see eqs. (\ref{eq93v2}),(\ref{def1_LC})--(\ref{def2_LC})\\

\textbf{Note:} Einstein summation convention applies for repeated indices unless in parentheses.

\end{tabbing} 
} 
\appendix

\section{Bitensorial formulation of the Stokes unbounded singularities}
\label{App_Stokeslet}
In this appendix the bitensorial formulation of the Stokeslet and higher order singularities in generic coordinate systems is obtained. There are several methods to obtain the Green function of the unbounded Stokes flow developed in the literature \cite{pozrikidis_1992,ladyzhenskaia_mathematical_2014,kim_1991,giona_hydrodynamic_2022}. In the following the Pozrikidis method \cite[p. 22]{pozrikidis_1992}, developed in Cartesian coordinates, will be applied considering generic curvilinear coordinate systems at both the pole and field point. 

As obtained in Section \ref{setting} and in \cite{procopio_bitensorial_2022}, the bitensorial formulation of the Green problem of the unbounded Stokes flow is
\begin{equation}
\begin{cases}
\Delta_x\, S^b_{\, \, \beta} ({\pmb x},{\pmb \xi})
-
\nabla^b {\rm P}_\beta ({\pmb x},{\pmb \xi})
=
-8 \pi g^b_{\,\, \beta}({\pmb x},{\pmb \xi})\, \delta({\pmb x},{\pmb \xi})
\\
\nabla_b\,  S^b_{\, \, \beta} ({\pmb x},{\pmb \xi})=0
\\
S^b_{\, \, \beta} ({\pmb x},{\pmb \xi}) \rightarrow {0} \quad \text{for} \quad {\pmb x} \rightarrow \infty
\end{cases}
\label{eqB1}
\end{equation}
By applying the covariant derivative operator $\nabla_b$
to the moment balance equation in eqs. (\ref{eqB1}) and considering the incompressibility equation, a Poisson equation is obtained for the bitensorial pressure
\begin{equation}
\Delta_x {\rm P}_\beta ({\pmb x},{\pmb \xi})= 8 \pi \nabla_b \,g^b_{\, \, \beta} ({\pmb x},{\pmb \xi})\, \delta({\pmb x},{\pmb \xi})
\label{eqB2}
\end{equation}
In the Euclidean space, the covariant derivative of the parallel propagator vanishes. In fact, by the definition eq. (\ref{eq3})
\begin{equation}
\nabla_c \,g_{b \, \beta} ({\pmb x},{\pmb \xi})\, =\, 
\dfrac{\partial Y_i}{\partial \xi^\beta}\, \nabla_c \,\left(  
\dfrac{\partial Y_i}{\partial x^b} \right) \, = \,
\dfrac{\partial Y_i}{\partial \xi^\beta}\, \left(  
\dfrac{d^2 Y_i}{\partial x^b\, \partial x^c}- \Gamma^a_{b\, c} \, \dfrac{\partial Y_i}{\partial x^a}  \right) =0
\label{eqB3}
\end{equation}
where the relation  
\begin{equation}
\dfrac{\partial^2 Y_i}{\partial x^b\, \partial x^c}=
\Gamma^a_{b\, c} \, \dfrac{\partial Y_i}{\partial x^a}
\label{B4}
\end{equation}
valid for Euclidean spaces \cite{de_souza_2016}, has been considered.
Therefore, since $ \nabla_b \,g^b_{\, \, \beta} ({\pmb x},{\pmb \xi})\,=\, g^{b\, c}({\pmb x})\, \nabla_c \,g_{b \, \beta} ({\pmb x},{\pmb \xi})$, eq. (\ref{eqB2}) becomes
\begin{equation}
\Delta_x {\rm P}_\beta ({\pmb x},{\pmb \xi})= 8 \pi\,g^b_{\, \, \beta} ({\pmb x},{\pmb \xi}) \, \nabla_b \, \delta({\pmb x},{\pmb \xi})
\label{eqB5}
\end{equation}
By using the distributional identity
\begin{equation}
\delta({\pmb x}, {\pmb \xi})=
\Delta_x \left( \dfrac{1}{- 4 \pi\, r} \right)
\label{eqB6}
\end{equation}
Eq. (\ref{eqB5}) provides
\begin{equation}
{\rm P}_\beta({\pmb x},{\pmb \xi})
=-2 \, g^b_{\, \, \beta} ({\pmb x},{\pmb \xi}) \, \nabla_b \, 
\left( \dfrac{1}{r} \right) 
\label{eqB7}
\end{equation}
Considering that $ \nabla_b \, r^{-1}=-r_b\, /\, r^{3}$
and that $g^b_{\, \, \beta}({\pmb x},{\pmb \xi})\, r_b=-r_\beta$
(and, more generally, $g^b_{\, \, \beta}({\pmb x},{\pmb \xi}) \nabla_b f(r) = - \nabla_\beta f(r)$ for any function of the distance $f(r)$ 
) \cite{poisson_motion_2011}
\begin{equation}
P_\beta({\pmb x},{\pmb \xi})
=-2 \,   
\dfrac{r_\beta}{r^3} 
\label{eqB8}
\end{equation}
By using eqs. (\ref{eqB6}) and (\ref{eqB8}), the moment balance in eqs. (\ref{eqB1}) becomes
\begin{equation}
\Delta_x\, S^b_{\, \, \beta}({\pmb x},{\pmb \xi})
=
2 \left(\,
g^b_{\, \, \beta} ({\pmb x},{\pmb \xi}) \Delta_x
+
\nabla^b \, \nabla_\beta\, 
\right) \dfrac{1}{r}
\label{eqB9}
\end{equation}
Let us try, for $S^b_{\, \, \beta}({\pmb x},{\pmb \xi})$, a solution
a solution with the form
\begin{equation}
S^b_{\, \, \beta}({\pmb x},{\pmb \xi})
=
\left(\,
g^b_{\, \, \beta} ({\pmb x},{\pmb \xi}) \Delta_x
+
\nabla^b\, \nabla_\beta\, 
\right) H(r)
\label{eqB10}
\end{equation}
where $H(r)$ is a scalar function of $r$ to be determined.
Substituting eq. (\ref{eqB10}) into eq. (\ref{eqB9})
\begin{equation}
\Delta_x\, H(r)
=
\dfrac{2}{r}
\label{eqB11}
\end{equation}
Applying the Laplacian operator $\Delta_x$ to eq. (\ref{eqB11})
 and using eq. (\ref{eqB6})
\begin{equation}
\Delta_x\, \Delta_x\,  H(r)
=
- 8 \pi \, \delta({\pmb x},{\pmb \xi})
\label{eqB12}
\end{equation}
Equation (\ref{eqB12}) is the biharmonic Green problem having solution $H(r)=r$, hence
\begin{equation}
S^b_{\, \, \beta}({\pmb x},{\pmb \xi})
=
\left(\,
g^b_{\, \, \beta} ({\pmb x},{\pmb \xi}) \Delta_x
+
\nabla^b \, \nabla_\beta\, 
\right) r
\label{eqB13}
\end{equation}
By using the relation $\Delta_x\, r = 2/r$ expressed in eq. (\ref{eqB11})
and the definitions $r_b=\nabla_b \, r^2 /2$ and $r_\beta =\nabla_\beta\, r^2/2$, it is possible to obtain from eq. (\ref{eqB13}) 
\begin{equation}
S^b_{\, \, \beta}({\pmb x},{\pmb \xi})
= \dfrac{g^b_{\, \, \beta}({\pmb x},{\pmb \xi})}{r}- \dfrac{r_b\, r_\beta}{r^3}
\label{eqB14}
\end{equation}
or alternatively
\begin{equation}
S^b_{\, \, \beta}({\pmb x},{\pmb \xi})
=
\left(\,
g^b_{\, \, \beta}({\pmb x},{\pmb \xi})
+
r^b\, \nabla_\beta\, 
\right) \dfrac{1}{r}
\label{eqB15}
\end{equation}
Higher order singularities are obtained by differentiating the Stokeslet eq. (\ref{eqB14}) at its pole ${\pmb \xi}$. The bitensorial expression for the Stokes dipole then becomes
\begin{equation}
\nabla_\alpha\, S^b_{\, \, \beta}({\pmb x},{\pmb \xi})
= 
-\dfrac{r^b\, g_{\alpha\, \beta}({\pmb \xi})}{r^3}
+
3\dfrac{r^b\, r_\alpha\, r_\beta}{r^5}
+
\dfrac{r_\beta\, g^b_\alpha ({\pmb x}.{\pmb \xi}) \, - \, r_\alpha \,g^b_\beta({\pmb x},{\pmb \xi})}{r^3}
\label{eqB16}
\end{equation}
obtained considering that 
$\nabla_\beta \, \nabla_\alpha\, r^2/2\, =\, g_{\alpha\, \beta}({\pmb \xi})$, $\nabla^b \, \nabla_\beta \, r^2/2\, =\,- g^b_{\,\, \beta}({\pmb x},{\pmb \xi})$, $\nabla_\alpha\, r^{-3}\, =\, -3\, r_\alpha \, r^{-5}$.
 
As in eq. (\ref{eq71_}), the Stokeslet dipole can be expressed in term of its symmetric and antisymmetric part 
\begin{equation}
\nabla_\alpha\, S^b_{\, \, \beta}({\pmb x},{\pmb \xi})
\,=\,
\dfrac{ \eta^{\gamma \, \delta}_{\alpha \, \beta} }{2}
\,
\nabla_\gamma\, S^b_{\, \, \delta}({\pmb x},{\pmb \xi})
\,+\,
\dfrac{ \varepsilon_{\zeta \alpha \beta}\, \varepsilon^{\zeta \gamma \delta} }{2}
\,
\nabla_\gamma\, S^b_{\, \, \delta}({\pmb x},{\pmb \xi})
\label{eqB18}
\end{equation}
The first term in eq. (\ref{eqB18}), referred to as the unbounded {Stresslet} is
\begin{equation}
\dfrac{ \eta^{\gamma \, \delta}_{\alpha \, \beta} }{2}
\,
\nabla_\gamma\, S^b_{\, \, \delta}({\pmb x},{\pmb \xi})=
\left( 
- \dfrac{g^{\gamma \, \delta}({\pmb \xi}) \, g_{\alpha \,\beta}({\pmb \xi}) }{3}
+
\dfrac{ \eta^{\gamma \, \delta}_{\alpha \, \beta}  }{2}
\right)
3\dfrac{r^b\, r_\gamma\, r_\delta}{r^5}
\label{eqB19}
\end{equation}
whereas, the so called unbounded Couplet at the second term of eq. (\ref{eqB18}) is
\begin{equation}
\dfrac{\varepsilon^{\beta \gamma \delta} }{2}
\,
\nabla_\gamma\, S^b_{\, \, \delta}({\pmb x},{\pmb \xi})
=
\dfrac{\varepsilon^{\beta \gamma \delta}\, g^b_{\, \, \gamma}({\pmb x},{\pmb \xi}) }{2 \sqrt{g({\pmb \xi}) } }
\, \dfrac{r_\delta}{r^3}
\label{eqB20}
\end{equation} 
The unbounded Sourcelet ${\rm M}^a({\pmb x},{\pmb \xi})$ is the solution of the system eqs. (\ref{eq9}) considering $\partial V_f$ being a surface at infinity. In this case, the corresponding pressure $\phi({\pmb x},{\pmb \xi})$ is a constant \cite{pozrikidis_1992} and, from eq. (\ref{eq14}) and the expression of pressure associated with the Green function in the unbounded domain eq. (\ref{eqB8})
\begin{equation}
{\rm M}^a({\pmb x},{\pmb \xi})= \dfrac{r^a}{r^3}
\label{eqB21}
\end{equation}
The Source dipole in the unbounded domain, obtained by differentiating ${\rm M}^a({\pmb x},{\pmb \xi})$ at its pole ${\pmb \xi}$, is
\begin{equation}
\nabla_\beta\, {\rm M}^b({\pmb x},{\pmb \xi})\,=\,
-
\left(
\dfrac{g^b_{\, \, \beta}({\pmb x},{\pmb \xi})}{r^3}
+
3
\dfrac{r^b\,r_\beta}{r^5}
\right) 
\label{eqB22}
\end{equation}

\section{Explicit expressions of entries}

\subsection{The entries of ${\pmb A}_\beta^{>}(\xi,R_o)$, defined in eqs. (\ref{eq50}), are}
\label{app_entries_A>}
\begin{eqnarray}
\nonumber
A^{>}_{(1\, 1)} (\lambda x,\lambda \xi, \lambda, n)=
\dfrac{1}{2} \left[\, 2 \lambda x K_{n}(\lambda x) I'_{n}(\lambda \xi)+
K_{n+1}(\lambda x) I_{n+1}(\lambda \xi)
+
K_{n-1}(\lambda x) I_{n-1}(\lambda \xi)
\right.
\\
\left.
-
\lambda \xi 
\left(
K_{n+1}(\lambda x) I'_{n+1}(\lambda \xi) 
+
K_{n-1}(\lambda x) I'_{n-1}(\lambda \xi) \,
\right)
\right]
\nonumber
\end{eqnarray}

\begin{equation}
A^{>}_{(1\, 2)}(\lambda x,\lambda \xi, \lambda, n) =
\dfrac{1}{2}\left[
2 n x K_n(\lambda x) I_n(\lambda \xi)
- \xi \left(
(n+2)K_{n+1}(\lambda x) I_{n+1}(\lambda \xi)
+
(n-2)K_{n-1}(\lambda x) I_{n-1}(\lambda \xi)
\, \right) \right]
\nonumber
\end{equation}

\begin{equation}
 A^{>}_{(1\, 3)}(\lambda x,\lambda \xi, \lambda, n) =
 \dfrac{\lambda }{2}\left[
2  x K_n(\lambda x) I_n(\lambda \xi)
- \xi \left(
K_{n+1}(\lambda x) I_{n+1}(\lambda \xi)
+
K_{n-1}(\lambda x) I_{n-1}(\lambda \xi)
\, \right) \right]
\nonumber
\end{equation}

\begin{eqnarray}
\nonumber
 A^{>}_{(2\, 1)}(\lambda x,\lambda \xi, \lambda, n) =
\dfrac{1}{2 x}\left[ K_{n+1}(\lambda x) I_{n+1}(\lambda \xi)
-K_{n-1}(\lambda x) I_{n-1}(\lambda \xi)
\right.
\\
\left.
- \lambda \xi \left(
K_{n+1}(\lambda x) I'_{n+1}(\lambda \xi)
-
K_{n-1}(\lambda x) I'_{n-1}(\lambda \xi)
 \right)
\,  \right]
\nonumber
\end{eqnarray}

\begin{equation}
 A^{>}_{(2\, 2)}(\lambda x,\lambda \xi, \lambda, n) =-
\dfrac{\xi}{2 x}\left[ 
(n+2)K_{n+1} (\lambda x) I_{n+1} (\lambda \xi)
-
(n-2) K_{n-1} (\lambda x) I_{n-1} (\lambda \xi)
\,  \right]
\nonumber
\end{equation}

\begin{equation}
 A^{>}_{(2\, 3)}(\lambda x,\lambda \xi, \lambda, n) =
\dfrac{\lambda \xi}{2 x}\left[ 
 K_{n-1} (\lambda x) I_{n-1} (\lambda \xi)
 -
 K_{n+1} (\lambda x) I_{n+1} (\lambda \xi)
\,  \right]
\nonumber
\end{equation}

\begin{equation}
 A^{>}_{(3\, 1)}(\lambda x,\lambda \xi, \lambda, n) =- \left[
\lambda x K'_{n} (\lambda x) I'_{n} (\lambda \xi)
+
\lambda \xi K_{n} (\lambda x) I''_{n} (\lambda \xi)
+
K_{n} (\lambda x) I'_{n} (\lambda \xi)
 \right]
\nonumber
\end{equation}

\begin{equation}
 A^{>}_{(3\, 2)}(\lambda x,\lambda \xi, \lambda, n) =- \left[
n x K'_{n} (\lambda x) I_{n} (\lambda \xi)
+
n \xi K_{n} (\lambda x) I'_{n} (\lambda \xi) \right]
\nonumber
\end{equation}

\begin{equation}
 A^{>}_{(3\, 3)}(\lambda x,\lambda \xi, \lambda, n) = -\left(
2 K_{n} (\lambda x) I_{n} (\lambda \xi)
+
\lambda x K'_{n} (\lambda x) I_{n} (\lambda \xi)
+
\lambda \xi K_{n} (\lambda x) I'_{n} (\lambda \xi) \, \right)
\nonumber
\end{equation}

\subsection{The entries of   ${\pmb A}_\beta^{<}(\xi,R_o)$, defined in eqs. (\ref{eq50}), are}
\label{app_entries_A<}

\begin{eqnarray}
\nonumber
A^{<}_{(1\, 1)}(\lambda x,\lambda \xi, \lambda, n) =
\dfrac{1}{2} \left[\, 2 \lambda x K'_{n}(\lambda \xi) I_{n}(\lambda x)+
K_{n+1}(\lambda \xi) I_{n+1}(\lambda x)
+
K_{n-1}(\lambda \xi) I_{n-1}(\lambda x)
\right.
\\
\left.
-
\lambda \xi 
\left(
K'_{n+1}(\lambda \xi) I_{n+1}(\lambda x) 
+
K'_{n-1}(\lambda \xi) I_{n-1}(\lambda x) \,
\right)
\right]
\nonumber
\end{eqnarray}

\begin{equation}
 A^{<}_{(1\, 2)}(\lambda x,\lambda \xi, \lambda, n) =
\dfrac{1}{2}\left[
2 n x K_n(\lambda \xi) I_n(\lambda x)
- \xi \left(
(n+2)K_{n+1}(\lambda \xi) I_{n+1}(\lambda x)
+
(n-2)K_{n-1}(\lambda \xi) I_{n-1}(\lambda x)
\, \right) \right]
\nonumber
\end{equation}

\begin{equation}
 A^{<}_{(1\, 3)} (\lambda x,\lambda \xi, \lambda, n)=
 \dfrac{\lambda }{2}\left[
2  x K_n(\lambda \xi) I_n(\lambda x)
- \xi \left(
K_{n+1}(\lambda \xi) I_{n+1}(\lambda x)
+
K_{n-1}(\lambda \xi) I_{n-1}(\lambda x)
\, \right) \right]
\nonumber
\end{equation}

\begin{eqnarray}
\nonumber
 A^{<}_{(2\, 1)}(\lambda x,\lambda \xi, \lambda, n) =
\dfrac{1}{2 x}\left[ K_{n+1}(\lambda \xi) I_{n+1}(\lambda x)
-K_{n-1}(\lambda \xi) I_{n-1}(\lambda x)
\right.
\\
\left.
- \lambda \xi \left(
K'_{n+1}(\lambda \xi) I_{n+1}(\lambda x)
-
K'_{n-1}(\lambda \xi) I_{n-1}(\lambda x)
 \right)
\,  \right]
\nonumber
\end{eqnarray}

\begin{equation}
 A^{<}_{(2\, 2)} =-
\dfrac{\xi}{2 x}\left[ 
(n+2)K_{n+1} (\lambda \xi) I_{n+1} (\lambda x)
-
(n-2) K_{n-1} (\lambda \xi) I_{n-1} (\lambda x)
\,  \right]
\nonumber
\end{equation}

\begin{equation}
 A^{<}_{(2\, 3)}(\lambda x,\lambda \xi, \lambda, n) =
\dfrac{\lambda \xi}{2 x}\left[ 
 K_{n-1} (\lambda \xi) I_{n-1} (\lambda x)-
 K_{n+1} (\lambda \xi) I_{n+1} (\lambda x)
\,  \right]
\nonumber
\end{equation}

\begin{equation}
 A^{<}_{(3\, 1)}(\lambda x,\lambda \xi, \lambda, n) =- \left[
\lambda x K'_{n} (\lambda \xi) I'_{n} (\lambda x)
+
\lambda \xi K''_{n} (\lambda \xi) I_{n} (\lambda x)
+
K'_{n} (\lambda \xi) I_{n} (\lambda x) 
 \, \right]
\nonumber
\end{equation}

\begin{equation}
 A^{<}_{(3\, 2)}(\lambda x,\lambda \xi, \lambda, n) =- \left[
n x K_{n} (\lambda \xi) I'_{n} (\lambda x)
+
n \xi K'_{n} (\lambda \xi) I_{n} (\lambda x) \right]
\nonumber
\end{equation}

\begin{equation}
 A^{<}_{(3\, 3)}(\lambda x,\lambda \xi, \lambda, n) = - \left(2
K_{n} (\lambda \xi) I_{n} (\lambda x)
+
\lambda x K_{n} (\lambda \xi) I'_{n} (\lambda x)
+
\lambda \xi K'_{n} (\lambda \xi) I_{n} (\lambda x) \,
\right)
\nonumber
\end{equation}

\subsection{The entries of  ${\bf C} (x) $, defined in eq. (\ref{eq53}), are}
\label{app_entries_C}

\begin{equation}
\begin{array}{cc}
{C}_{(1,1)} (\lambda\, x, \lambda, n)= (\lambda)^2 x I''_n(\lambda x)
&
{C}_{(1,4)} (\lambda\, x, \lambda, n)= (\lambda)^2 x K''_n(\lambda x) 
\\
[10pt]
{C}_{(1,2)} (\lambda\, x, \lambda, n)=  \lambda I'_n(\lambda x)
&
{C}_{(1,5)} (\lambda\, x, \lambda, n)= \lambda K'_n(\lambda x)
\\
[10pt]
{C}_{(1,3)} (\lambda\, x, \lambda, n)= \dfrac{\lambda n}{\lambda x} I_n(\lambda x)
&
{C}_{(1,6)} (\lambda\, x, \lambda, n)= \dfrac{\lambda n}{\lambda x} K_n(\lambda x)
\\
[10pt]
{C}_{(2,1)} (\lambda\, x, \lambda, n)=  \dfrac{(\lambda)^2 n}{\lambda x} \left[
\dfrac{I_n(\lambda x)}{\lambda x}
-I'_n(\lambda x)
\right]
& \quad
{C}_{(2,4)} (\lambda\, x, \lambda, n)=
 \dfrac{(\lambda)^2 n}{\lambda x} \left[
\dfrac{K_n(\lambda x)}{\lambda x}
-K'_n(\lambda x)
\right]
\\
[10pt]
{C}_{(2,2)} (\lambda\, x, \lambda, n)= - \dfrac{(\lambda)^2 n \, I_n (\lambda x)}{(\lambda x)^2}
&
{C}_{(2,5)} (\lambda\, x, \lambda, n)=  - \dfrac{(\lambda)^2 n \, K_n (\lambda x)}{(\lambda x)^2}
\\
[10pt]
{C}_{(2,3)} (\lambda\, x, \lambda, n)=  - \dfrac{(\lambda)^2 I'_n (\lambda x)}{\lambda x}
&
{C}_{(2,6)} (\lambda\, x, \lambda, n)=  - \dfrac{(\lambda)^2 K'_n (\lambda x)}{\lambda x}
\\
[10pt]
{C}_{(3,1)} (\lambda\, x, \lambda, n)= - \lambda \left[
\lambda x I'_n(\lambda x)+ I_n(\lambda x)
\right]
&
\quad
{C}_{(3,4)} (\lambda\, x, \lambda, n)=- \lambda \left[
\lambda x K'_n(\lambda x)+ K_n(\lambda x)
\right]
\\
[10pt]
{C}_{(3,2)} (\lambda\, x, \lambda, n)= - \lambda I_n(\lambda x)
&
{C}_{(3,5)} (\lambda\, x, \lambda, n)= - \lambda K_n (\lambda x)
\\
[10pt]
{C}_{(3,3)} (\lambda\, x, \lambda, n)= 0
&
{C}_{(3,6)} (\lambda\, x, \lambda, n)=0
\end{array}
\nonumber
\end{equation}
{\color{black}
The matrix $({\pmb C}^o (x) )^{-1} $, with $ {\pmb C}^o (x) $ being defined in eqs. (\ref{eq52}), can be expressed as
\begin{equation}
({\pmb C}^o (x) )^{-1} 
=
\dfrac{ {\pmb \Lambda }^o (x) }{\det \big[ {\pmb C}^o (x)] }
\end{equation}
where
\begin{equation}
\det \big[ {\pmb C}^o (x)]=
\lambda^4
\left(
I_n'(\lambda x)^3 - I_n''(\lambda x) I_n'(\lambda x) I_n(\lambda x) - \dfrac{2 I_n(\lambda x)^3 n^2}{(\lambda x)^3} + \dfrac{I_n'(\lambda x)^2 I_n(\lambda x)}{\lambda x }
\right)
\end{equation}
and $ {\pmb \Lambda}^o (x) $ is the adjugate matrix \cite{bernstein_scalar_2018} of $ {\pmb C}^o (x) $ with entries
\begin{equation}
\begin{array}{cc}
\Lambda^o_{1,1} (x)=
-\dfrac{\lambda^2 I'_n(\lambda x) I_n(\lambda x)}{x}
\\ [10pt]
\Lambda^o_{1,2} (x)=-\dfrac{n \lambda I_n(\lambda x)^2 }{x}
\\ [10pt]
\Lambda^o_{1,3} (x)=\dfrac{n^2 I_n(\lambda x)^2 }{x^3} - \dfrac{ \lambda^2 I_n'(\lambda x)^2}{x}
\\ [10pt]
\Lambda^o_{2,1} (x)= \dfrac{\lambda^2 I_n'(\lambda x) I_n(\lambda x)}{x} + \lambda^3 I_n'(\lambda x)^2
\\ [10pt]
\Lambda^o_{2,2} (x)= \dfrac{n \lambda I_n(\lambda x)^2}{x} + n \lambda^2 I_n'(\lambda x) I_n(\lambda x)
\\ [10pt]
\Lambda^o_{2,3} (x)=\dfrac{n^2 I_n(\lambda x)^2}{x^3} - \dfrac{n^2 \lambda I_n'(\lambda x) I_n(\lambda x)}{x^2}
\\ [10pt]
\Lambda^o_{3,1} (x)=-\dfrac{2 n \lambda I_n(\lambda x)^2}{x^2}
\\ [10pt]
\Lambda^o_{3,2} (x)= -\lambda^2 I_n'(\lambda x) I_n(\lambda x) - x \lambda^3 I_n'(\lambda x)^2 + x \lambda^3 I_n''(\lambda x) I_n(\lambda x)
\\ [10pt]
\Lambda^o_{3,3} (x)=-\dfrac{n \lambda I_n'(\lambda x) I_n(\lambda x)}{x^2} + \dfrac{n \lambda^2 I_n'(\lambda x)^2}{x} - \dfrac{n \lambda^2 I_n''(\lambda x) I_n(\lambda x)}{x}
\\ [10pt]
\end{array}
\end{equation}
The matrix $({\pmb C}^i (x) )^{-1} $, with $ {\pmb C}^i (x) $ being defined in eqs. (\ref{eq52}), can be expressed as
\begin{equation}
({\pmb C}^i (x) )^{-1} 
=
\dfrac{ {\pmb \Lambda }^i (x) }{\det \big[ {\pmb C}^i (x)] }
\end{equation}
where
\begin{equation}
\begin{array}{cc}
\det \big[ {\pmb C}^i (x)]=
\\ [10pt]
\lambda^4
\left(
K_n'(\lambda x)^3 - K_n''(\lambda x) K_n'(\lambda x) K_n(\lambda x) - \dfrac{2 K_n(\lambda x)^3 n^2}{(\lambda x)^3} + \dfrac{K_n'(\lambda x)^2 K_n(\lambda x)}{\lambda x }
\right)
\end{array}
\end{equation}
and $ {\pmb \Lambda}^i (x) $ is the adjugate matrix of $ {\pmb C}^i (x) $ with entries
\begin{equation}
\begin{array}{cc}
\Lambda^i_{1,1} (x)=
-\dfrac{\lambda^2 K'_n(\lambda x) K_n(\lambda x)}{x}
\\ [10pt]
\Lambda^i_{1,2} (x)=-\dfrac{n \lambda K_n(\lambda x)^2 }{x}
\\ [10pt]
\Lambda^i_{1,3} (x)=\dfrac{n^2 K_n(\lambda x)^2 }{x^3} - \dfrac{ \lambda^2 K_n'(\lambda x)^2}{x}
\\ [10pt]
\Lambda^i_{2,1} (x)= \dfrac{\lambda^2 K_n'(\lambda x) K_n(\lambda x)}{x} + \lambda^3 K_n'(\lambda x)^2
\\ [10pt]
\Lambda^i_{2,2} (x)= \dfrac{n \lambda K_n(\lambda x)^2}{x} + n \lambda^2 K_n'(\lambda x) K_n(\lambda x)
\\ [10pt]
\Lambda^i_{2,3} (x)=\dfrac{n^2 K_n(\lambda x)^2}{x^3} - \dfrac{n^2 \lambda K_n'(\lambda x) K_n(\lambda x)}{x^2}
\\ [10pt]
\Lambda^i_{3,1} (x)=-\dfrac{2 n \lambda K_n(\lambda x)^2}{x^2}
\\ [10pt]
\Lambda^i_{3,2} (x)= -\lambda^2 K_n'(\lambda x) K_n(\lambda x) - \lambda^3 x K_n'(\lambda x)^2 + \lambda^3 x K_n''(\lambda x) K_n(\lambda x)
\\ [10pt]
\Lambda^i_{3,3} (x)=-\dfrac{n \lambda K_n'(\lambda x) K_n(\lambda x)}{x^2} + \dfrac{n \lambda^2 K_n'(\lambda x)^2}{x} - \dfrac{n \lambda^2 K_n''(\lambda x) K_n(\lambda x)}{x}
\\ [10pt]
\end{array}
\end{equation}

\subsection{Entries of the inverse matrix of  $\overline{\bf C} (R_o,R_i) $, defined in eq. (\ref{eq54}).}
 \label{app_entries_B4}
The inverse of the matrix $ \overline{\bf C} (R_o,R_i)  $, expressed in terms of the adjugate matrix $ {\pmb \Lambda }$, reads
\begin{equation}
( \overline{\bf C} (R_o,R_i)  )^{-1} 
=
\dfrac{ {\pmb \Lambda} }{\det \big[ \overline{\bf C} (R_o,R_i) ] }
\end{equation}
By defining the quantities
\begin{equation}
\begin{array}{ll}
\alpha=I_n(\lambda R_o)K_n(\lambda R_i)-I_n(\lambda R_i)K_n(\lambda R_o);  & \quad \beta=I_n'(\lambda R_o)K_n(\lambda R_i)-I_n(\lambda R_i)K_n'(\lambda R_o) 
\\
\gamma= I_n(\lambda R_o)K_n'(\lambda R_i)-I_n'(\lambda R_i)K_n(\lambda R_o); & \delta=I_n'(\lambda R_o)K_n'(\lambda R_i)-I_n'(\lambda R_i)K_n'(\lambda R_o) 
\\
\eta= \dfrac{\lambda^2}{R_o^3 R_i^3}
\end{array}
\end{equation}
the determinant of $ \overline{\bf C} (R_o,R_i) $ reads
\begin{equation}
\begin{array}{l}
\det \big[ \overline{\bf C} (R_o,R_i) ]= 
\eta \Big( 4 n^2 \alpha (2+n^2 \alpha^2) \\
+ \left(2 n^2 R_o (2+n^2 \alpha^2) \beta + 2 n^2 R_i (2+n^2 \alpha^2) \gamma\right) \lambda \\
+ \left(-4 n^2 R_o^2 \alpha \beta^2 - 4 n^2 R_i^2 \alpha \gamma^2 + n^2 R_i R_o (2+n^2 \alpha^2) \delta\right) \lambda^2 \\
+ \left(2 n^2 R_o^3 \beta (\alpha^2-\beta^2) - 2 n^2 R_i R_o^2 \beta^2 \gamma - 2 n^2 R_i^2 R_o \beta \gamma^2 + 2 n^2 R_i^3 \gamma (\alpha^2-\gamma^2)\right) \lambda^3 \\
+ \left(R_i R_o^3 (1+n^2 (\alpha^2-\beta^2)) \delta + R_i^3 R_o (1+n^2 (\alpha^2-\gamma^2)) \delta + 4 R_i^2 R_o^2 \alpha \delta^2\right) \lambda^4 \\
+ \left(2 R_i^2 R_o^3 \delta (\beta \delta-\alpha \gamma) + 2 R_i^3 R_o^2 \delta (\gamma \delta-\alpha \beta)\right) \lambda^5 \\
+ R_i^3 R_o^3 \delta (\alpha^2-\beta^2-\gamma^2+\delta^2) \lambda^6 \Big)
\end{array}
\end{equation}
The adjugate matrix ${\pmb \Lambda}$ is expressed as a block matrix as follows
\begin{equation}
{\bf \Lambda} = 
\left(
\begin{array}{cc}
{\bf H} &  
- \mathcal{G}\left[ {\bf H} \right] 
\\
- \mathcal{G}\left[  \mathcal{F}\left[ {\bf H} \right] 
\right] 
  & \mathcal{F}\left[ {\bf H} \right] 
\end{array}
\right)
\end{equation}
where the operator $ \mathcal{G}\left[ \right] $
transforms $R_o \rightarrow R_i$ and $R_i \rightarrow R_o$
such that, for an arbitrary function $f(R_i,R_o)$,
\begin{equation}
\mathcal{G}\left[ f(R_i,R_o)\right] =
f(R_o,R_i)
\end{equation}
and 
the operator $ \mathcal{F}\left[ \right] $
transforms $I_n \rightarrow K_n $ and $K_n \rightarrow -I_n$
such that, for an arbitrary function $f(I_n(x),K_n(y))$,
\begin{equation}
\mathcal{F}\left[ f(I_n(x),K_n(y)) \right] =
f(K_n(x),-I_n(y) ) 
\end{equation}
Therefore
\begin{equation}
\mathcal{G}\left[ \alpha \right]=-\alpha; \quad
\mathcal{G}\left[ \beta \right]=-\beta; \quad
\mathcal{G}\left[ \gamma \right]=-\gamma; \quad
\mathcal{G}\left[ \delta \right]=-\delta; \quad
\end{equation}
and
\begin{equation}
\mathcal{F}\left[ \alpha \right]=\alpha; \quad
\mathcal{F}\left[ \beta \right]=\beta; \quad
\mathcal{F}\left[ \gamma \right]=\gamma; \quad
\mathcal{F}\left[ \delta \right]=\delta; \quad
\end{equation}

 \newpage
 
The entries of the matrix ${\bf H}$ read
\begin{equation}
\begin{array}{ll}
H_{1,1}=
\eta R_o \Big( - 2 n^2 \alpha K_n(\lambda R_o) \\
+ 2 n^2 \alpha \beta K_n(\lambda R_i) R_o \lambda \\
+ n^2 \alpha \delta K_n(\lambda R_i) R_i R_o \lambda^2 \\
- \delta (2 \alpha K_n'(\lambda R_i) R_i + K_n'(\lambda R_o) R_o) R_i R_o \lambda^3 \\
+ \delta (\alpha K_n(\lambda R_i) - \gamma K_n'(\lambda R_i)) R_i^3 R_o \lambda^4 \Big)
\\[20pt]
H_{1,2}=
\eta R_o^2 \Big( 2 n^3 \alpha^2 K_n(\lambda R_i) \\
+ n \alpha (n^2 \alpha K_n'(\lambda R_i) R_i - 2 K_n'(\lambda R_o) R_o) \lambda \\
- n (\alpha K_n(\lambda R_o) R_o^2 - \delta K_n(\lambda R_o) R_i R_o \\
\quad + 2 \alpha \gamma K_n'(\lambda R_i) R_i^2 + 2 \gamma K_n'(\lambda R_o) R_i R_o + \beta K_n'(\lambda R_o) R_o^2) \lambda^2 \\
- n \gamma (\gamma K_n'(\lambda R_i) - \alpha K_n(\lambda R_i)) R_i^3 \lambda^3 \Big)
\\[20pt]
H_{1,3}=
\dfrac{\eta}{\lambda} \Big( -2 n^2 (2 + n^2 \alpha^2) K_n(\lambda R_i) \\
- n^2 (2 + n^2 \alpha^2) K_n'(\lambda R_i) R_i \lambda \\
+ n^2 (2 \gamma^2 K_n(\lambda R_i) R_i^2 + 2 \beta^2 K_n(\lambda R_i) R_o^2 - \alpha K_n(\lambda R_o) R_o^2) \lambda^2 \\
+ n^2 ( \gamma^2 K_n'(\lambda R_i) R_i^2 + \beta^2 K_n'(\lambda R_i) R_o^2-\alpha \gamma K_n(\lambda R_i) R_i^2 ) R_i \lambda^3 \Big)
\\[20pt]
 H_{2,1}=
\eta R_o \Big( n^2 K_n(\lambda R_i) + 2 n^2 \alpha K_n(\lambda R_o) \\
+ 2 n^2 (\gamma K_n(\lambda R_o) R_i + \beta K_n(\lambda R_o) R_o - \alpha \beta K_n(\lambda R_i) R_o) \lambda \\
+ n^2 (
-2 \beta^2 K_n(\lambda R_i) R_o^2 
- \beta \gamma K_n(\lambda R_i) R_i R_o 
+ \delta K_n(\lambda R_o) R_i R_o)
 \lambda^2 \\
- (n^2 \beta \delta K_n(\lambda R_i) R_i R_o^2 - 2 \alpha \delta K_n'(\lambda R_i) R_i^2 R_o) \lambda^3 \\
+ (\delta^2 K_n(\lambda R_i) R_i^2 R_o^2 + \beta \delta K_n'(\lambda R_i) R_i^2 R_o^2 - \alpha \delta K_n(\lambda R_i) R_i^3 R_o + \delta K_n(\lambda R_o) R_i^3 R_o + \gamma \delta K_n'(\lambda R_i) R_i^3 R_o) \lambda^4 \Big)
\\[20pt]
 H_{2,2}=
\eta\, n R_o^2 \Big( n^2 \alpha (-2 \alpha K_n(\lambda R_i) + K_n(\lambda R_o)) \\
- \alpha (n^2 \gamma K_n(\lambda R_i) R_i + 2 n^2 \beta K_n(\lambda R_i) R_o - 2 K_n'(\lambda R_o) R_o) \lambda \\
+ (2 \alpha \gamma K_n'(\lambda R_i) R_i^2 - K_n(\lambda R_i) R_o^2 - \alpha K_n(\lambda R_o) R_o^2 + \beta K_n'(\lambda R_o) R_o^2 \\
\quad - n^2 \alpha \delta K_n(\lambda R_i) R_i R_o + \gamma K_n'(\lambda R_o) R_i R_o) \lambda^2 \\
+ ((\beta \gamma + \alpha \delta) K_n'(\lambda R_i) R_i^2 R_o - \alpha \gamma K_n(\lambda R_i) R_i^2 + \gamma K_n(\lambda R_o) R_i^2 \\
\quad + \gamma^2 K_n'(\lambda R_i) R_i^2 - \gamma K_n(\lambda R_o) R_o^2 + \delta K_n'(\lambda R_o) R_o^2) R_i \lambda^3 \Big)
\end{array}
\nonumber
\end{equation}

\begin{equation}
\begin{array}{l}
 H_{2,3}=
\dfrac{\eta n^2}{\lambda} \Big( -2 (2 + n^2 \alpha^2) K_n(\lambda R_i) \\
- (2 + n^2 \alpha^2) K_n'(\lambda R_i) R_i \lambda \\
+ (- K_n(\lambda R_i) R_i^2 + 2 \gamma^2 K_n(\lambda R_i) R_i^2 + 2 \beta^2 K_n(\lambda R_i) R_o^2 + \alpha K_n(\lambda R_o) R_o^2) \lambda^2 \\
+ (- \alpha \gamma K_n(\lambda R_i) R_i^3 - 2 \alpha \beta K_n(\lambda R_i) R_o^3 - \gamma K_n(\lambda R_o) R_i^3 + \gamma K_n(\lambda R_o) R_i R_o^2 \\
\quad + \gamma^2 K_n'(\lambda R_i) R_i^3 + \beta^2 K_n'(\lambda R_i) R_i R_o^2) \lambda^3 \Big)
\\[20pt]
 H_{3,1}=
2  \eta \, n R_o \Big( 2 n^2 \alpha^2 K_n(\lambda R_i) \\
+ \alpha (n^2 \alpha K_n'(\lambda R_i) R_i - 2 K_n'(\lambda R_o) R_o) \lambda \\
+ (-2 \alpha \gamma K_n'(\lambda R_i) R_i^2 - \beta K_n'(\lambda R_o) R_o^2 + \delta K_n(\lambda R_o) R_i R_o - 2 \gamma K_n'(\lambda R_o) R_i R_o) \lambda^2 \\
+ (\alpha^2 - \gamma^2) K_n'(\lambda R_i) R_i^3 \lambda^3 \Big)
\\[20pt]
 H_{3,2}=
\eta R_o^2 \Big( -2 n^2 (1 + n^2 \alpha^2) K_n(\lambda R_i) - 4 n^2 \alpha K_n(\lambda R_o) \\
+ n^2 (4 \alpha \beta K_n(\lambda R_i) R_o - (2 + n^2 \alpha^2) K_n'(\lambda R_i) R_i) \lambda \\
+ 2 n^2 (\gamma^2 K_n(\lambda R_i) R_i^2 + \beta \gamma K_n(\lambda R_i) R_i R_o + (-\alpha^2 + \beta^2) K_n(\lambda R_i) R_o^2) \lambda^2 \\
+ ((-1 + n^2 (-\alpha^2 + \gamma^2)) K_n'(\lambda R_i) R_i^3 - 4 \alpha \delta K_n'(\lambda R_i) R_i^2 R_o \\
\quad + (-1 + n^2 (-\alpha^2 + \beta^2)) K_n'(\lambda R_i) R_i R_o^2) \lambda^3 \Big)
\\[20pt]
 H_{3,3}=
\dfrac{\eta \,n}{\lambda} \Big( 2 n^2 (2 + n^2 \alpha^2) K_n(\lambda R_i) \\
+ n^2 (2 + n^2 \alpha^2) K_n'(\lambda R_i) R_i \lambda \\
+ 2 n^2 (-\gamma^2 K_n(\lambda R_i) R_i^2 + (\alpha^2 - \beta^2) K_n(\lambda R_i) R_o^2) \lambda^2 \\
+ ((1 + n^2 (\alpha^2 - \gamma^2)) K_n'(\lambda R_i) R_i^3 - 2 \alpha K_n'(\lambda R_o) R_o^3 \\
\quad + (1 + n^2 (\alpha^2 - \beta^2)) K_n'(\lambda R_i) R_i R_o^2 - 2 \delta K_n'(\lambda R_o) R_i R_o^2) \lambda^3 \Big)
\end{array}
\end{equation}

}
\section{Useful geometric relations for the cylindrical coordinate system}
\label{app_useful_geo}
The metric tensor of the cylindrical coordinate system at the point ${\pmb x}$ is obtained by using eqs. (\ref{eq20}) and the expression \cite{synge_tensor_1978}
\begin{equation}
g_{a\, b}({\pmb x})= \dfrac{\partial \overline{x}_i}{\partial x^a}\, \dfrac{\partial \overline{x}_i}{\partial x^b}
\label{eqA1}
\end{equation}
From which
\begin{equation}
g_{1\, 1}({\pmb x})= 1 ; \quad g_{2\, 2}({\pmb x})= (x^1)^2 ; \quad g_{3\, 3}({\pmb x})= 1 ;\quad g_{a\,b}({\pmb x}) = 0\quad \text{for}\quad a \neq b
\label{eqA2}
\end{equation}
the contravariant metric tensor $ g^{a\, b}({\pmb x})$, defined by the relation $g^{a\, c}({\pmb x})\, g_{c\, b}({\pmb x})= \delta^a_b$ reads

\begin{equation}
g^{1\, 1}({\pmb x})= 1 ; \quad g^{2\, 2}({\pmb x})= \dfrac{1}{ (x^1)^2} ; \quad g^{3\, 3}({\pmb x})= 1 ;\quad g^{a\,b}({\pmb x}) = 0\quad \text{for}\quad a \neq b
\label{eqA3}
\end{equation}
Similarly, at the point ${\pmb \xi}$, the metric tensor of the cylindrical coordinate system is
\begin{equation}
g_{1\, 1}({\pmb \xi})= 1 ; \quad g_{2\, 2}({\pmb \xi})= (\xi^1)^2 ; \quad g_{3\, 3}({\pmb \xi})= 1 ;\quad g_{\alpha\,\beta}({\pmb \xi}) = 0\quad \text{for} \quad \alpha \neq \beta
\label{eqA4}
\end{equation}
and
\begin{equation}
g^{1\, 1}({\pmb \xi})= 1 ; \quad g^{2\, 2}({\pmb \xi})= \dfrac{1}{ (\xi^1)^2} ; \quad g^{3\, 3}({\pmb \xi})= 1 ;\quad g^{\alpha \,\beta}({\pmb \xi}) = 0\quad \text{for}\quad \alpha \neq \beta
\label{eqA5}
\end{equation}
The
non vanishing entries of the
Christoffel symbols $\Gamma^a_{b \,c}({\pmb x})$ of second kind at the point ${\pmb x}$
are
\begin{equation}
\Gamma^{1}_{2\, 2}({\pmb x})\,=\, -x^1; \quad  \Gamma^{2}_{1\, 2}({\pmb x})\,=\,\Gamma^{2}_{2\, 1}({\pmb x})\, =\, \dfrac{1}{x^1}
\end{equation}
and of the Christoffel symbols $\Gamma^\alpha_{\beta\, \gamma}({\pmb \xi})$
at the point ${\pmb \xi}$
\begin{equation}
\Gamma^{1}_{2\, 2}({\pmb \xi})\,=\, -\xi^1; \quad  \Gamma^{2}_{1\, 2}({\pmb \xi})\,=\,\Gamma^{2}_{2\, 1}({\pmb \xi})\, =\, \dfrac{1}{\xi^1}
\label{eqA7ch}
\end{equation}
The transformation matrix $\dfrac{\partial \overline{x}_i}{\partial x^a}$,
that transforms the contravariant entries of a vector ${\pmb v}({\pmb x})$ from the cylindrical coordinates into the Cartesian coordinates via
\begin{equation}
\overline{v}_i({\pmb x})= \dfrac{\partial \overline{x}_i}{\partial x^a} \, v^a({\pmb x})
\label{eqA7}
\end{equation}
admits entries 
\begin{equation} 
\begin{array}{l}
\dfrac{\partial \overline{x}_1}{\partial x^1}= \cos{(x^2)}; \quad
\dfrac{\partial \overline{x}_1}{\partial x^2}= - x^1 \sin{(x^2)}; \quad
\dfrac{\partial \overline{x}_2}{\partial x^1}= \sin{(x^2)}; \quad
\dfrac{\partial \overline{x}_2}{\partial x^2}=  x^1 \cos{(x^2)}; 
\\
[15 pt]
\dfrac{\partial \overline{x}_3}{\partial x^3}= 1;
\quad 
\dfrac{\partial \overline{x}_1}{\partial x^3}= 
\dfrac{\partial \overline{x}_3}{\partial x^1}= 
\dfrac{\partial \overline{x}_2}{\partial x^3}= 
\dfrac{\partial \overline{x}_3}{\partial x^2}= 0
\end{array}
\label{eqA8}
\end{equation}
Its inverse matrix $\dfrac{\partial x^a}{\partial \overline{x}_i}$ is defined componentwise
\begin{equation}
\begin{array}{l}
\dfrac{\partial x^1}{\partial \overline{x}_1}= \cos{(x^2)} ; \quad
\dfrac{\partial x^1}{\partial \overline{x}_2}= \sin{(x^2)}; \quad
\dfrac{\partial x^2}{\partial \overline{x}_1}= - \dfrac{ \sin{(x^2)} }{x^1}; \quad
\dfrac{\partial x^2}{\partial \overline{x}_2}= \dfrac{ \cos{(x^2)} }{x^1} ;
\hspace{1cm} 
\\
[15 pt]
\dfrac{\partial x^3}{\partial \overline{x}_3}= 1;
\quad 
\dfrac{\partial x^1}{\partial \overline{x}_3}=  
\dfrac{\partial x^3}{\partial \overline{x}_1}= 
\dfrac{\partial x^2}{\partial \overline{x}_3}= 
\dfrac{\partial x^3}{\partial \overline{x}_2}= 0
\end{array}
\label{eqA9}
\end{equation}
Similarly, at the point ${\pmb \xi}$, the transformation matrix 
$ \dfrac{\partial \overline{\xi}_i}{\partial \xi^a}$
\begin{equation} 
\begin{array}{l}
\dfrac{\partial \overline{\xi}_1}{\partial \xi^1}= \cos{(\xi^2)}; \quad
\dfrac{\partial \overline{\xi}_1}{\partial \xi^2}= - \xi^1 \sin{(\xi^2)}; \quad
\dfrac{\partial \overline{\xi}_2}{\partial \xi^1}= \sin{(\xi^2)}; \quad
\dfrac{\partial \overline{\xi}_2}{\partial \xi^2}=  \xi^1 \cos{(\xi^2)}; 
\\
[15 pt]
\dfrac{\partial \overline{\xi}_3}{\partial \xi^3}= 1;
\quad 
\dfrac{\partial \overline{\xi}_1}{\partial \xi^3}= 
\dfrac{\partial \overline{\xi}_3}{\partial \xi^1}= 
\dfrac{\partial \overline{\xi}_2}{\partial \xi^3}= 
\dfrac{\partial \overline{\xi}_3}{\partial \xi^2}= 0
\end{array}
\label{eqA10}
\end{equation}
and the inverse matrix $ \dfrac{\partial \xi^a}{\partial \overline{\xi}_i}$
\begin{equation}
\begin{array}{l}
\dfrac{\partial \xi^1}{\partial \overline{\xi}_1}= \cos{(\xi^2)} ; \quad
\dfrac{\partial \xi^1}{\partial \overline{\xi}_2}= \sin{(\xi^2)}; \quad
\dfrac{\partial \xi^2}{\partial \overline{\xi}_1}= - \dfrac{ \sin{(\xi^2)} }{\xi^1}; \quad
\dfrac{\partial \xi^2}{\partial \overline{\xi}_2}= \dfrac{ \cos{(\xi^2)} }{\xi^1} ;
\hspace{1cm} 
\\
[15 pt]
\dfrac{\partial \xi^1}{\partial \overline{\xi}_1}= 1;
\quad 
\dfrac{\partial \xi^1}{\partial \overline{\xi}_3}=  
\dfrac{\partial \xi^3}{\partial \overline{\xi}_1}= 
\dfrac{\partial \xi^2}{\partial \overline{\xi}_3}= 
\dfrac{\partial \xi^3}{\partial \overline{\xi}_2}= 0
\end{array}
\label{eqA11}
\end{equation}

\bibliography{biblio.bib}

\end{document}